\newif\iftwocolumn\twocolumnfalse
\newif\iffootinbib
\iftwocolumn\footinbibtrue\else\footinbibfalse\fi
\PassOptionsToPackage{colorlinks,
  linkcolor = blue!60!black,
  urlcolor = blue!60!black,
  citecolor = blue!60!black, unicode, hyperfootnotes=false}{hyperref}
\documentclass[aps,prd,11pt,\iftwocolumn two\else one\fi column,groupedaddress,unsortedaddress, superscriptaddress,tightenlines,\iffootinbib\else no\fi footinbib,longbibliography,preprintnumbers,floatfix,amssymb,amsmath]{revtex4-2}%
\ifx\texorpdfstring\undefined\newcommand\texorpdfstring[2]{#1}\fi
\ifx\pdfminorversion\undefined\else\pdfminorversion=7\fi

\usepackage{textgreek}
\usepackage{mathtools}
\allowdisplaybreaks

\usepackage{subfigure} 
\ifnum\paperheight=0\relax\paperheight=11in\relax\fi
\usepackage{graphicx,xcolor,cancel,slashed}
\graphicspath{{FIGS/}}
\usepackage{multirow}

\makeatletter
\newcommand\showtitleinbib{{\escapechar=`\\ \immediate\write\@auxout{%
\csname citation{REVTEX42Control}\endcsname^^J%
\csname citation{apsrev42Control}\endcsname
}}}
\makeatother

\newcommand\Dslash{\slashed D}
\newcommand\pslash{\slashed p}

\newcommand{\Tr}{\mathop{\rm Tr}\nolimits}

\newcommand{\CPV}{$\cancel{\text{CP}}$}
\newcommand\slashnext[1]{{\setbox0\hbox{\(#1\)}\setbox1\hbox to \wd0{\hss/\hss}\rlap{\box1}{\box0}}}
\newcommand\Aslash{{\slashnext A}}

\makeatletter
\ifx\@bibitemShut\undefined\let\@bibitemShut\relax\fi
\makeatother

\usepackage[autogenerated]{ucs}
\usepackage{orcidlink}
\usepackage[capitalize]{cleveref}
\begin{document}


\title{Quark Chromo-Electric Dipole Moment Operator on the Lattice}
\date{\today}
\author{Tanmoy Bhattacharya\,\orcidlink{0000-0002-1060-652X}}
\email{tanmoy@lanl.gov}
\affiliation{Group T-2, Los Alamos National Laboratory, Los Alamos, NM 87545}
\author{Vincenzo Cirigliano\,\orcidlink{0000-0002-9056-754X}}
\email{cirigv@uw.edu}
\affiliation{Department of Physics, University of Washington,
Seattle, WA 98195-1560}
\author{Rajan Gupta\,\orcidlink{0000-0003-1784-3058}}
\email{rg@lanl.gov}
\affiliation{Group T-2, Los Alamos National Laboratory, Los Alamos, NM 87545}
\author{Emanuele Mereghetti\,\orcidlink{0000-0002-8623-5796}}
\email{emereghetti@lanl.gov}
\affiliation{Group T-2, Los Alamos National Laboratory, Los Alamos, NM 87545}
\author{Jun-Sik Yoo\,\orcidlink{0000-0001-6272-4790}}
\email{junsik@lanl.gov}
\affiliation{Group T-2, Los Alamos National Laboratory, Los Alamos, NM 87545}
\author{Boram Yoon\,\orcidlink{0000-0003-0069-6514}}
\email{byoon@nvidia.com}
\affiliation{NVIDIA Corporation, Santa Clara, CA 95050, USA}
\preprint{LA-UR-21-21913,  INT-PUB-23-013}
\begin{abstract}
We present a lattice QCD study of the contribution of the isovector quark chromo-electric dipole moment (qcEDM) operator to the nucleon electric dipole moments (nEDM). The calculation was carried out on four 2+1+1-flavor of highly improved staggered quark (HISQ) ensembles using Wilson-clover quarks to construct correlation functions. This clover-on-HISQ formulation is not fully $O(a)$ improved, and gives rise to  additional systematics over and above those due to removing excited state contributions to getting ground-state matrix elements, and the final chiral and continuum extrapolations to get the physical result.  We use the non-singlet axial Ward identity including corrections up to \(O(a)\) to show how to control the power-divergent mixing of the isovector qcEDM  operator with the lower dimensional pseudoscalar operator. The residual corrections are observed to give rise to $O(25\%)$ violations in relations arising from the axial Ward identity. We devise three  methods attempting to control the resulting uncertainty in the CP violating form factor; each of these, however, can have large $O(a^2)$ corrections.  Preliminary results for the nEDM due to qcEDM are presented choosing the method giving the most uniform behavior. 
\end{abstract}
\maketitle

\iffootinbib
\newcommand\punctfootnote[2]{\footnote{#2}#1}
\let\origfootnote\footnote\def\footnote{~\origfootnote}
\else
\newcommand\punctfootnote[2]{{#1\edef\box{\ifnum`.=`#1\relax#1\else\ifnum`,=`#1\relax#1\fi\fi}\setbox0\hbox{\box}\toks0={\footnote{#2}}\edef\tmp{\kern-\the\wd0\relax\the\toks0\spacefactor=\the\spacefactor\relax}\expandafter}\tmp}
\fi


\section{Introduction}
\label{sec:intro}

The observation of permanent electric dipole moments (EDMs) in nondegenerate systems
requires the simultaneous breaking of  parity (P) and time reversal (T), or,
equivalently, the combination of charge
conjugation and parity (CP)~\cite{Luders:1954zz}.  Given the smallness of CP-violating (\CPV) contributions induced by quark mixing described by the Cabibbo-Kobayashi-Maskawa (CKM) matrix in the Standard Model (SM)~\cite{Kobayashi:1973fv}, CP-violation in nucleon, nuclear, and atomic/molecular~\cite{Abel:2020gbr,Andreev:2018ayy,Cairncross:2017fip,Graner:2016ses,Afach:2015sja,Baron:2013eja,Baker:2006ts} systems
provide very strong constraints on the SM $\Theta$-term (currently constrained at
the level of $\Theta \sim 10^{-10}$) and  new sources of CP
violation arising from physics beyond the Standard Model (BSM)~\cite{Pospelov:2005pr,Chupp:2017rkp}.

New \CPV\ interactions are ubiquitous 
in BSM models and may play a key role  in relatively low-scale baryogenesis mechanisms, 
such as electroweak baryogenesis (see \cite{Morrissey:2012db} and references
therein).  
Probing
them through hadronic electric dipole moments (EDMs), however,  
requires including possible large corrections due to the strong interactions 
between quark and gluon fields. These are analyzed  using low-energy effective operators and require nonperturbative treatment.  For hadronic systems such as the neutron,  lattice QCD (LQCD) 
has emerged as the tool of choice to compute the contribution of these  \CPV\ operators to the EDMs. 
Furthermore, it has been shown quantitatively that improving the precision of these 
hadronic matrix elements will drastically improve the constraints that EDMs provide on \CPV\ BSM interactions of the Higgs particle~\cite{Chien:2015xha,Cirigliano:2016nyn,Cirigliano:2019vfc}.

The outline of this paper is as follows.  In \cref{sec:BAU}, we review the need for new sources of \CPV. The parameterization of 
\CPV\ operators at the hadronic scale using effective field theory methods is presented in \cref{sec:operators}. In this study, we only 
calculate the EDMs of the neutron (nEDM) and proton (pEDM) induced by the 
isovector component of the quark chromo-electric dipole moment (qcEDM). 
 The decomposition of the matrix elements of \CPV\ operators within ground state nucleons into 
 vector form-factors of the nucleons, and the phase conventions are 
 given in \cref{sec:FF}.
In \cref{sec:SSM}, we describe the method for calculating these form factors on the lattice. The use of the axial Ward identity to control the power law divergence due to mixing with the lower dimension pseudoscalar operator is described in \cref{sec:AWI} and some 
preliminary numerical results are presented in \cref{sec:Num}. In \cref{sec:mixing}, 
we discuss the multiplicative renormalization of the qcEDM operator and the connection to the \(\overline{\rm MS}\) scheme. Our conclusions are given in \cref{sec:conclusions}.
Four appendices present technical details: \Cref{sec:AWI_app} discusses the nonsinglet Ward Identity, \cref{sec:operatorbasis} provides a complete list of dimension-5 CPV operators, \cref{sec:KX1} details the nonperturbative procedure for the extraction of the coefficient, \(K_{X1}\), needed to control the power-divergent mixing of the qcEDM and pseudoscalar  operators, and \cref{sec:alphadet} gives the chiral perturbation theory determination of the chiral phase \(\alpha_N\) for the nucleon.

\subsection{Baryogenesis and the need for new sources of  CP violation}
\label{sec:BAU}

The observed universe has \(6.1^{+0.3}_{-0.2}\times 10^{-10}\)
baryons for every black body photon~\cite{Bennett:2003bz}, whereas in a baryon
symmetric universe, we expect no more that about \(10^{-20}\) baryons and antibaryons
for every photon~\cite{Kolb:1990vq}.  It is difficult to include such a
large excess of baryons as an initial condition in an inflationary
cosmological scenario~\cite{Coppi:2004za}.  The way out of the impasse
lies in generating the baryon excess (baryogenesis) dynamically during the evolution
of the universe.  

In the early history of the universe, if the matter-antimatter
asymmetry was generated post inflation and reheating, then one has to satisfy 
Sakharov's three necessary conditions~\cite{Sakharov:1967dj}: the
process has to violate baryon number, evolution has to occur out of
equilibrium, and CP (or, equivalently, time reversal invariance if CPT
remains unbroken) has to be violated.

To probe sources of \CPV, a very
promising approach is to search for static EDMs of
elementary particles, atoms and nondegenerate states of molecules, all of which are necessarily
proportional to their spin. Since under time-reversal, the direction of
spin reverses but the electric dipole moment does not, a nonzero
measurement would imply T, or equivalently CP, violation.  Of the
elementary particles, atoms and nuclei that are being investigated,
nEDM and pEDM are the cleanest to analyze using lattice QCD. 

CP violation exists in the electroweak sector of the SM of particle
interactions due to a phase in the CKM quark
mixing matrix~\cite{Kobayashi:1973fv}, and possibly by a similar phase
in the leptonic sector~\cite{Maki:1962mu,Nunokawa:2007qh}, given that the neutrinos have mass and mix.  The contribution of the \CPV\ phase in the CKM quark mixing matrix~\cite{Kobayashi:1973fv} to nEDM is $O(10^{-32})$ e-cm~\cite{Seng:2014lea}, 
much smaller than the current experimental bound $d_n < 1.8 \times
10^{-26}$ e-cm (90\% CL)~\cite{Abel:2020gbr}. This \CPV\ is too small to explain baryogenesis~\cite{Shaposhnikov:1987tw,Farrar:1993sp,Gavela:1993ts,Gavela:1994dt,Gavela:1994ds,Huet:1994jb}. 
Similarly, \CPV\ due to a possible topological term~\cite{Dolgov:1991fr} is unlikely to lead to appreciable baryon asymmetry~\cite{Gross:1980br}. For  baryogenesis, BSM \CPV\ would, therefore, need to have played a major role.
Most extensions of the SM have new sources of \CPV. 
Each of these contributes to the nEDM and for some models it can be as
large as $ 10^{-26}$ e-cm. Planned experiments  aim to reach 
a sensitivity of  $d_n \sim 3 \times 10^{-28}$ e-cm~\cite{Chupp:2017rkp}.

In order to connect the reduction in the upper bounds or actual values from EDM searches 
to new sources of \CPV\ and models of baryogenesis, robust calculations of the  hadronic EDMs induced by low-energy effective quark 
and gluon operators are needed. 
Lattice QCD offers the most promising method with 
control over all uncertainties to provide the matrix elements of novel \CPV\ operators between nucleon states that are needed to connect the experimental bound (or value!) of the EDMs 
to the CP violating couplings in a given BSM theory. Here, we present 
the calculation of the isovector part of the qcEDM operator.

\subsection{CP violation at low energy up to dimension-five}
\label{sec:operators}

At the hadronic scale (${}\lesssim 2$~GeV), the effects of BSM theories that involve heavy degrees of freedom at mass scales greater than the weak scale,   $\Lambda > M_W$, 
can be described in terms of effective
local operators composed of quarks and gluons.  Using 
effective field theory techniques, one can organize all
such \CPV\ interactions
based on symmetry and dimension. In
general, operators with higher dimension are suppressed by increasing inverse 
powers of $\Lambda$. The couplings
associated with these low-energy operators encode information about the BSM model
at $\Lambda \sim$~TeV scale with the renormalization group providing their evolution from $\Lambda$
to the hadronic scale. 
The nEDM induced by any \CPV\ interaction 
can be obtained from the \CPV\
form factor $F_3$ of the electromagnetic current, $J_\mu^{\rm EM}$, within the nucleon state, 
as discussed below. 

At dimension five and lower, only three \CPV\
local operators arise:
\begin{align}
&{\cal L}_{\rm QCD}
    \mathbin{{\longrightarrow}} {\cal L}_{\rm
      QCD}^{\cancel{\text{CP}}} = {\cal L}_{\rm QCD} 
+ \frac{i}{32 \pi^2} \Theta\ 
    G_{\mu\nu} {\tilde G_{\mu\nu}} \nonumber\\
 \quad\span\!{} - \left. \frac{i}{2} \sum_q d_q \  \overline{q} \sigma^{\mu\nu} {\tilde F_{\mu\nu}} q \right. 
- \left. \frac{i}{2} \sum_q d_q \  \overline{q} \sigma^{\mu\nu} {\tilde G_{\mu\nu}} q  \right.\,,
\label{eq:Lcpv}
\end{align}
where $\tilde F^{\mu \nu} = \epsilon^{\mu \nu \alpha \beta} F_{\alpha \beta}/2$ is the dual of
the electromagnetic field-strength tensor, 
$\tilde G^{\mu \nu} = \epsilon^{\mu \nu \alpha \beta} G_{\alpha \beta}/2$ is the dual
of the QCD field-strength tensor,
and $\sigma_{\mu \nu} = (i/2) [\gamma_\mu, \gamma_\nu]$\punctfootnote.{We use the Euclidean
notation throughout, and normalize the kinetic term for the gauge fields
as \(F_{\mu\nu} F^{\mu\nu}/4e^2\) and \(G_{\mu\nu}^aG^{a\mu\nu}/4g^2\). Please refer to our earlier 
work~\cite[Appendix A]{Bhattacharya:2021lol} for further details of the convention choice. In particular,
\(\sigma_{\mu \nu} \gamma_5 G^{a\mu\nu} = - \sigma_{\mu\nu} \tilde G^{a\mu \nu}\) and the Lagrangian
of \cref{eq:Lcpv} corresponds to
\protect\begin{align*}
  {\cal L}_{\rm QCD} 
&- \frac{g^2}{32 \pi^2} \Theta\ 
    G_{\mu\nu} {\tilde G_{\mu\nu}} \\
 &- \left. i\frac{e}{2} \sum_q d_q \  \overline{q} \sigma^{\mu\nu} \gamma_5 {F_{\mu\nu}} q \right. \\
&- \left.  i\frac{g}{2}  \sum_q \ \tilde{d}_q \ \overline{q} \sigma^{\mu\nu} \gamma_5 G_{\mu\nu}
      q  \right.\,,
\protect\end{align*}
in Minkowski space with the conventional normalization for the gauge fields.}
These three \CPV\ opertors are the $d=4$ $\Theta$-term and the $d=5$  quark EDM (qEDM) and the qcEDM with dimensionful coefficients 
$d_q$ and $\tilde d_q$, respectively. Recent work on the dimension six gluonic operator (the Weinberg operator)~\cite{Weinberg:1989dx} can be found in Refs.~\citep{Cirigliano:2020msr,Rizik:2020naq}, while there has been less work done on the \CPV\ four-fermion operators~\cite{Grzadkowski:2010es,Buhler:2023gsg}.

To the lowest order, the calculation of the qEDM is special; it reduces to the calculation of the flavor diagonal tensor
charges of the neutron.  The methodology for this
calculation, including disconnected contributions from
up, down, strange, and charm  quark
loops, is mature and first lattice results obtained by us are
given in Ref.~\citep{Bhattacharya:2015wna} and phenomenological consequences
for a particular BSM theory (Split SUSY) were analyzed
in Ref.~\citep{Bhattacharya:2015esa}. These results were 
updated in Ref.~\citep{Gupta:2018lvp}, and the status of various lattice calculations are reviewed by the 
Flavor Lattice Averaging Group (FLAG) in Refs.~\citep{Aoki:2019cca,Aoki:2021kgd}. 

The calculation of nEDM induced by the $\Theta$-term requires the matrix elements of the product of the gluonic operator with the $J_\mu^{\rm EM}$ within the ground state of the  nucleon. While the computation is only slightly more expensive than of the 3-point function with just $J_\mu^{\rm EM}$, the calculation is still not 
under control due to both statistical errors and lack of a 
clear methodology 
to fully remove excited state contributions (ESC), 
especially from the low-lying tower of nucleon-pion states. 
Recent progress has been reported in 
Refs.~\citep{Dragos:2019oxn,Syritsyn:2019vvt,Alexandrou:2020mds,Bhattacharya:2021lol,Liang:2023jfj}.

Note that because of the anomaly in the axial Ward identity, the
$\Theta$-term can be rotated into a pseudoscalar mass term $i
m_\ast({\Theta}) \sum_q \overline{q} \gamma_5 q$ under a chiral
transformation~\cite{Crewther:1979pi}, and conversely, any phase arising in
the determinant of the quark mass-matrix can be traded for \(\Theta\). Since the nonanomalous
axial Ward identities allow us to remove the rest of the phases in the mass-matrix, we will, henceforth, treat all
quark masses as real.
The $\Theta$-term is part
of the SM, but is usually neglected under the assumption that some
form of a Peccei-Quinn mechanism that promotes $\Theta$ to a dynamical field relaxes the minimum of its
effective action to $\Theta = 0$~\cite{Peccei:1977hh} in the absence of other \CPV\ sources
in the action. It is, however, important to note that in the presence of other \CPV\ operators from BSM,  the minimum, called \(\Theta_{\rm induced}\)~\cite{Bigi:1990kz}, of this effective potential is, in general, shifted by the disconnected contributions, i.e., those in which the quark fields in the operator are contracted to form a quark loop. 

Furthermore, the contribution to the minimum, induced by the qEDM and qcEDM operators vanishes for the isovector combination in an isospin-symmetric theory. We, therefore, do not consider this effect here as we present only the connected contributions of the qcEDM operator.

Calculations of even the bare qcEDM operator are  the most challenging computationally of the three $D \le 5$ operators. Furthermore, to get finite results in the continuum limit, one must resolve its divergent 
mixing with the $D=3$ pseudoscalar operator $i \sum_q \bar q \gamma_5 q$ as
discussed in Refs.~\citep{Bhattacharya:2015rsa,Rizik:2020naq,Kim:2021qae}.
Our analysis starts with the  Hermitian flavor-diagonal  and isovector qcEDM operators defined as
\begin{align}
    C^{(q)} &\equiv     - \frac{i}{2}   \, \overline{q} \sigma_{\mu\nu}   \tilde G_{\mu\nu} q \ = \  
    \frac{i}{2}  \, \overline{q} \sigma_{\mu\nu}   \gamma_5   G_{\mu\nu} q
    \nonumber \\
    C^{(a)} &\equiv i \, \overline{\psi} \sigma_{\mu\nu}   \gamma_5   G_{\mu\nu}  T^a \psi~,
    \label{eq:qcEDMdef}
\end{align}
where $q$ denotes the quark field of a given flavor while 
$\psi$ denotes the $SU(N_f)$ multiplet  in the fundamental 
representation and $T^a$ represents the generic hermitian $SU(N_f)$ generator,
normalized as \(\Tr [(T^a)^2] = \frac12\). 

To the lowest order in \(\tilde d_q\),  the nEDM induced by the qcEDM operator 
requires calculating
\begin{align}
\langle n \mid J^{\rm EM}_\mu \mid n \rangle  \Big|_{\not{{\rm CP}}}^{\rm qcEDM} \!\!\!\!\! &=
       \langle n | J^{\rm EM}_\mu \  \times {}\nonumber\\
       \quad\span\int d^4 x\ (- \frac{i}{2})
       \sum_{q\in u,d,s}  \tilde d_q \,\bar q \sigma_{\alpha \beta} q\ 
         {\tilde G}^{\alpha\beta} | n \rangle\,,
\label{eq:chromoEDM}
\end{align}
where  effects of the heavier quarks are ignored. This is a 4-point function--the volume integral of 
the qcEDM operator correlated with  the electromagnetic current inserted on each time slice between the nucleon source and sink. This can be calculated in two ways using lattice QCD: directly as a 4-point function~\cite{Abramczyk:2017oxr} or using the Schwinger source method discussed  
in~\cite{Bhattacharya:2016oqm,Gupta:2017anz}. Here, we continue to develop the latter.

\begin{figure}[b]   
  \subfigure{
    \includegraphics[width=0.25\linewidth]{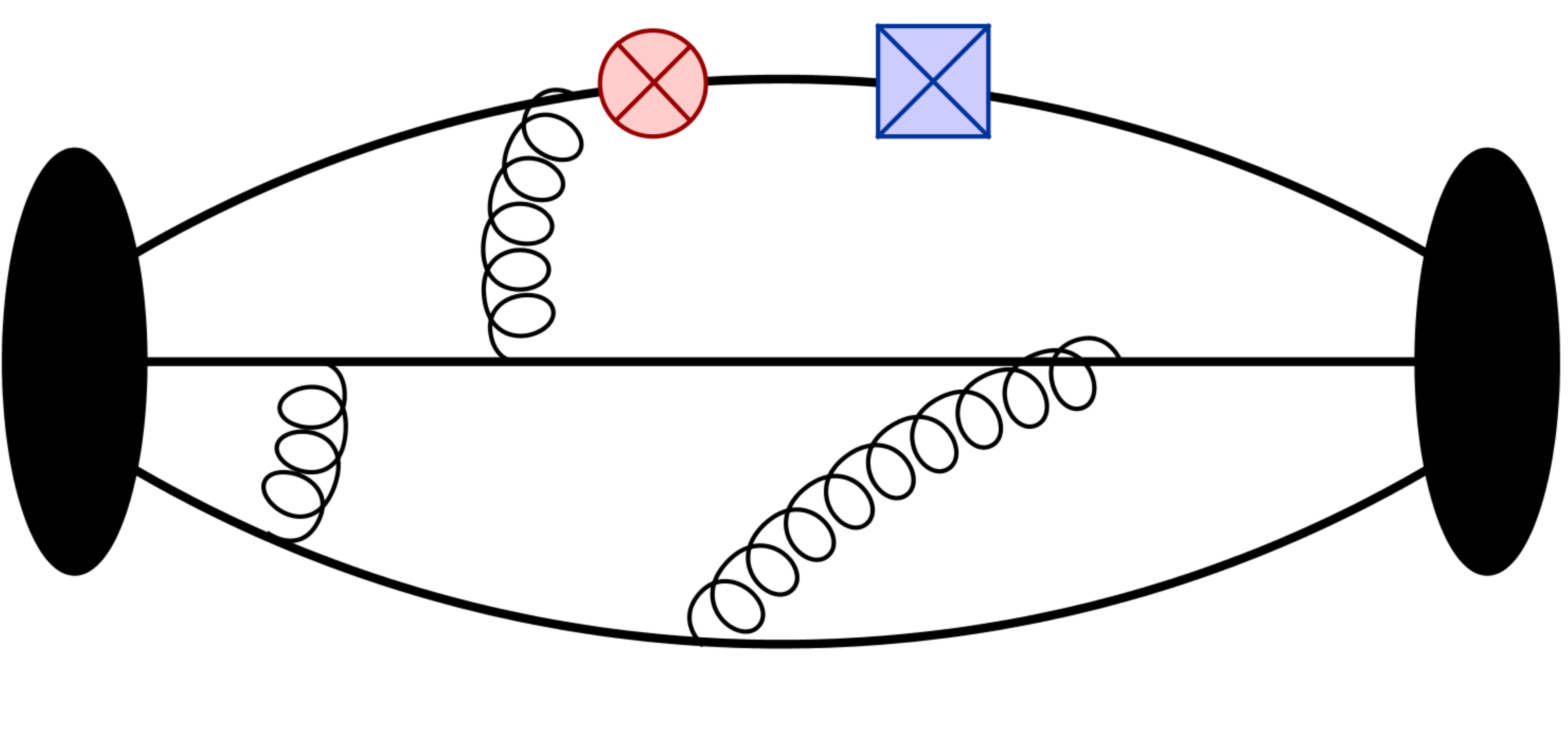}
  }
  \hspace{0.04\linewidth}
  \subfigure{
    \includegraphics[width=0.25\linewidth]{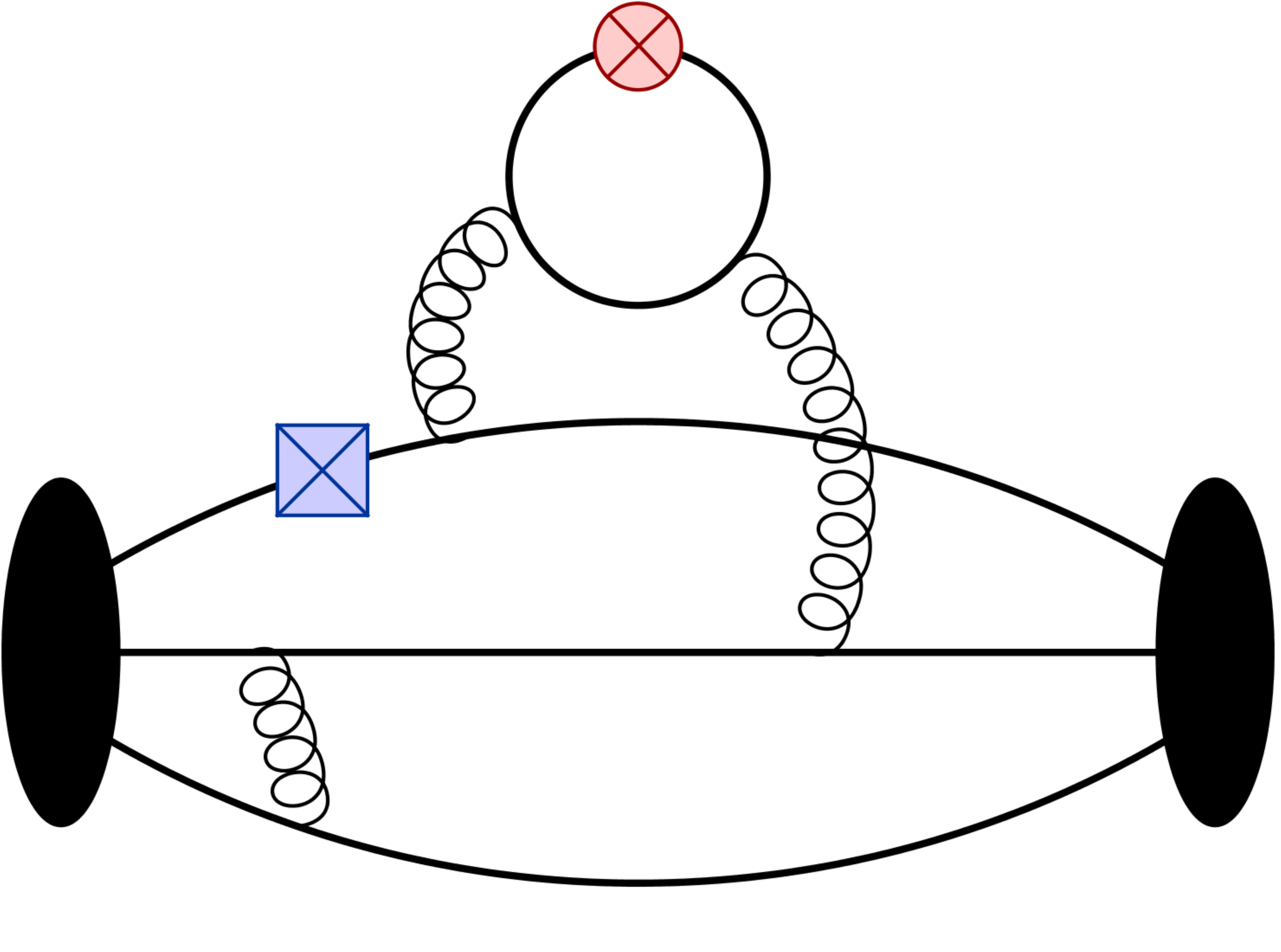}
  }
  \hspace{0.04\linewidth}
  \subfigure{
    \includegraphics[width=0.25\linewidth]{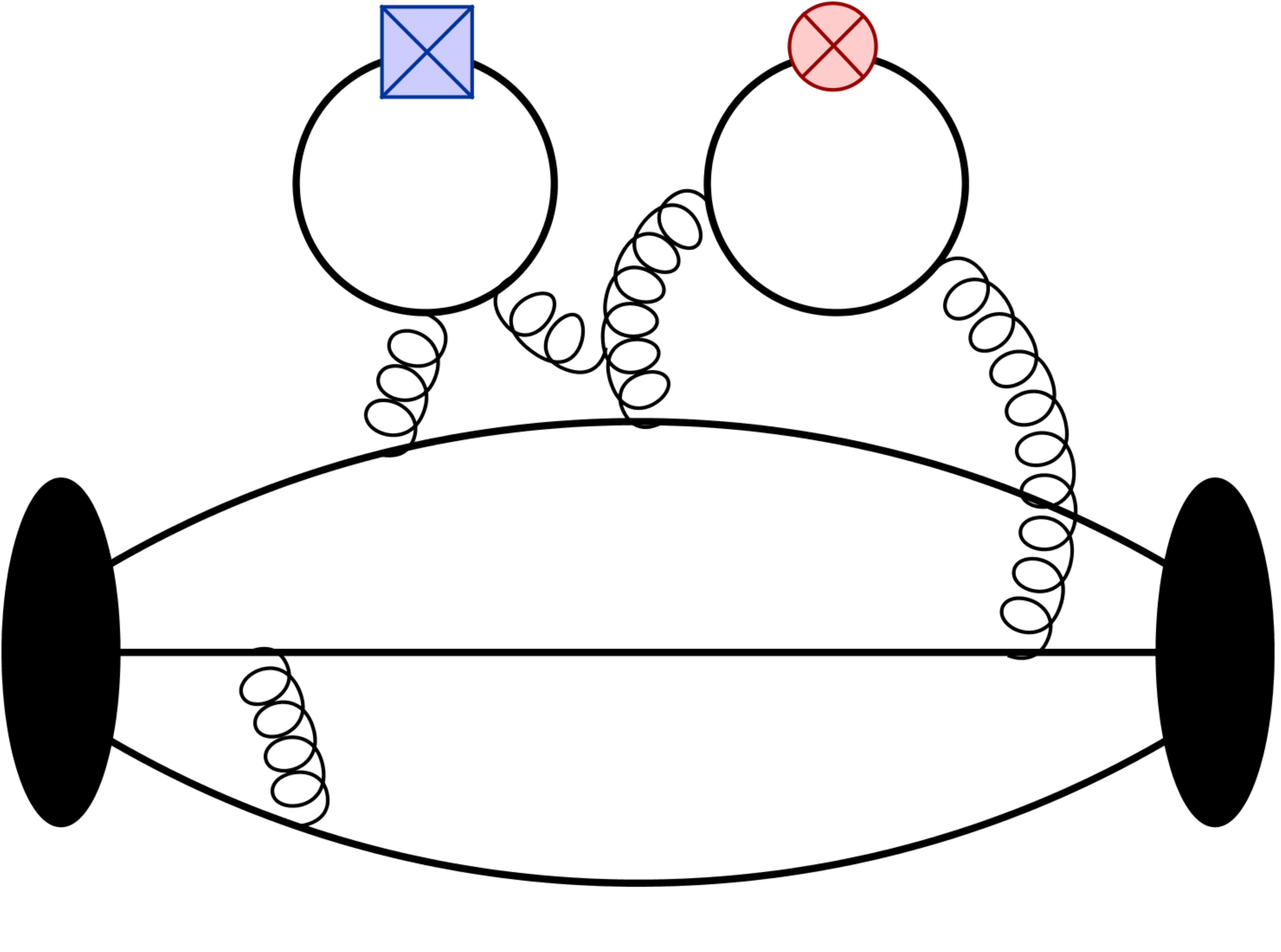}
  }
\caption{Illustration of one of the connected quark line diagram contributing to
  the qcEDM (left), the one loop disconnected diagram
  (middle) and the two loop disconnected diagram (right). The circle
  indicates the insertion of the electromagnetic current, and the
  square the qcEDM operator. 
\label{fig:qChromo}}
\end{figure}

In the Schwinger source method, the qcEDM operator, a bilinear in the quark fields, is added as a source term to the QCD action. Correlation functions with its insertion can then be calculated by taking derivatives with respect to its coupling  \(\tilde d_q\). We divide this calculation into two steps: 
first, regular, $P$, and modified,  $P_\epsilon$, propagators are calculated by inverting the Dirac operator without and with the qcEDM term. Second, these two propagators are used
to construct the three-point function with the insertion of the vector current
between nucleon states, i.e., the quark-line diagrams in~\cref{fig:qChromo,fig:qcEDMfull}.

Since the modified propagator inserts 
arbitrary powers of the qcEDM operator, one gets 
uncontrolled divergences if the continuum limit is taken holding \(\tilde d_q\) fixed.
We, therefore, scale \(\tilde d_q\) appropriately to keep the contribution  
in the linear regime as we take the continuum limit.  For convenience, we will express most quantities in terms of the dimensionless ratios\looseness-1
\begin{equation}
  \epsilon_{{q}} \equiv -\frac{2 \tilde d_q}{a r}\,  \qquad q=u,d,s ,\label{eq:epsilon}
\end{equation}
where \(a\) is the lattice spacing and \(r\) is a dimensionless parameter in the Wilson discretization of the fermion action, as described in \cref{eq:modifiedD}.

\section{Form Factors decomposition of the Electromagnetic Current in Presence of CP violation}
\label{sec:FF}

The nucleon matrix element of the electromagnetic current  $J^{\rm EM}_\mu = \sum_q  e_q\, \bar{q} \gamma_\mu q$, where \(e_q\) is the charge of the quark,
in the presence of parity violating interactions can be parameterized in terms of the most general 
set of form factors consistent with the symmetries of the theory. Working in the Euclidean 
space\footnote{We use the DeGrand-Rossi basis~\cite{DeGrand:1990dk} for our Euclidean gamma matrices. For details of the connection between our Euclidean and Minkowski, please refer to Appendix~A of Ref.~\citep{Bhattacharya:2021lol}.} we have:
\begin{align}
\langle N (p^\prime)  | J_\mu^{\rm EM} | N (p)  \rangle  &=
   \overline {u}_N (p') \big[ 
 \gamma_\mu   F_1  + {}\hspace*{0.10\textwidth}\nonumber\\
 \qquad\qquad\quad\span \frac{1}{2 M_N} \sigma_{\mu \nu} q_\nu \left( F_2  - i F_3 \gamma_5 \right)   + {}\nonumber\\
 \qquad\qquad\quad\span \frac{F_A}{M_N^2}  (\slashed q   q_\mu  - q^2 \gamma_\mu) \gamma_5~    \big] u_N (p) \  ,
\label{eq:FFdef}
\end{align}
where \(M_N\) is the neutron mass, \(q=p^\prime -p\) is the momentum carried
by the electromagnetic current, $\sigma_{\mu \nu} = (i/2) [\gamma_\mu, \gamma_\nu]$, 
and \(u_N(p)\) represents the free
    neutron spinor of momentum \(p\) obeying \(( i \pslash + M_N) u_N (p)=0\).
\(F_1\) and \(F_2\) are the Dirac and Pauli form factors, in terms of
which the Sachs electric and magnetic form factors are
\(G_E = F_1 - (q^2/4M_N^2) F_2\) 
and \(G_M = F_1 + F_2\), respectively.
The anapole form factor \(F_A\) and the electric dipole form factor
\(F_3\) violate parity P; and \(F_3\) violates CP as well.  The zero
$q^2$ limit of these form factors gives the charges and dipole
moments: the electric charge is \(G_E(0) = F_1(0) \) and the
magnetic dipole moment is
\( F_2(0) / 2 M_N\).
The nEDM is obtained from $F_3(q^2)$  as follows:
\begin{equation}
 d_n = \lim_{q^2\to0} \frac{F_3(q^2)}{2 M_N}~. 
\label{eq:dnF3}
\end{equation}

In what follows we will specialize to the isovector qcEDM
$\tilde{d}  \equiv \tilde{d}_u = - \tilde{d}_d$ 
(implying $\epsilon \equiv \epsilon_u = - \epsilon_d$),
which corresponds to 
\begin{equation}
{\cal L}^{\cancel{\text{CP}}} = 
\tilde d  \Big( C^{(u)} - C^{(d)} \Big)  = \tilde{d}  \, C^{(3)}
= - \frac{a r \epsilon}{2} \, C^{(3)} \,.
\label{eq:Civ}
\end{equation}
All lattice results will be presented in terms of the 
dimensionless coefficient $X_c$ relating the nEDM to $\tilde{d} \equiv \tilde d_u$,
{
\begin{equation}
    d_n = X_c  \ \tilde{d}  = - X_c \frac{ar}2\epsilon~. 
    \label{eq:Xc1}
\end{equation}
}
where
\begin{equation}
    X_c   = \frac{-F_3(q^2=0)}{a r  \epsilon M_N}. 
    \label{eq:Xc2}
\end{equation}
with $r=1$ in the Wilson-clover action we use.

Subtleties related to the phase convention for the neutron interpolating field  in presence of \CPV\ have been discussed and clarified in  Refs.~\citep{Pospelov:2000bw,Abramczyk:2017oxr,Bhattacharya:2021lol}. 
For completeness, we discuss the relevant issues here from a slightly different perspective. 

The usual representation of the free Dirac Equation
\((i \pslash + m) u^s (p)= 0\) is invariant under the Lorentz
transformation and the discrete symmetries C, P, and T with the
familiar expressions for their generators\punctfootnote.{For example, in
  the Dirac representation, the Lorentz generators are proportional to
  \(\sigma^{\mu\nu}\), whereas the matrices implementing P, C, and T
  are intrinsic phase factors multiplied by \(\gamma^0\),
  \(i\gamma^2\) and \(i\gamma^1\gamma^3\),
  respectively.}  The asymptotic in and out states of an interacting
field theory are free states, and hence obey the  symmetries of the free theory  even when the underlying theory does not preserve these.  There is,
however, an important distinction between the cases when the full
theory preserves the symmetry and when the symmetries only appear
asymptotically.

There is a freedom of representation in the free Dirac equation:
an arbitrary transformation \(u \to X u\),
\(\Gamma \to \Gamma_X \equiv X \Gamma X^{-1}\), with X a fixed
arbitrary matrix and \(\Gamma\) an element of the Clifford algebra
of the $\gamma$-matrices, preserves the form of the free Dirac
equation; but all the symmetry generators need to be written using 
\(\Gamma_X\) instead of \(\Gamma\).  In a theory where the symmetries
are preserved, the same generators can be chosen to implement the
symmetries on all states, and, hence, interpolating operators can be
chosen so that \(X=1\) for all asymptotic fermionic states without solving for the dynamics.  In a general theory,
however, the symmetry operations on each asymptotic state that an interpolating operator couples
to will have a different \(X\), and the interpolating operators cannot
be chosen to make all of them unity.

In particular, consider an interpolating field that produces
asymptotic states described by the conventional \(u^s\) spinors when
parity is conserved. When parity is broken, the asymptotic states that
this operator couples to will instead be solutions of the rotated
equation \((i \pslash + m) \exp(i\alpha\gamma_5) u^s = 0\), for some {\it
  a priori} unknown \(\alpha\).  To determine the
value of this constant, one can study the propagator, i.e., two-point function of the interpolating field $N$.  Because of Lorentz
symmetry, a spin-$\frac12$ field \(N\) will have a propagator given by\footnote{The expansion of this gives \(A(p^2) \times\allowbreak(\cosh 2 \mathop{\rm Im} \alpha(p^2) - i \gamma_5\sinh 2 \mathop{\rm Im} \alpha(p^2))\) as the coefficient of \(i\pslash\). There is, however, no reason that the coefficient of \(i\gamma_5\) has to be smaller in magnitude than the coefficient of unity in this expression. Since in our PT symmetric theory, \(\alpha(p^2)\) is real, we ignore this subtlety except to note that this might necessitate extra factors of \(\gamma_5\) for some states in \cref{eq:LSZ,eq:rotation} when PT symmetry is broken.}
\begin{align}
  \langle T N \bar N \rangle &=\hspace*{0.38\textwidth}\nonumber\\
  \quad\span e^{i\alpha(p^2)\gamma_5}
          \left[iA(p^2)\pslash+\Sigma(p^2)\right]^{-1}
          e^{i\alpha^*(p^2)\gamma_5}\,.
          \label{eq:alphaN}
\end{align}
The asymptotic large-time behavior of this propagator is given by the
residues of its poles.  The residue of the pole at \(p^2=-M_N^2\) (i.e.,
\(p^2_{\rm Minkowski} = M_N^2\)) is given by
\begin{align}
Z e^{i\alpha(-m_N^2)\gamma_5} (-i \pslash + m) e^{i\alpha^*(-m_N^2)\gamma_5} &=\hspace*{0.1\textwidth}\nonumber\\
\qquad\qquad\qquad\qquad\span
Z \sum_s \tilde u^s(p) \overline{\tilde u^s} (p)\,,
          \label{eq:LSZ}
\end{align}
where
\begin{equation}
  \tilde u^s(p) \equiv e^{i\alpha(-m_N^2)\gamma_5} u^s(p)
  \label{eq:rotatedspinor}
\end{equation}
is the spinor in the rotated representation.

After obtaining \(\alpha_N\equiv\alpha(-M_N^2)\) in this way, one has two
options: First, continue to use this representation cognizant
of the fact that this leads to a `rotated' Dirac equation, and hence
all the symmetry operators and projectors need to be written using the
rotated $\gamma$-matrices.  In particular, in the coefficients of the
\(F_2\) and \(F_3\) form factors in \cref{eq:FFdef} we need to replace
\([\gamma_\mu,\gamma_\nu]\) by
\(e^{i\alpha_N^*\gamma_5}[\gamma_\mu,\gamma_\nu]e^{i\alpha_N\gamma_5}\). We follow this strategy 
since we do not calculate the full \(4\times4\) Dirac structure of the three-point function needed for the second option. 

Conceptually, the 
much simpler alternative is to rotate the interpolating field
\(N\) to \(N_{\alpha_N} \equiv e^{-i\alpha_N\gamma_5}N\), or equivalently,
rotate all the correlation functions constructed from \(N\) as
\begin{align}
  \langle N {\cal O} \bar N \rangle &\to \langle N_{\alpha_N} {\cal O} \bar
  N_{\alpha_N} \rangle \nonumber\\
  &= e^{-i\alpha_N\gamma_5} \langle N {\cal O} \bar
  N\rangle e^{-i\alpha_N^*\gamma_5}\,.
  \label{eq:rotation}
\end{align}
{The residue of the two-point function of the rotated field $N_{\alpha_N}$ at the pole  $p^2=-M_N^2$ can be written in terms of the standard spinors $u^s(p)$ on which the discrete symmetries $P$, $T$, and $C$ are realized in the usual form. Then the analysis of the three-point function
$\langle N_{\alpha_N} {\cal O} \bar  N_{\alpha_N} \rangle$ proceeds 
in the standard ways and, in particular, the coefficient $F_3$ of 
$\sigma_{\mu \nu} q_\nu$ is the $CP$-odd form factor. \looseness-1
}

It is important to note that, in general, the ground state
and each excited state will have a different value of \(\alpha\), i.e.,
the rotation depends on the state we choose to
study\punctfootnote.{The choice 
  \(N_{\alpha(p^2)}\equiv \exp(-i\alpha(p^2)\gamma_5) N\) as an
  interpolating field is not admissible, since even though it has the usual propapagator
  and no $\gamma_5$-phases for any asymptotic state, this operator is
  not local in general.} Here, we will need $\alpha_N$ and $\alpha_N^5$ corresponding to the nucleon ground state with the insertion of qcEDM and pseudoscalar operators, respectively. 

In summary, as explained in Ref.~\citep{Bhattacharya:2021lol}, the rotation phase \(\alpha_N\) can be determined from
the long-distance behavior of the neutron propagator:\looseness-1
\begin{align}
\langle\Omega|N(\vec p,t)\bar N(\vec p,0)|\Omega \rangle &= \sum_N
|A_N|^2 e^{-E_N t} \tilde u_N \overline {\tilde u}_N \nonumber\\
\quad\span=\span \sum_N |A_N|^2 e^{-E_N t} {e^{i\alpha_N \gamma_5}} (- i\pslash + m_N) {e^{i\alpha_N \gamma_5}}\,,\nonumber\\
\label{eq:phase}
\end{align}
where the vacuum-to-\(N\) transition matrix element of the interpolating operator $N$, \(A_N\equiv\langle\Omega|N|N\rangle\), and the phase angles \(\alpha_N\) depend on the state, the interpolating operator, and CP violating couplings. Here, and henceforth, we have also assumed PT symmetry, so that \(\alpha_N\) is real. These phases can then be used to 
rotate the three-point functions as specified in \cref{eq:rotation}, 
to the standard
basis in which the form factors are given by \cref{eq:FFdef}. As explained above, we use the first method in which  the coefficient of 
\(F_2\) and \(F_3\) in \cref{eq:FFdef} are rotated. For further details, we refer the reader to Ref.~\citep{Bhattacharya:2021lol}.

%

\section{Method for calculating the Matrix Elements of the {qcEDM} operator}
\label{sec:SSM}

\begin{figure*}[tp]   
\begin{align*}
&\vcenter{\hbox{\includegraphics[width=0.15\textwidth]{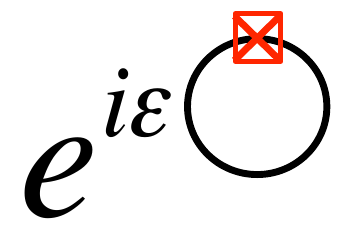}}} \times {}&\\[3\jot]
&\left(\;\vcenter{\hbox{\includegraphics[width=0.4\textwidth]{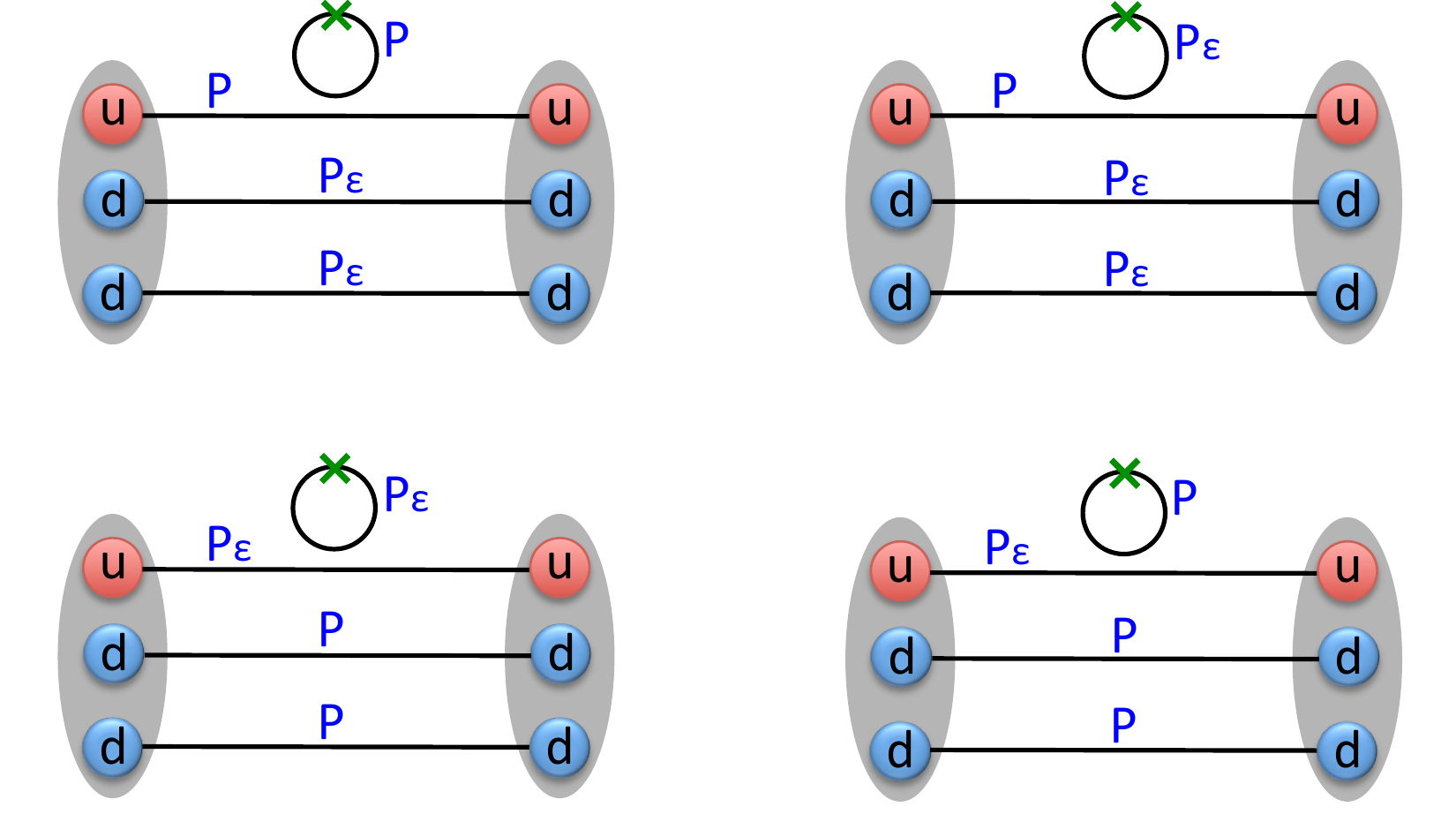}}} + 
 \vcenter{\hbox{\includegraphics[width=0.4\textwidth]{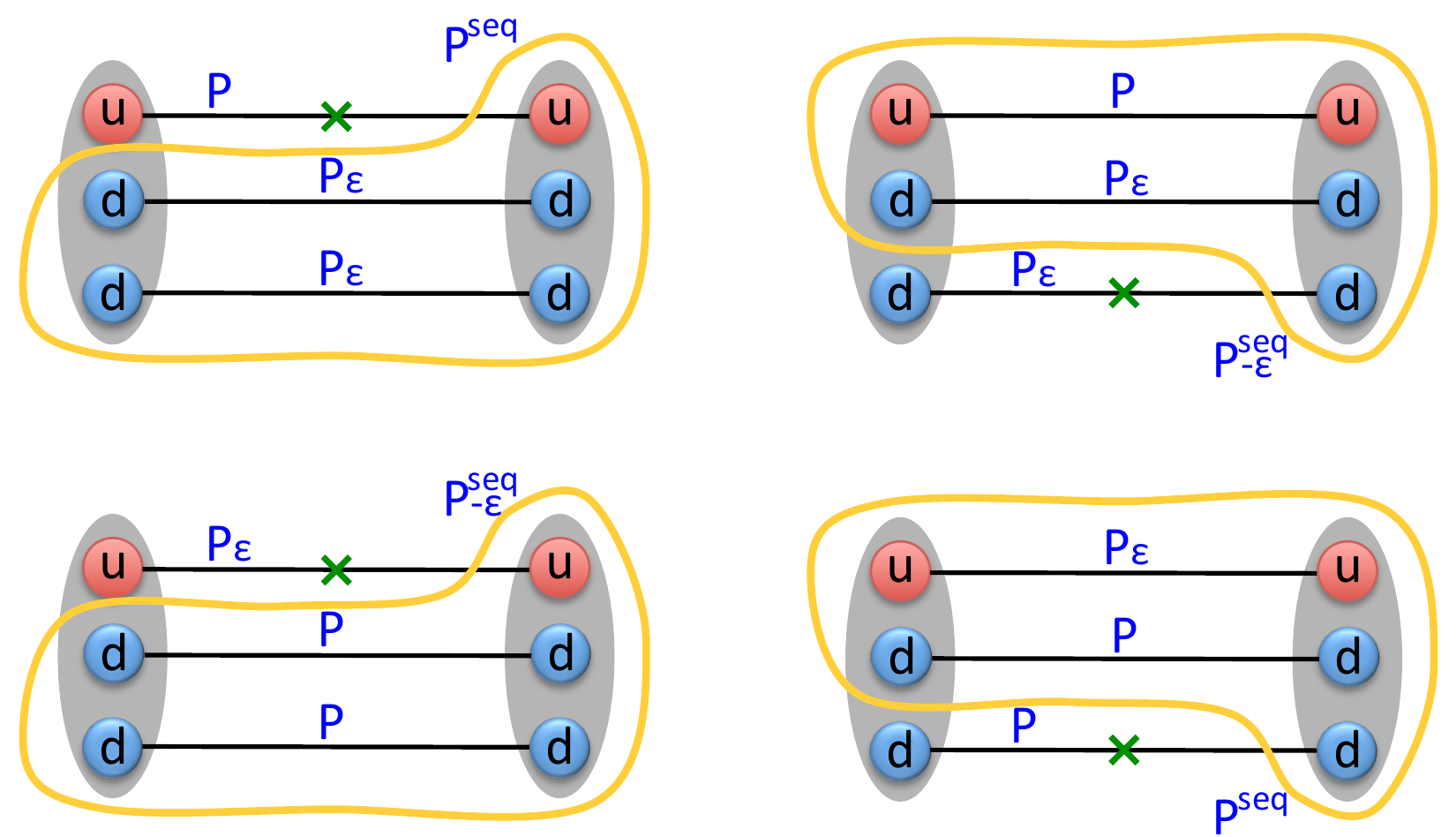}}}\;\right)&
\end{align*}
\caption{The full calculation requires the reweighting factor times
  the sum of the connected and disconnected diagrams. }
\label{fig:qcEDMfull}
\end{figure*}

\begin{table}[tbhp] 
\centering
\begin{tabular}{ l|llcr }
 \multicolumn1{c|}{ID} & 
 \multicolumn1c{$a$ (fm)}     & 
 \multicolumn1c{$M_\pi$ (MeV)} & 
 \multicolumn1c{$L^3\times T$}  & 
 \multicolumn1c{$N_{conf}$}   \\
\hline 
$a12m310$   & $0.1207(11)$ & $310.2(2.8)$ & $24^3\times64$  & 1013 \\
$a12m220L$  & $0.1189(09)$ & $227.6(1.7)$ & $40^3\times64$  &  475 \\
$a09m310$   & $0.0888(08)$ & $313.0(2.8)$ & $32^3\times96$  &  447 \\
$a06m310$   & $0.0582(04)$ & $319.3(0.5)$ & $48^3\times144$ &   72 \\
\end{tabular}
\caption{MILC HISQ ensembles and the simulation parameters used in the analysis. For each lattice configuration, the observables are measured on 128 source positions.}
\label{tab:ensembles}
\end{table}

The calculations presented here use the nonunitary clover-on-HISQ formulation, in which correlation functions are calculated using Wilson-clover fermions on four ensembles of background gauge configurations generated with  \(2+1+1\)-flavors highly improved staggered quarks (HISQ) by the MILC collaboration~\cite{Bazavov:2012xda}. The lattice parameters are specified in \cref{tab:ensembles} and the gauge fields are smoothed using HYP-smearing~\cite{Hasenfratz:2001hp} before calculating the correlation functions.  Three of these ensembles, \(a12m310\), \(a09m310\), and \(a06m310\), have a pion mass of \(M_\pi\approx315\rm~MeV\) and three different lattice spacings, \(a\approx0.12\), \(0.09\), and \(0.06\rm~fm\), respectively, to study the continuum limit.  The fourth ensemble, \(a12m220L\), at \(a\approx0.12\rm~fm\), \(M_\pi\approx225\rm~MeV\) provides a first check on the dependence on $M_\pi^2$.  For the valence quarks, the clover coefficient was fixed at its tadpole-improved value \(c_{sw}=1/u_0^3\), where \(u_0\) is the fourth root of the plaquette on the smoothed lattices. Further details of the ensembles, the quark mass parameters, methodology, statistics, and the interpolating operator used in the construction of two- and three-point correlators are given in our previous publications~\cite{Yoon:2016dij,Gupta:2018qil,Jang:2019jkn,Bhattacharya:2021lol}.

Working within the framework of the Schwinger source method 
to calculate the contribution of  qcEDM to the nEDM~\cite{Bhattacharya:2016oqm} allows us to recast the challenging calculation of the 4-point function given in \cref{eq:chromoEDM} and depicted in \cref{fig:qChromo} 
to a still difficult but well-exercised calculation of three-point
functions~\cite{Bhattacharya:2016oqm}.  The quark-level diagrams needed to
calculate $\langle N | J_\mu^{\rm EM}(q) | N \rangle $ in the presence
of qcEDM interactions are illustrated in
\cref{fig:qcEDMfull}.  The steps in each measurement are as follows\punctfootnote:{Additional complications due to mixing between operators
  and using a lattice discretizations that breaks chiral symmetry are
  discussed in \cref{sec:mixing}.}
\begin{itemize}
\item
Calculate propagators, labeled $P$ in \cref{fig:qcEDMfull}, using
the standard Wilson-clover Dirac operator, which 
in the continuum EFT notation reads
\begin{align}
O_D &=  D_L  + m_W
\nonumber\\
D_L &=  \slashed{D} - a \,  \left( \frac{r}{2}  D^2  +
\kappa_{SW}  \sigma \cdot G \right) \nonumber\\
\qquad\kappa_{SW}&= \frac{r}{4}  \, c_{SW}  ~,
\label{eq:DO1}
\end{align}
with \(m_W \equiv \frac{1}{2\kappa}-4\) the Wilson mass and $c_{SW} = 1 + O(g^2)$ the Sheikholeslami-Wohlert coefficient. This calculation of $P$ uses the same methodology as in our previous publications~\cite{Gupta:2018qil}.
We assume isospin symmetry
so the propagators for $u$ and $d$ quarks are numerically the same. 
\item
Calculate a second set of propagators that include the qcEDM term  
with coefficient {$\epsilon_q \equiv - (2 \tilde d_q)/(a r)$}. This is done by modifying the clover term in the Dirac matrix: 
\begin{align}
\qquad\frac{ra c_{SW}}4 \sigma^{\mu\nu} G_{\mu\nu} &\mathbin{{\longrightarrow}}\nonumber\\
\qquad\qquad\span\frac {ra} 4 \sigma^{\mu\nu} ( c_{SW} + i \epsilon_q \gamma_5 ) G_{\mu\nu}\,.
\label{eq:modifiedD}
\end{align}
We have allowed the qcEDM operator insertion to have different coefficients for the two flavors to make explicit what would need to be 
done to study the flavor diagonal qcEDM insertions in future work.
Choosing $\epsilon_u = - \epsilon_d = \epsilon$ corresponds to inserting the isovector qcEDM operator $C^{(3)}$ with coefficient $\tilde d= - (a r \epsilon)/2$, see \cref{eq:Civ}. 
These propagators, labeled $P_\epsilon$, include the full effect of
inserting the qcEDM operator at all possible intermediate points. 
Naively, the cost of this inversion is larger by about $7\%$ with respect to
$P$, however, using $P$ as the starting guess in the inversion for
$P_\epsilon$ reduces the number of iterations by 20--40\%
depending on the quark mass. Overall, the average cost of calculating $P_\epsilon$ is  about 80\% of $P$.

\item As discussed below, in order to remove power divergences from isovector qcEDM operator, we need to also calculate insertions of the isovector pseudoscalar density operator $P^{(3)} = i \bar \psi \gamma_5 T^3 \psi$. 
We implement this with the replacement 
\begin{equation}
   a m_W  \to     a m_W - 2 i \epsilon_5  \gamma_5 \ T_3 \end{equation}
in the up and down quark propagators. 
This prescription corresponds to inserting the operator $P^{(3)}$ with 
coefficient given by $(-2 \epsilon_5/a)$. 

\item
Using $P$ and $P_\epsilon$, we construct four kinds of sequential
sources, labeled $P_u^{\rm seq}$, $P_d^{\rm seq}$,
$P_{-\epsilon,u}^{\rm seq}$, and $P_{-\epsilon,d}^{\rm seq}$. These sources are at the
sink time-slice and include the insertion of a neutron at zero
momentum and the spin projection operator $(1+\gamma_4)(1+i\gamma_5 \gamma_3)/2$. The subscripts $u$ or $d$ in $P_u^{\rm seq}$ and $P_d^{\rm
  seq}$, and similarly in $P_{-\epsilon,u}^{\rm seq}$ and
$P_{-\epsilon,d}^{\rm seq}$, denote the flavor of the free spinor in this
neutron source. For the backward moving sequential propagators with qcEDM insertion, the coupling gets a minus sign, i.e., $-\epsilon$. \looseness-1

\item
In our implementation, a number of calculations are done in the same computer job by placing independent sources with maximal separation in time. The corresponding sequential sources are added together  to obtain the coherent sequential source~\cite{LHPC:2010jcs,Yoon:2016dij}, which is then used in the construction of the four types of sequential propagators listed above and illustrated in the four
correlation functions shown in the right half of \cref{fig:qcEDMfull}.

\item
The connected three-point function is then calculated using the two original and the four sequential propagators, and with the insertion of $J_\mu^{\rm EM}$ separately on the $u$ and $d$ quark lines. This construction is similar to those used in our study of the CP-conserving form-factors~\cite{Yoon:2016dij,Gupta:2018qil,Jang:2019jkn}. The difference here is the combinations involve propagators with and without the insertion of \CPV\ term as shown in the
four three-point functions in the right half of \cref{fig:qcEDMfull}.

\item
Looking to the future, to construct the disconnected quark loop contribution, the  electromagnetic current would be inserted in the quark loop,  with and without the qcEDM term, and correlated with the appropriate nucleon two-point 
correlation functions as illustrated in the left half of
\cref{fig:qcEDMfull}. The loop term is integrated over the time-slice $t$ and should be calculated for each of the quark flavors, $u,\ d, \, s,\ c$ and $b$. In this first study, neglect these disconnected diagrams since they  
are expensive to simulate and their effect is expected to be small.
\item
We correct for not having included the qcEDM operator in the action in the 
generation of the gauge configurations by re-weighting the configurations by the ratio of the determinants of the Dirac operators for the two theories: 
{\begin{widetext}\begin{align}
\frac {\det ( 
 \Dslash + m_W - \frac {ra}2 D^2 - \frac {ra} 4 \sigma^{\mu\nu} ( c_{SW} + i \epsilon_q \gamma_5 ) G_{\mu\nu} )  )}
          {\det ( 
           \Dslash + m_W - \frac r2 aD^2 - \frac{rc_{SW}}4 \sigma^{\mu\nu} aG_{\mu\nu}
          )}\span\omit\hfill\nonumber\\
&=
\exp \Tr \ln \left[1 - i \epsilon_q \frac{r a}{4} \, {\sigma^{\mu\nu} \gamma_5  G_{\mu\nu}}  
\left(\Dslash + m_W - \frac r2 aD^2 - \frac{rc_{SW}}4 \sigma^{\mu\nu} aG_{\mu\nu}  \right)^{-1} 
 \right]
\nonumber\\
&=
\exp \left[- i {\epsilon_q}  \frac{r a}{4} \Tr {\sigma^{\mu\nu}  \gamma_5  G_{\mu\nu}} 
\left(\Dslash + m_W - \frac r2 aD^2 - \frac{rc_{SW}}4 \sigma^{\mu\nu} aG_{\mu\nu}  \right)^{-1} 
\right] + O(\epsilon^2)\,.
\label{eq:reweight}
\end{align}\end{widetext}}%
The trace in the exponential cancels between the up and down quark contributions when the inserted chromo-EDM is isovector, {\it i.e.,} $\epsilon_u = - \epsilon_d = \epsilon$.
In general, this factor, the volume
integral of the quark loop with the qcEDM insertion,  has to be calculated for each active flavor. To implement re-weighting, the sum of the connected and disconnected
contributions has to be multiplied by the exponential of the 
sum of the qcEDM loop with the appropriate coupling $i\epsilon_q$ for each flavor.  
This is shown by the overall factor in
\cref{fig:qcEDMfull}. In short, for the isovector qcEDM operator analyzed here, all disconnected contributions are either neglected or cancel.

\begin{table}[tbp]   
\setlength{\tabcolsep}{1pt}
  \begin{tabular}{l|c|c|c|c}
    \hline
    \multirow{2}{*}{Ensemble} & \multicolumn2{c|}{qcEDM}            & \multicolumn2{c}{\(\gamma_5\)}\\\cline{2-5}
             & \(\epsilon\) & \(\alpha/\epsilon\) & \(\epsilon_5\) & \(\alpha_5/4\epsilon_5\)    \\
    \hline
    a12m310 & 0.0080       & $-$10.835(55)      & 0.0024         & $-$8.908(45) \\
    a12m220L& 0.0010       & $-$21.80(31)       & 0.0003         & $-$17.88(24) \\
    a09m310 & 0.0080       & $-$12.360(36)      & 0.0024         & $-$12.052(34) \\
    a06m310 & 0.0080       & $-$15.00(12)       & 0.0012         & $-$19.82(16) \\
    \hline
  \end{tabular}
  \caption{The couplings \(\epsilon\) and \(\epsilon_5\) used in the simulations with the qcEDM and pseudoscalar operators, and the corresponding  neutron phases, \(\alpha\) and  \(\alpha_5\), obtained on each ensemble from fits shown in \cref{fig:alpha_fit}.}
  \label{tab:alpha}
\end{table}

\item
The above calculation is repeated for different values of $\epsilon$ to  
extract $F_3 (0)$ as a function of $\epsilon$. The contribution to the 
neutron electric dipole moment of the qcEDM operator is then given 
by the slope versus $\epsilon$.  
In the dimensionless parameter $X_c$ defined in \cref{eq:Xc2}, taking
the limit is not required for sufficiently small \(\epsilon\), which is 
in the linear regime. 
\end{itemize}

\begin{figure*}[p]   
  \centering
  \includegraphics[width=0.45\textwidth]{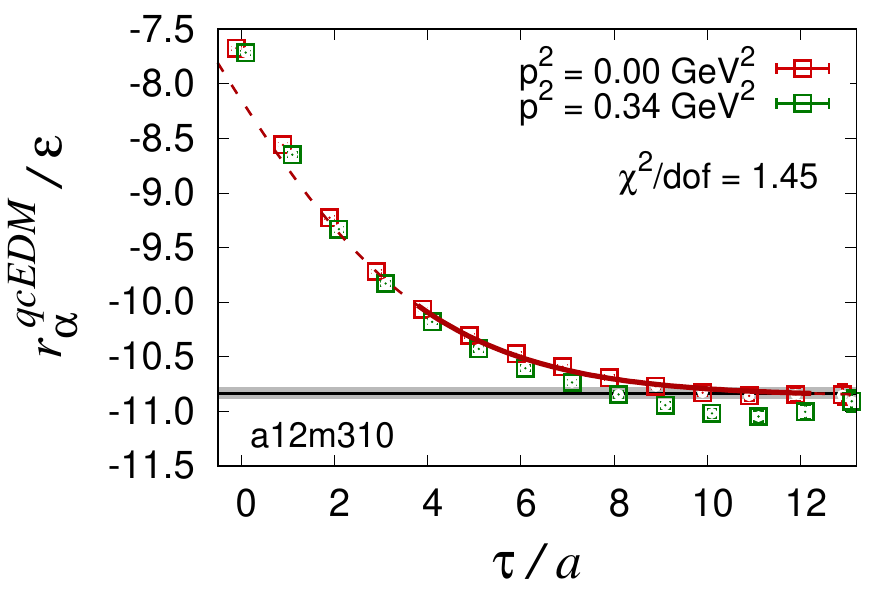} \qquad
  \includegraphics[width=0.45\textwidth]{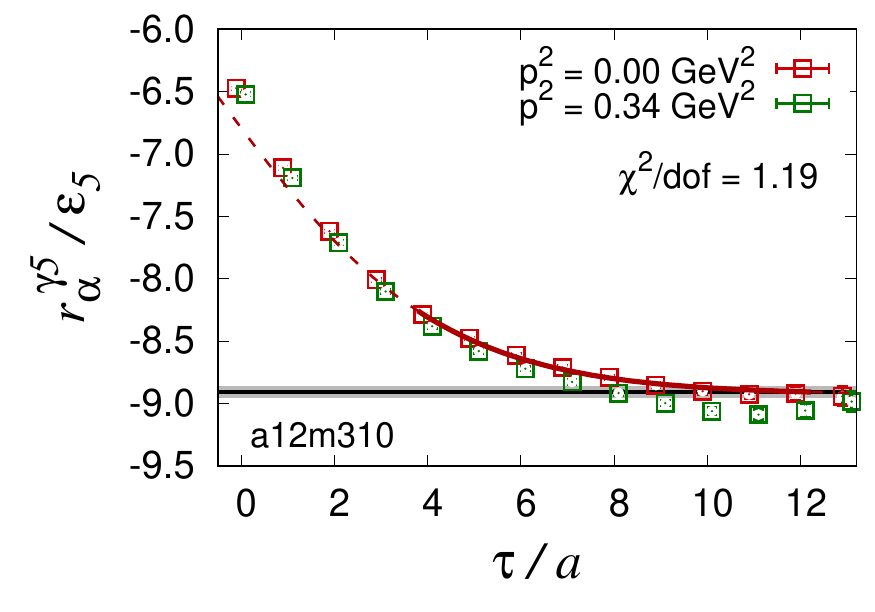}\\
  \includegraphics[width=0.45\textwidth]{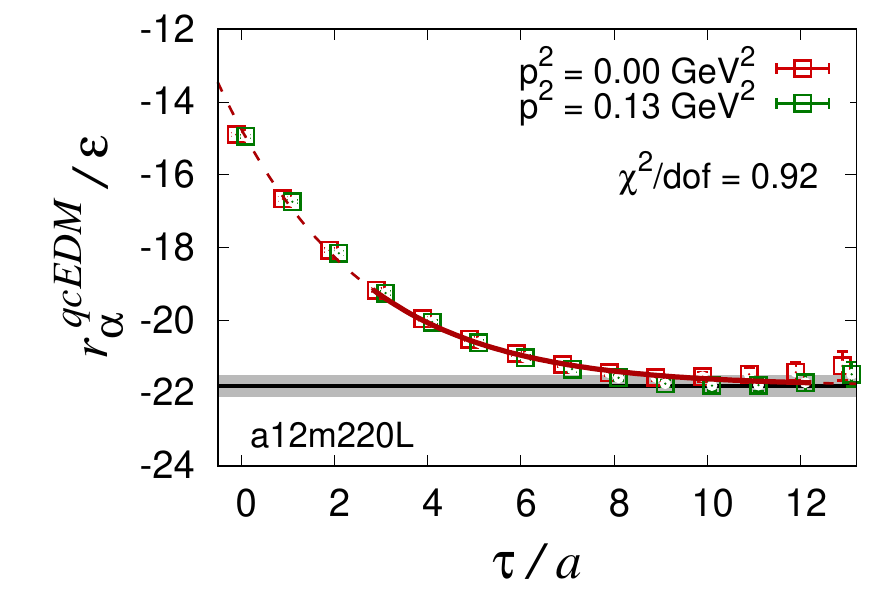} \qquad
  \includegraphics[width=0.45\textwidth]{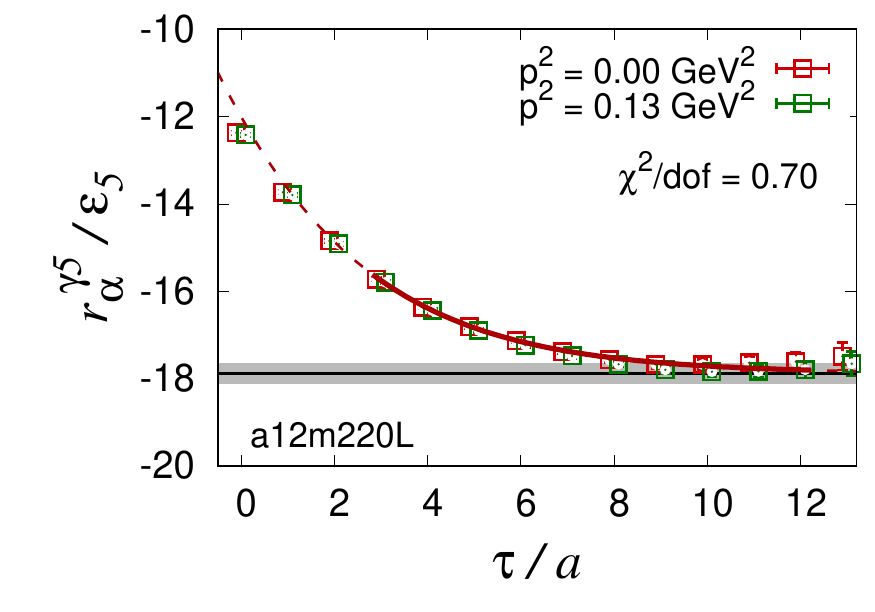}\\
  \includegraphics[width=0.45\textwidth]{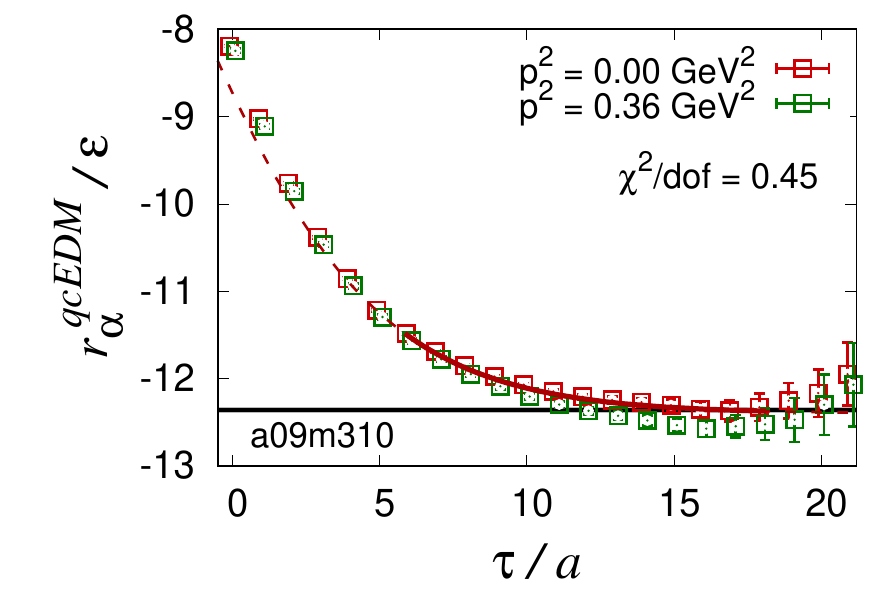} \qquad
  \includegraphics[width=0.45\textwidth]{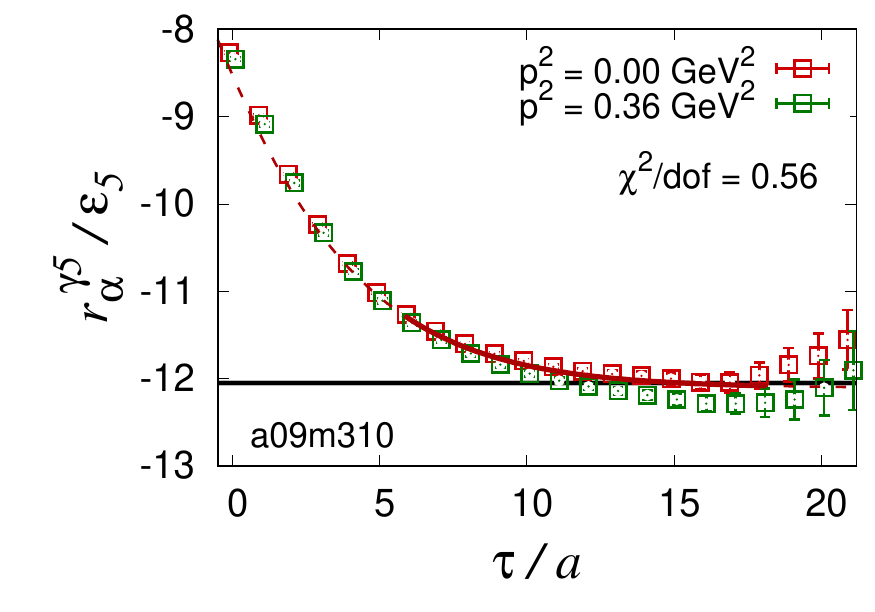}\\
  \includegraphics[width=0.45\textwidth]{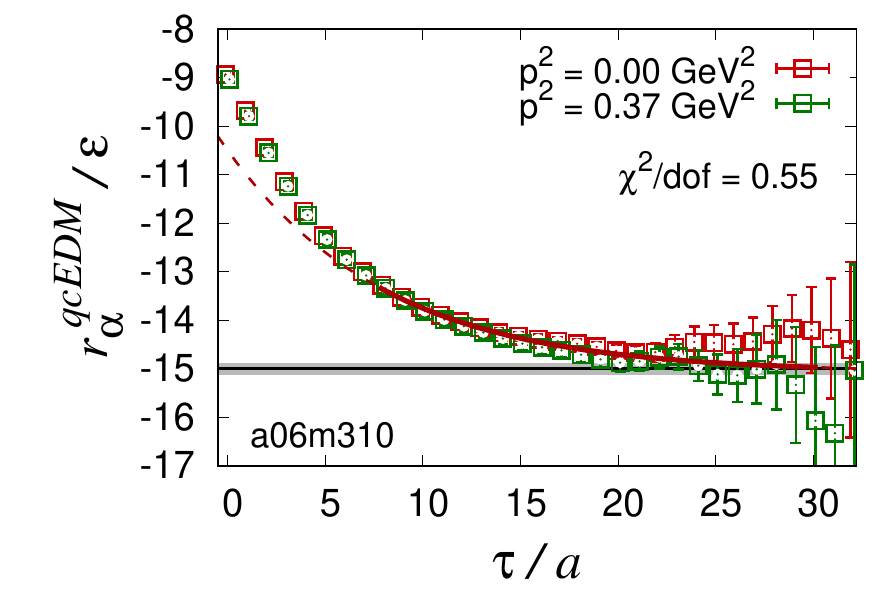} \qquad
  \includegraphics[width=0.45\textwidth]{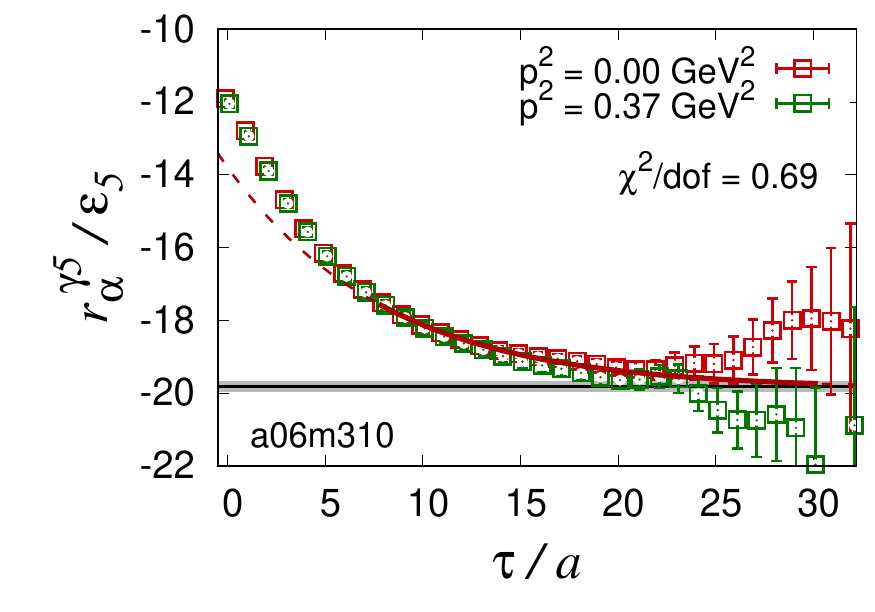}  
  \caption{A 2-state fit to the ratio $r_\alpha$ defined in \cref{eq:ralpha} used to extract the isovector CP violating phase $\alpha$ for the insertion of the qcEDM (left)  and pseudoscalar (right) operators. The phase is independent of the momentum of the state and the results are summarized 
  in \cref{tab:alpha}. }
  \label{fig:alpha_fit}
\end{figure*}

\begin{figure}[tbp]   
  \begin{center} 
  \hbox{\includegraphics[width=0.99\linewidth]{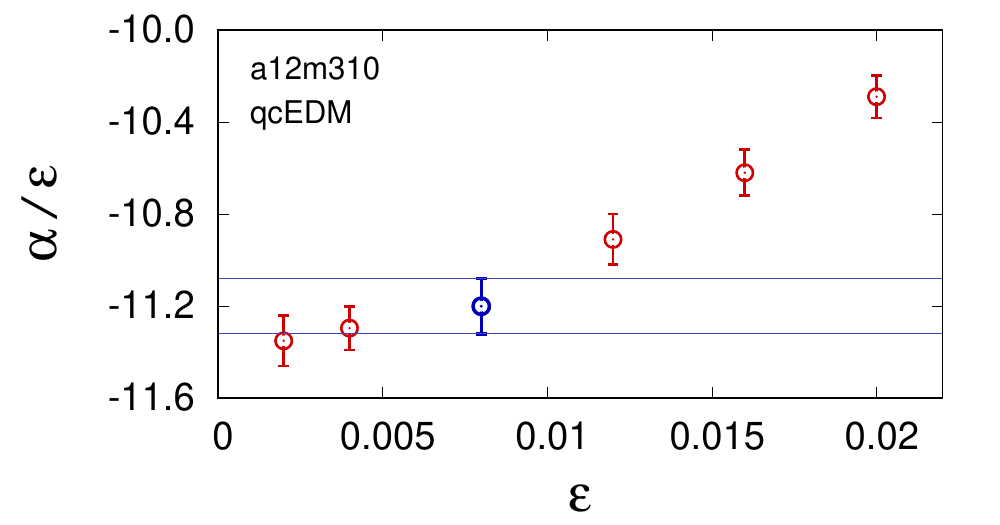}}
  \vspace*{3\baselineskip}
  \hbox{\includegraphics[width=0.99\linewidth]{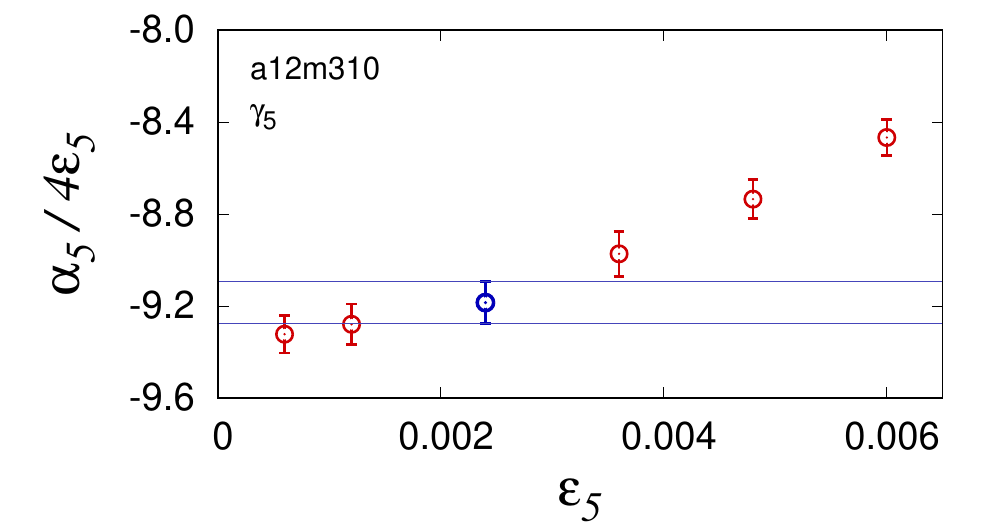}}
    \end{center}
  \caption{The value of the phase $\alpha$ ($\alpha_5$) as a function of the qcEDM (pseudoscalar) coupling $\epsilon$ ($\epsilon_5$). For small enough $\epsilon$, the $\mathcal{O}(\epsilon^2)$ contributions become negligible and $\alpha/\epsilon$ should be a constant for different choices of $\epsilon$. The blue data points within the linear region show the $\epsilon$ used in the main simulation. The data shown are obtained using 50 configurations of the a12m310 ensemble. 
}
\label{fig:alphaVSeps}
\end{figure}

\subsection{The extraction of the \texorpdfstring{\CPV}{CPV} phase \texorpdfstring{\(\alpha_N\)}{\textalpha\_N}}

The first step in the calculation is to extract the \CPV\ phase \(\alpha_N\),  defined through \cref{eq:alphaN}, from the nucleon two-point functions.  Since this phase is state-dependent, its value for the ground state nucleon has to be extracted at large source-sink separations where ESC have died out as described in our previous work~\cite{Bhattacharya:2021lol}. The data and the fits for the four ensembles and with the insertion of $\epsilon$ and $\epsilon_5$ are shown in \cref{fig:alpha_fit} and the results summarized in \cref{tab:alpha}. The behavior of  \(\alpha_N\) versus \(\epsilon\) is 
shown in \cref{fig:alphaVSeps}, which we use to select the  value of \(\epsilon\)  that is small enough to lie in the linear regime and yet large enough to give a good statistical signal. This value is highlighted in \cref{fig:alpha_fit,tab:alpha}. 

To extract the phase, we calculate the ratio~\cite{Bhattacharya:2021lol}
\begin{align}
    r_\alpha (\tau) &\equiv\frac{\Im \Tr \gamma_5 \frac{1+\gamma_4}2 \langle N(0) \overline N(\tau)\rangle}
         {\Re \Tr\frac{1+\gamma_4}2 \langle N(0) \overline N(\tau)\rangle}\nonumber\\
&\approx
    \tan \alpha_0 \times \frac{1 + \frac{\sin (2 \alpha_1)}{\sin (2 \alpha_0)} |\tilde{\cal A}_1|^2 \ e^{- (M_1 - M_0)\tau}}{1 + \frac{\cos^2 (\alpha_1)}{\cos^2 ( \alpha_0)} |\tilde{\cal A}_1|^2 \ e^{- (M_1 - M_0)\tau}}\,,
\label{eq:ralpha}
\end{align}
which approaches  \(\alpha_N\equiv\alpha_0\) at large \(\tau\) and allows us to extract \(\alpha_N\) from a fit~\cite{Bhattacharya:2021lol}.

\section{Pseudoscalar density versus  quark chromo-EDM insertions}
\label{sec:AWI}

In this section we discuss the connection 
between insertions of the isovector qcEDM operator and the isovector pseudoscalar density at {\it finite} lattice spacing $a$.
The discussion is based on the framework of a continuum EFT  for the lattice action and the 
axial Ward Identities, following Refs.~\citep{Bochicchio:1985xa, Testa:1998ez}.
{We first discuss the nonsinglet  axial Ward identity 
and the relation between chromo-EDM and pseudoscalar density  that follows from it. We then present the lattice analysis to determine the relevant nonperturbative coefficients arising in the mixing.  }

\subsection{Nonsinglet axial Ward identity and implications}
We will  denote by $O_n^{(d)}$,   $\tilde{O}_n^{(d)}$,  $O_n^{(d),{\rm ren}}$,  
 the set of bare, subtracted, and renormalized  operators of dimension $d$, respectively. 
Subtracted operators, i.e. operators free of power divergences,  are defined as 
\begin{equation}
\tilde O_{n'}^{(d)} =  O_{n'}^{(d)}  -  \sum_{d'<d , k}   \frac{\beta_{n' k}^{(d)}}{a^{d- d'}} \, \tilde O_k^{(d')}
\label{eq:Osub}
\end{equation}
with the sum over $k$ running over all operators of dimension $d'$. The finite (renormalized) operators are given by
\begin{equation}
O_{n}^{(d),{\rm ren}} =   Z_{n n'}   \  \tilde O_{n'}^{(d)}  ~.
\label{eq:Zd}
\end{equation}
The presence of $\tilde O_k^{(d')}$ and not  $O_k^{(d')}$ in \cref{eq:Osub}
 is needed to avoid ambiguities in the definition of coefficients 
$\beta_{n' k}^{(d)}$ of lower-dimensional operators.

{As derived in \cref{sec:AWI_app},} under the 
axial transformation on the quark fields collectively defined by $\psi^T = (u,d,s,c)$  
\begin{align}
\psi (x) &\to (1 + i \xi^{(a)} (x)  T^a \gamma_5 ) \psi (x)  \nonumber\\
\bar \psi (x)  &\to \bar \psi  (x) (1 + i \xi^{(a)} (x)  T^a \gamma_5)\,,  
\label{eq:axialtrans}
\end{align} 
where $\xi^{(a)} (x)$ is the local chiral transformation parameter and $T^a$ are the generators of flavor $SU(4)$, 
the flavor nonsinglet axial Ward Identity (AWI) for the expectation value of $O$ is given by
\begin{align}
 \span\int d^4x  \Big\langle O(x_1,..., x_n)  \times{}\hspace*{0.23\textwidth}\nonumber\\
  \span
  \Big[-  \bar{\psi} (x)  \{ m , T^a \} \gamma_5 \psi  (x)   \Big( 1 + O(a m) \Big)
 - \nonumber\\
 \span\qquad\qquad\qquad a i  K_{X1} \ \tilde C^{(a)}
 \Big]
\Big \rangle
\nonumber \\
\qquad\qquad&= - \int d^4x \left \langle   \frac{\delta O (x_1, ..., x_n)}{\delta (i \xi^{(a)} (x))}  
\right \rangle\,.
\label{eq:AWI4.5n}
\end{align}
This is correct up to \(O(a^2)\) corrections when applied to the Wilson-clover fermion action that includes \(O(a)\) hard breaking of chiral symmetry. Here, \(K_{X1}\) is a nonperturbative constant that characterizes the \(O(a)\) breaking of chiral symmetry, and vanishes if the theory is fully \(O(a)\) improved.
The RHS of  \cref{eq:AWI4.5n} does not contribute to on-shell matrix elements, 
and hence to the form factor $F_3$. In this case, insertions of the iso-vector  pseudoscalar density 
are proportional to insertions of the subtracted iso-vector quark chromo-EDM operator. 
In the next subsection, we show how an analysis of the unintegrated version of this equation given in \cref{eq:AWI4n} allows us to 
determine the needed nonperturbative factor $K_{X1}$ relating the two operators.

\subsection{Determining the nonperturbative parameters}

We now specialize to the isovector case,  corresponding to the flavor index $a=3$. 
We also work in the isospin limit and denote the common light quark masses by $m_l$. Taking the pion field to be $P^{(3)}\equiv i \overline \psi \gamma_5 T^3 \psi$, and specializing to $O(x_1, ..., x_n) \to O(z)$, 
\cref{eq:AWI4n} becomes 
\begin{align}
\Big\langle O(z)  
  \Big[
   Z_A \, ( 1 + b_A  m_l a ) \,  \partial_\mu A_\mu^{(3)} (x)  + {}\span\nonumber\\
   \qquad\span i a Z_A c_A \, \partial^2  P^{(3)}(x)
 + 2 m_l i  P^{(3)} (x)  
- {}\nonumber\\
\qquad\span\qquad \qquad a i  K_{X1} \ \tilde C^{(3)}(x)
 \Big] \Big \rangle 
  \nonumber\\
  \qquad\qquad\qquad&= - 
\Big \langle   \frac{\delta O (z)}{\delta (i \xi^3 (z))}  
\Big \rangle\,, 
\label{eq:NP1}
\end{align}
up to \(O(a^2)\) corrections. In order to determine $K_{X1}$, we need to first define $\tilde C^{(3)}$. 

  \begin{table}  
    {
    \begin{tabular}{c|cc|cc}
      Ensemble&\(c_{SW}\)&\(a\) (fm) &
      \(t\)-range&\(A\)\\ \hline
      a12m310 & 1.05094 & 0.1207(11) &
      6--14 & \(1.21374(62)\)\\
      a12m220L& 1.05091 & 0.1189(09) &
      7--14 & \(1.21800(33)\)\\
      a09m310 & 1.04243 & 0.0888(08) &
      8--22 & \(0.99621(30)\)\\
      a06m310 & 1.03493 & 0.0582(04) &
      14--30 & \(0.77917(24)\)\\
    \end{tabular}}
    \caption{Determination of 
    \(A\) defined in \cref{eq:sub1} from an average over the plateau region (\(t\)-range) of ratios of two-point functions.
    }
    \label{tab:Atab}
  \end{table}
  
\begin{figure}    
  \includegraphics[width=0.48\textwidth]{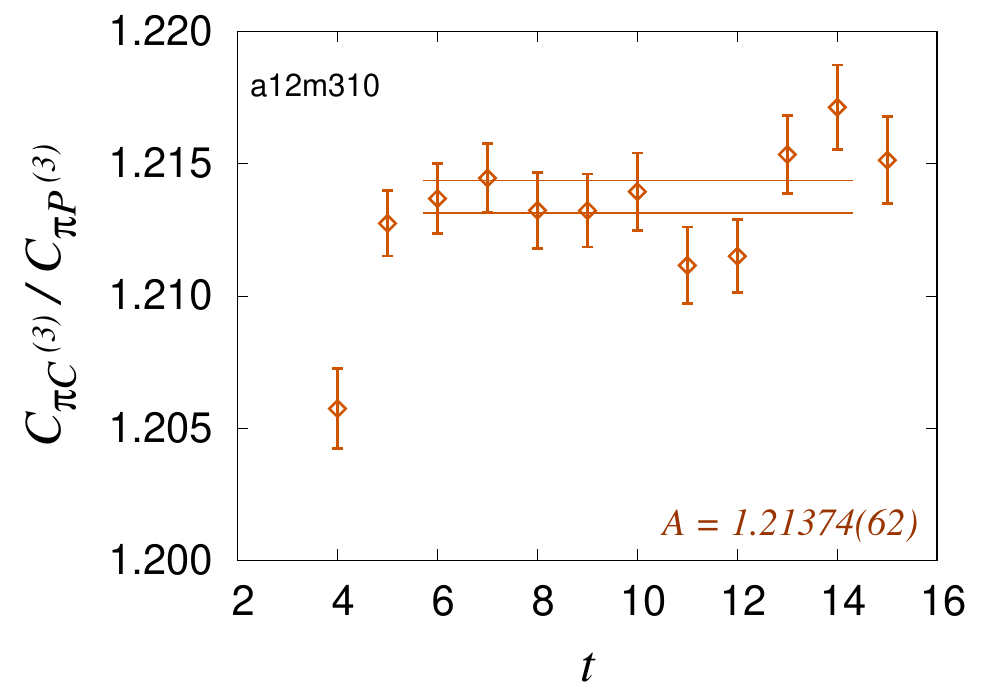}\\
  \includegraphics[width=0.48\textwidth]{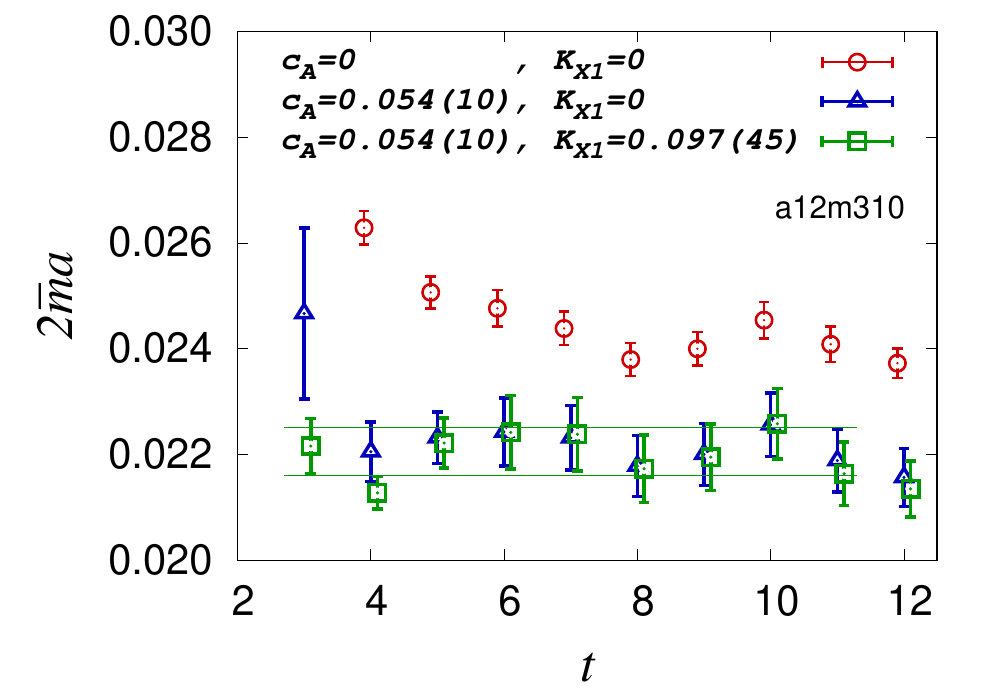}
\caption{Data from a12m310 ensemble showing the determination of the subtraction coefficient $A$  using \cref{eq:Adetermination} (upper panel), and  the coefficients \(\bar c_A\), \(\bar K_{X1}\) and \(2\bar m a\) using \cref{eq:FitFormula} (lower panel).}
  \label{fig:divergent}
  \label{fig:CoeffDetermination}
\end{figure}

\begin{figure*}   
  \includegraphics[width=0.48\textwidth]{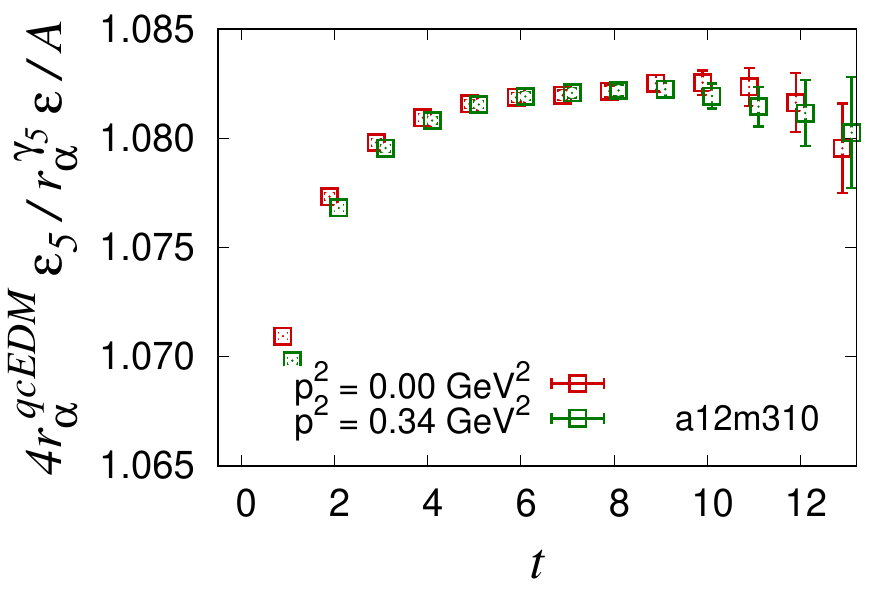} 
  \includegraphics[width=0.48\textwidth]{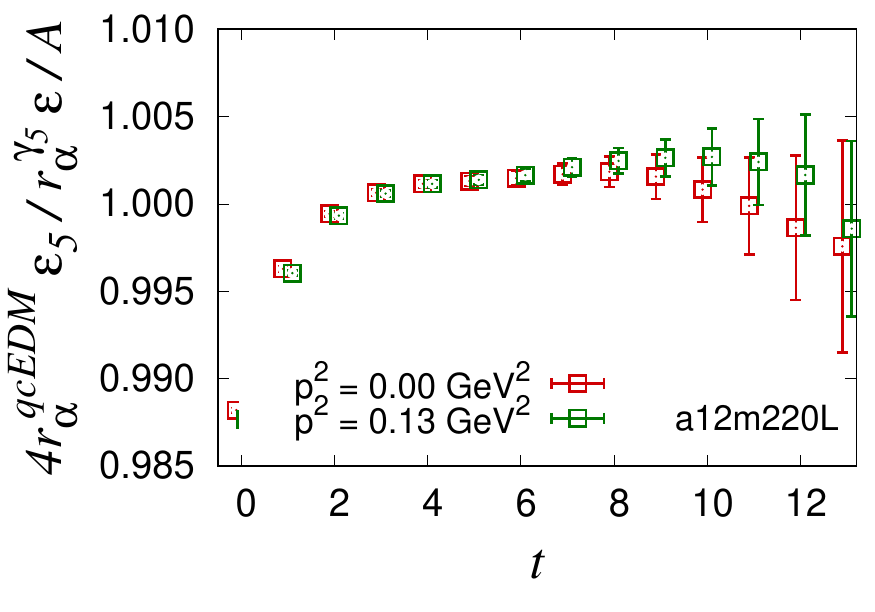} \\[1\jot]
  \includegraphics[width=0.48\textwidth]{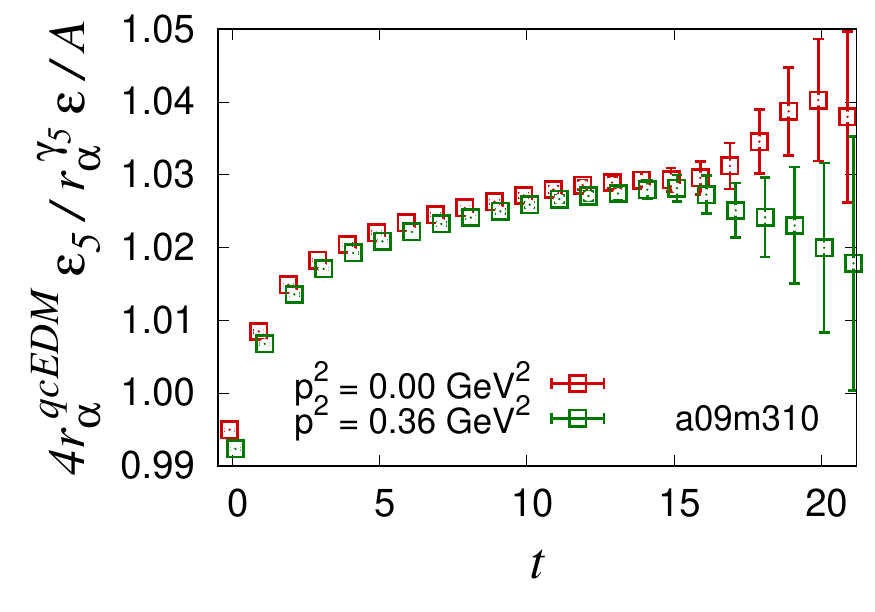}
  \includegraphics[width=0.48\textwidth]{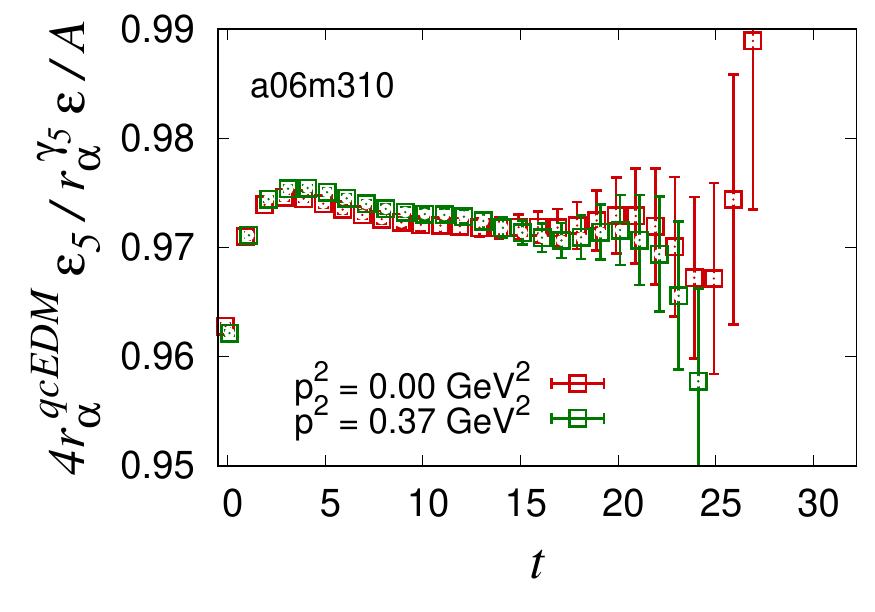} \\
  \caption{Test of the relation given in Eq.~\eqref{eq:Acheck}. The ratios  $r_\alpha$, defined in Eq.~\eqref{eq:ralpha}, give the CP violating phases $\alpha_N$.  The mixing coefficient $A$ is defined in Eq.~\eqref{eq:sub1}.}
  \label{fig:alpha_cedm_g5}
  \vspace*{2\baselineskip}
\end{figure*}

\subsubsection{Determination of parameter \texorpdfstring{$A$}{A}}
\label{sec:A}

As explained in \cref{sec:operatorbasis}, under isospin symmetry and when applied to on-shell zero momentum correlators, \cref{eq:Osub}, which relates the subtracted qcEDM operator to the  bare un-subtracted one used in the lattice calculation, reduces to
\begin{equation}
\tilde{C}^{(3)} (x)   = C^{(3)} (x)  - \frac{A}{a^2} P^{(3)} (x) ~, 
\label{eq:sub1}
\end{equation}
where \(A\) is \(O(\alpha_s)\) with \(O(am)\) and convention-dependent \(O(a^2)\) corrections. To determine $A$, we can, for example, define \(\tilde C^{(3)}  \) by demanding \(\langle \Omega | 
\tilde C^{(3)} | \pi(\vec p=0)\rangle = 0\).  This is particularly simple to implement on the lattice---we, in the two-point functions  \(C_{\pi P^{(3)} }(t) = \langle T P^{(3)} (t) \pi(0) \rangle\) and \(C_{\pi C^{(3)}}(t) = \langle T C^{(3)} (t) \pi(0) \rangle\), place the 
pseudoscalar and qcEDM interpolating operators at the sink and use the same pion source, \(\pi(0)\). Since the pion ground state dominates these two-point functions at long distances, the coefficient $A$ is given by the asymptotic behavior of their ratio: 
\begin{equation}
  A = \lim_{t\to\infty} \frac{a^2C_{\pi C^{(3)}} (t) }{C_{\pi P^{(3)}} (t)}\,.
  \label{eq:Adetermination}
\end{equation}
Other choices of correlation functions used to determine $A$ change only the convention-dependent \(O(a^2)\) contributions. As shown in \cref{tab:Atab,fig:divergent}, this construction gives a very precise determination of \(A\). Though formally \(O(\alpha_s)\), its value is close to unity at values of the lattice spacing where current simulations have been done.

We can use this determination of $A$ to perform a consistency check on the phases 
\(\alpha_N^{5}\) and  \(\alpha_N\) induced by the \CPV\ operators \(P^{(3)}\) and \(C^{(3)}\), respectively.  As described in \cref{sec:alphadet}, at leading order in chiral perturbation theory ($\chi$PT), \(\tilde C^{(3)}\) defined in \cref{eq:sub1} gives no contribution to \(\alpha_N\).  Then, from the right hand side we expect the relation (see \cref{eq:D8}).
\begin{equation}
\frac1A \frac{\alpha_N} {\alpha_N^5}= 1 + O\left(\frac{m_\pi^2}{\Lambda_\chi^2}\right) \,.
\label{eq:Acheck}
\end{equation}
Since \(\alpha\) is state dependent, the determination of \(\alpha_N\) (\(\alpha_N^5\)) are straightforward only when extracted at asymptotically long Euclidean times where ESC are negligible. Since the signal-to-noise in nucleon correlators degrades exponentially, this asymptotic region cannot be reached with current statistics. Instead, we analyze \(r_\alpha\) defined in \cref{eq:ralpha} that includes the lowest excited state contributions.
Data  in \cref{fig:alpha_cedm_g5} for the four ensembles show that the relation \cref{eq:Acheck} is satisfied to within ten percent by the $\alpha$ we  determine.

\subsubsection{Determination of parameter \texorpdfstring{$K_{X1}$}{K\unichar{"2092}\unichar{"2081}}}
\label{sec:KX1_subsec}
In terms of the unsubtracted  $C^{(3)}$, the nonsinglet AWI, \cref{eq:NP1}, can {now} be cast as follows: 

\begin{align}
\Big\langle O(z)  
\Big[    Z_A \, ( 1 + b_A  m a ) \,  \partial_\mu A_\mu^{(3)} (x)  + {}\span\nonumber\\
 \span i a
  Z_A c_A \, \partial^2  P^{(3)}(x)
  +  i  {2 m}  P^{(3)}(x)  \nonumber\\
  \span
  {} - i {K_{X1}} \left( a  \ C^{(3)}(x) {- \frac A a P^{(3)}}\right)  \Big] \Big \rangle \nonumber\\
\qquad\qquad\qquad\qquad &= - 
\Big \langle   \frac{\delta O (z)}{\delta (i \xi^{(3)} (x))}  
\Big \rangle\,. 
\label{eq:NP2}
\end{align}
To calculate {\(K_{X1}\) and \(m\)}, consider the two point functions 
\(C_{\pi A_\mu^{(3)}}(t)\), \(C_{\pi P^{(3)}}(t)\) and \(C_{\pi C^{(3)}}(t)\) of \(A_4^{(3)}(t)\), \(P^{(3)}(t)\) and \(a C^{(3)}(t)\) with a common source \(\pi(0)\).  Further, let \(\Delta\) and \(\Delta^2\) define the symmetric lattice first and second derivatives in the time direction: {\it e.g.,} \((\Delta C)(t) \equiv [C(t+1)-C(t-1)]/2\) and \((\Delta^2 C)(t) \equiv [C(t+1)+C(t-1)-2 C(t)]\).  Then, on-shell, {\it i.e.,} at \(t\not=0\), we have
\begin{widetext}
\begin{equation}
    \frac{i a \Delta C_{\pi A_4^{(3)}} (t)  - \bar{c}_A a^2 \Delta^2 C_{\pi P^{(3)}} (t) + {\bar{K}_{X1}}  \left(a^2C_{\pi C^{(3)}} (t){-AC_{\pi P^{(3)}}(t)}\right)}
         {C_{\pi P^{(3)}}(t)} = {2 \bar{m} a} \ + \  O(a^2 m^2, a^2 \Lambda_{QCD}^2)~,  
         \label{eq:FitFormula}
\end{equation}
\end{widetext}
where
\begin{subequations}
\begin{align}
 \bar c_A &= \frac{c_A}{(1 + b_A m a)}
 \\
{\bar{K}_{X1}} &= \frac{{K_{X1}}}{Z_A (1 + b_A m a)}
  \\
{\bar{m}} &= \frac{{m}}{Z_A (1 + b_A m a)}\,.
\end{align}
\end{subequations}
Thus, one can extract the coefficients \(\bar c_A\) and \({\bar{K}_{X1}} \) by fitting the left hand side and requiring it be independent of \(t\) and the pion interpolating field \(\pi\). Simultaneously, the fit provides  
\({2\bar{m}a} \), and thus  \({K_{X1}}/{2ma}\), which is required next.

We implemented the identity \cref{eq:FitFormula} numerically by choosing \(\pi\) to be the pseudoscalar interpolating operator \(P^{(3)}\) constructed using Wuppertal-smeared sources for quark propagators with various radii and at various momentum.  As discussed in \cref{sec:xydet}, these three-parameter fits are very unstable with our statistics.  We, therefore, proceeded by noting that the factor multiplying \(\bar{K}_{X1}\) vanishes once the contribution of the excited states vanish. Thus, we first make a two-parameter fit to the large-\(t\) region ignoring \(\bar{K}_{X1}\).  We then hold \(c_A\) fixed and determine \(\bar{K}_{X1}\) by extending the region to smaller \(t\).  The results are given in \cref{tab:xytab}.

  \begin{table*} 
    {\small
      \begin{tabular}{c|cc|cc|rrrrr}
        &\multicolumn{2}{c|}{{fit-range}}&\multicolumn{2}{c|}{{\(\chi^2/\rm d.o.f\)}}&&\\
      Ensemble&{\(c_A\)}&{\({\bar K}_{X1}\)}&{\(c_A\)}&{\({\bar K}_{X1}\)}&\multicolumn1c{\(c_A\)}&\multicolumn1c{\({\bar{K}_{X1}}\)}&\multicolumn1c{\({2\bar m a}\)}&\multicolumn1c{{\(\displaystyle\frac{2 m a}{K_{X1}}\)}}&\multicolumn1c{{\(\displaystyle\frac{2 m a} {2 m a + AK_{X1}\strut}\)}}\\\hline
      a12m310  & 4--11 & 3--11 & 0.66 & 0.88 & $0.054(10)$  & $0.097(45) $ & $0.02205(46) $ & $0.23(10) $ &  $0.158(58) $ \\
      a12m220L & 4--11 & 3--11 & 2.08 & 3.09 & $0.0342(77)$ & $0.183(35) $ & $0.01152(21) $ & $0.063(12)$ &  $0.0491(86)$ \\
      a09m310  & 5--15 & 4--15 & 0.99 & 1.09 & $0.0277(40)$ & $0.047(15) $ & $0.01684(15) $ & $0.35(11) $ &  $0.263(61) $ \\
      a06m310  & 6--20 & 5--20 & 0.29 & 1.53 & $0.0093(17)$ & $0.0272(60)$ & $0.010460(37)$ & $0.385(87)$ &  $0.331(50) $ \\
    \end{tabular}}
    \caption{{The determination of \({\bar K}_{X1}\) and \(2{\bar m}a\), and their ratios \(2 m a / K_{X1}\) and \(2 m a /(2 m a + AK_{X1})\) using Eq.~\protect\eqref{eq:FitFormula} and see the discussion below it.}}
    \label{tab:xytab}
  \end{table*}

\subsection{Implications for the {nEDM}}

The relation \cref{eq:AWI4.5n}  implies  that,   up to $O(a^2)$ effects, 
insertions of the subtracted isovector {qcEDM}  and  the pseudoscalar operators at zero four-momentum transfer and between on-shell states 
are proportional to each other. 
Furthermore, \cref{eq:NP2}, for zero-momentum on-shell matrix elements,  gives  
$\langle P^{(3)} \rangle = {[K_{X1}/(2ma+A K_{X1})]} \langle a^2 C^{(3)} \rangle$, a relation between unsubtracted operators. Thus, we have the following relations between subtracted and unsubtracted isovector operators up to \(O(a^2)\):
\begin{subequations}
\begin{align}
a \tilde C^{(3)}  &=   {\frac{2 a m}{K_{X1}}} \frac{P^{(3)}}{a} 
\label{eq:P-effect}\\
a \tilde C^{(3)}  &= 
\left({\frac{2am}{2am+A K_{X1}}}\right) a C^{(3)}\,.
\label{eq:C-effect}
\end{align}
\end{subequations}
{Note that even though both the quantities \(K_{X1}\sim O(\alpha_s)\) and \(2am\) are small, for values of the lattice spacing $a$ used in current simulations, their ratio is $O(1)$. In the continuum limit, \(P^{(3)}\) can be rotated away, but \cref{eq:P-effect} shows that at \(a\sim0.1\rm fm\), the effect of the lattice operators \(P^{(3)}/a\) and \(aC^{(3)}\) are comparable.

Furthermore, if \(c_{SW}\) is nonperturbatively tuned, there are no \(O(a)\) effects and \(K_{X1}\) vanishes, \(P^{(3)}\) gives no contributions and \(\tilde C^{(3)} = C^{(3)}\) to this order.  Finally, we remark that, in the continuum limit \(a\to0\), chiral symmetry is broken only by the {qcEDM} term as \(m\to0\), therefore, its entire effect can be rotated away, i.e., \(\tilde C^{(3)}\) gives no contribution to physics in these limits~\cite{Bhattacharya:2015rsa}.} At finite \(a\), the same holds true at \(O(a)\).\looseness-1

\begin{figure*}   
  \centering
  \includegraphics[width=0.24\textwidth]{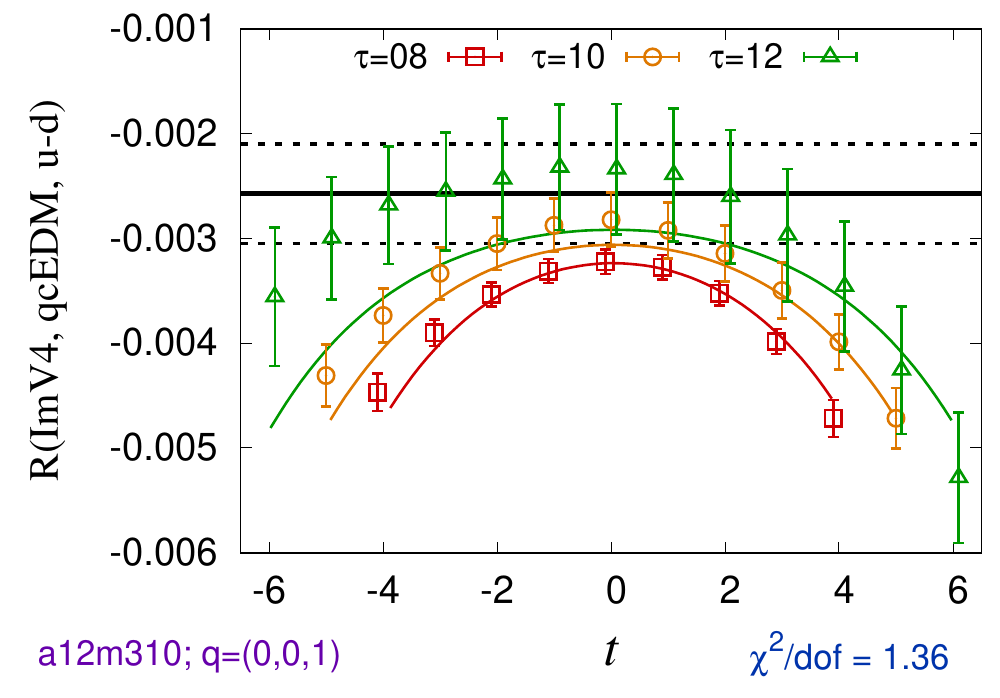}
  \includegraphics[width=0.24\textwidth]{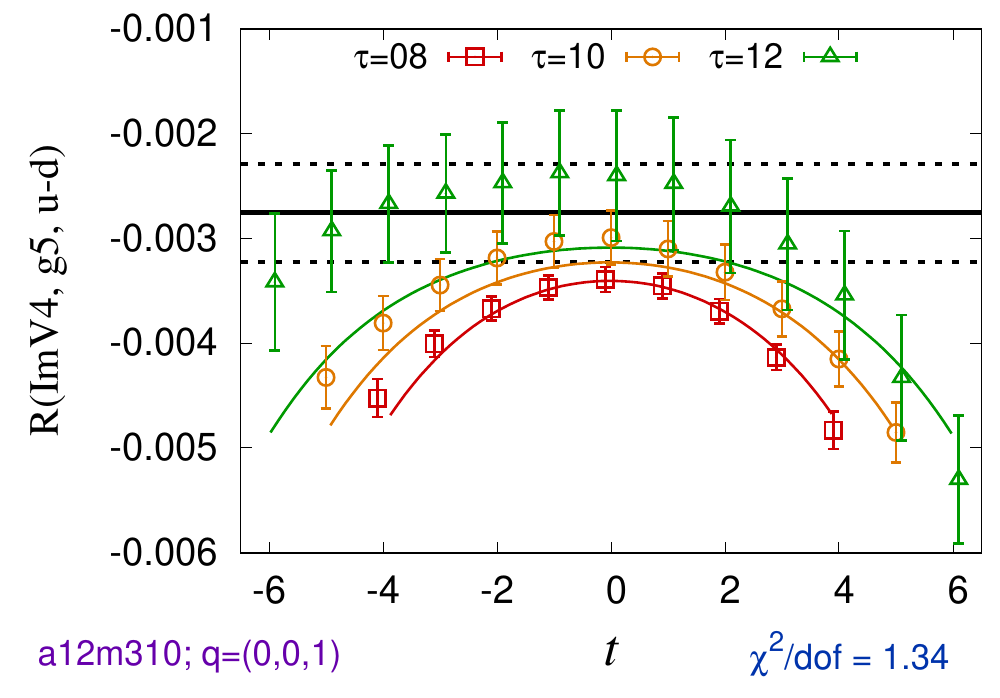}
  \includegraphics[width=0.24\textwidth]{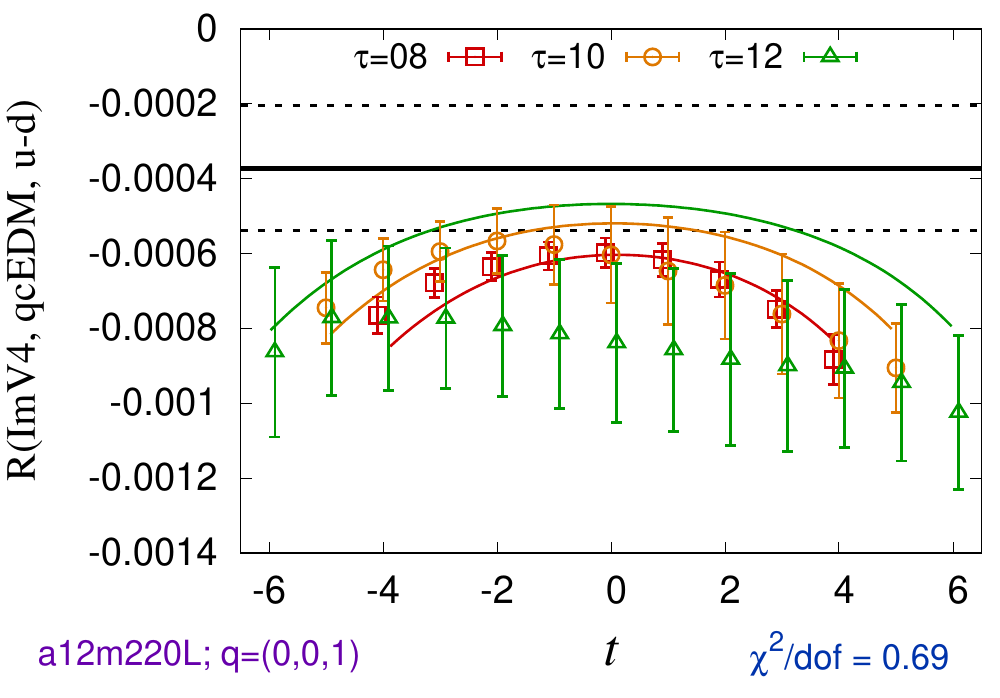}
  \includegraphics[width=0.24\textwidth]{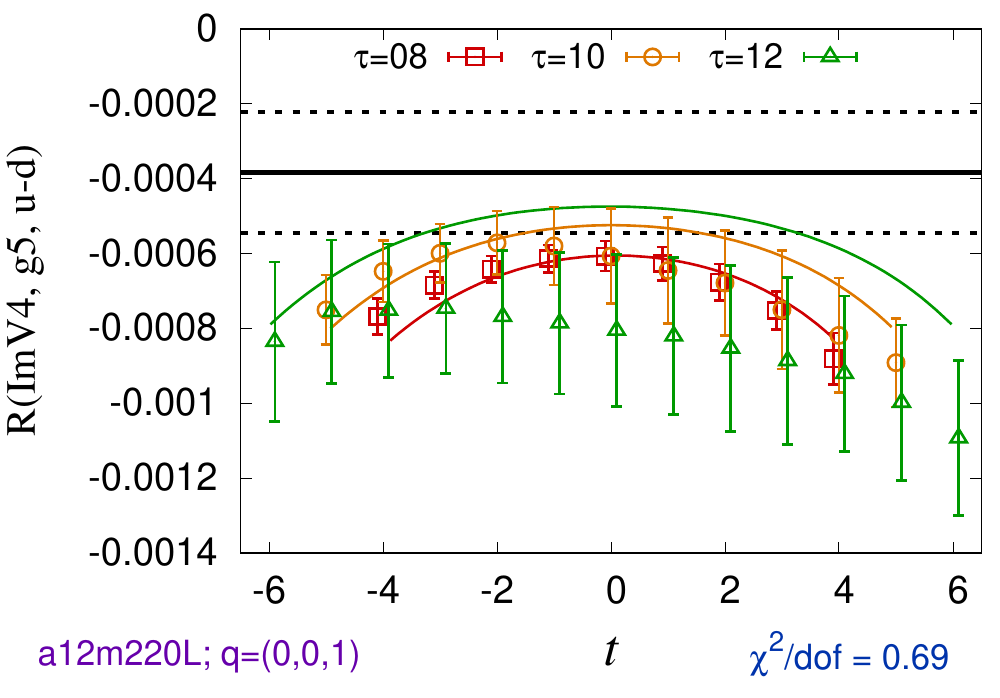} \\
  \label{ESCn}
  \includegraphics[width=0.24\textwidth]{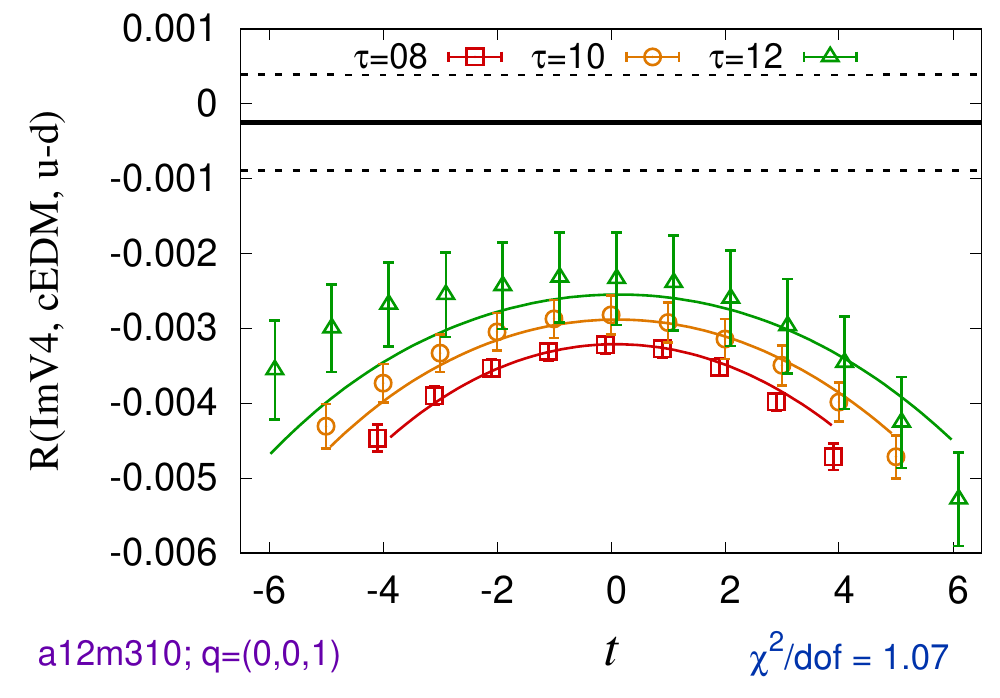}
  \includegraphics[width=0.24\textwidth]{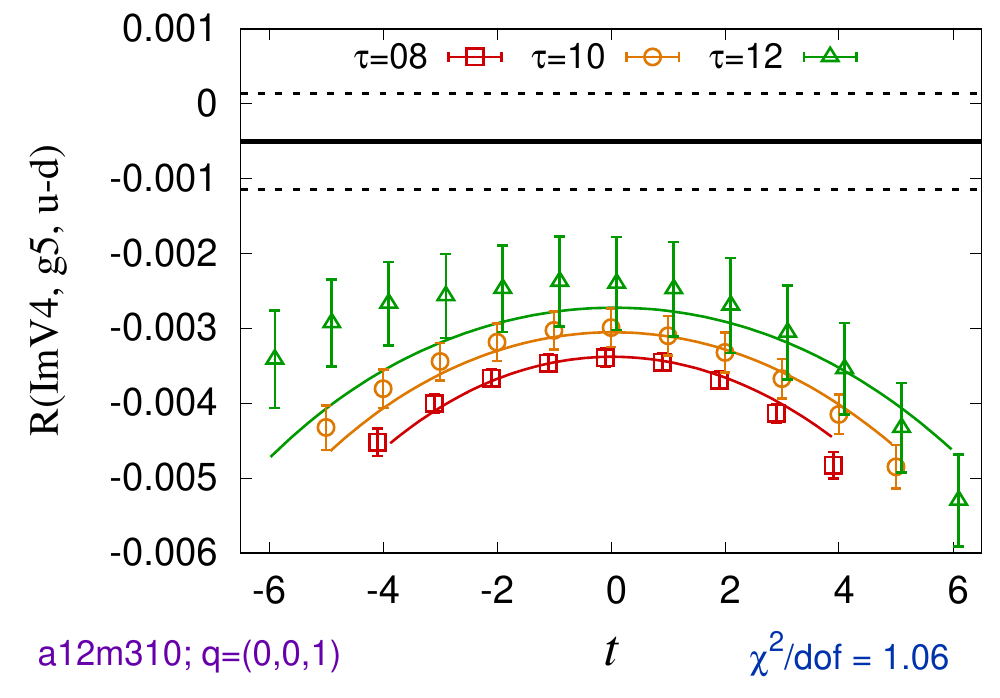}
  \includegraphics[width=0.24\textwidth]{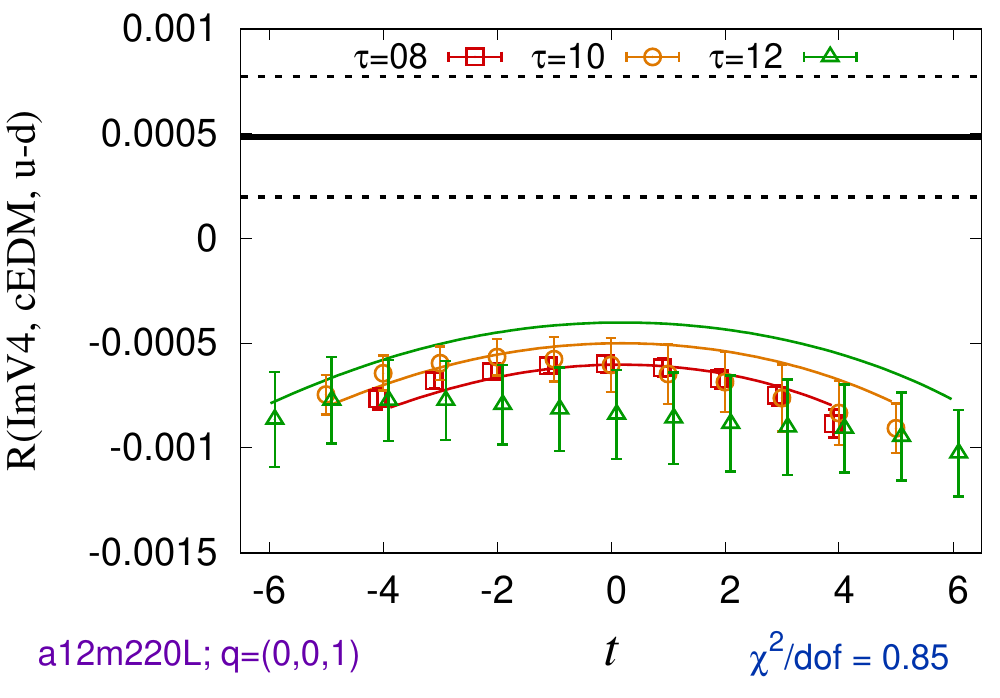}
  \includegraphics[width=0.24\textwidth]{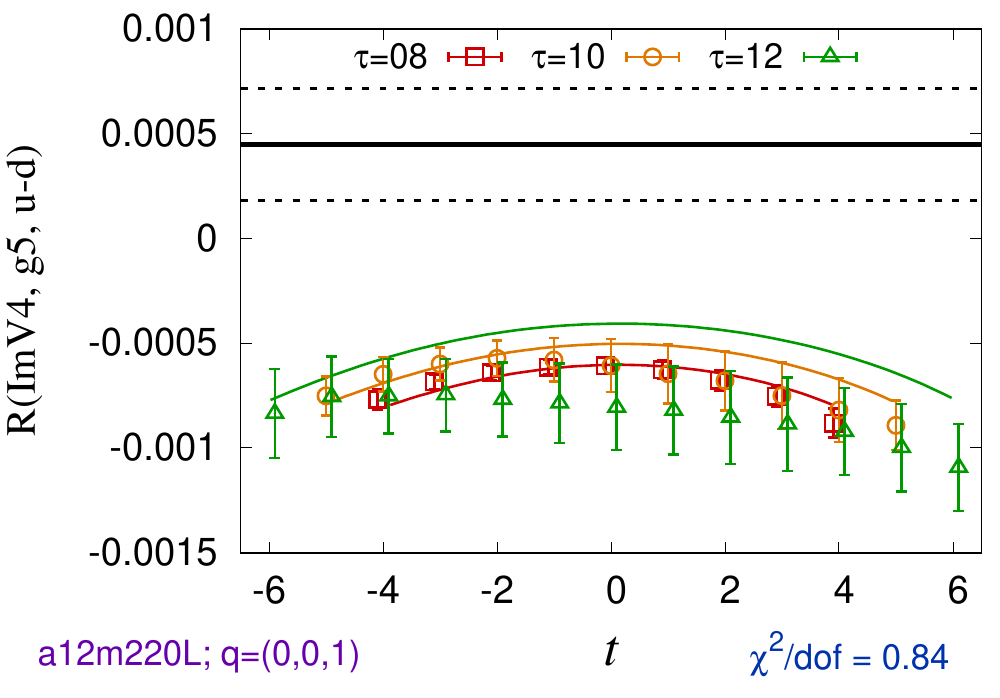}
  \caption{Fits to remove ESC and obtain ground state matrix elements  from correlation functions with the insertion of \(\rm qcEDM\) and $\gamma_5$ operators along with the $V_4$ current at momentum transfer $q=(0,0,1)$. Data are shown for the {a12m310 and a12m220L} ensembles.  (Top) ``Standard'' excited state fit with the mass gaps taken from fits to the two-point function. (Bottom) ``$N\pi$'' excited state fits assuming the first excited state is  \(N(0,0,1)\pi(0,0,0)\), i.e., the energy gap is \(M_\pi\). The form factor ${\widetilde F}_3$ is extracted from such matrix elements.} 
  \label{fig:ESCnpi}
\end{figure*}

\begin{figure*}[tp]    
  \centering
  \includegraphics[width=0.24\textwidth]{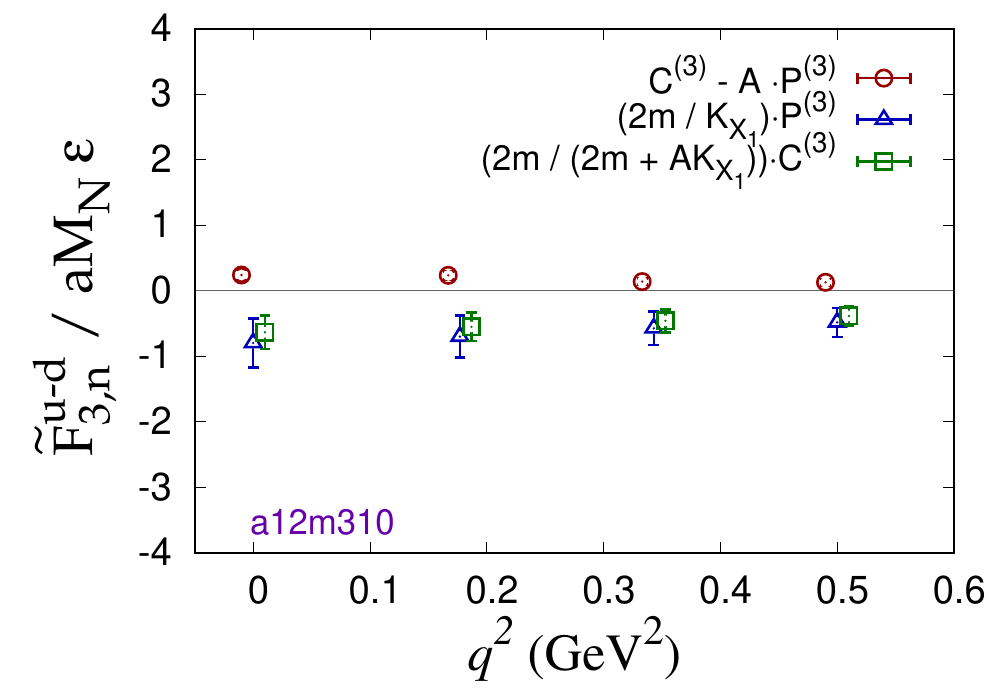}
  \includegraphics[width=0.24\textwidth]{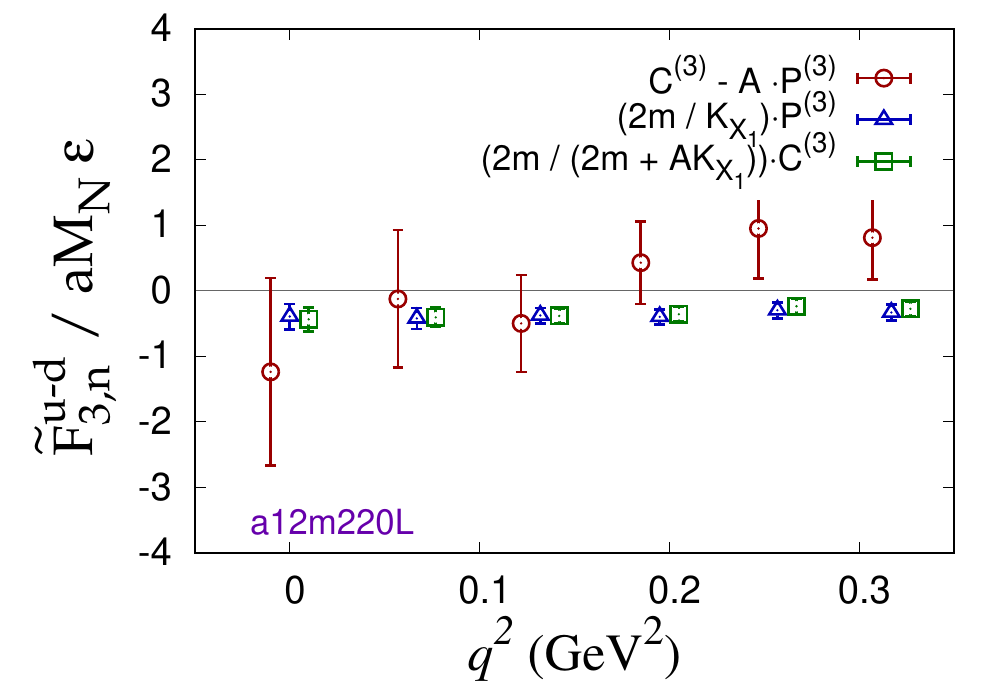}
  \includegraphics[width=0.24\textwidth]{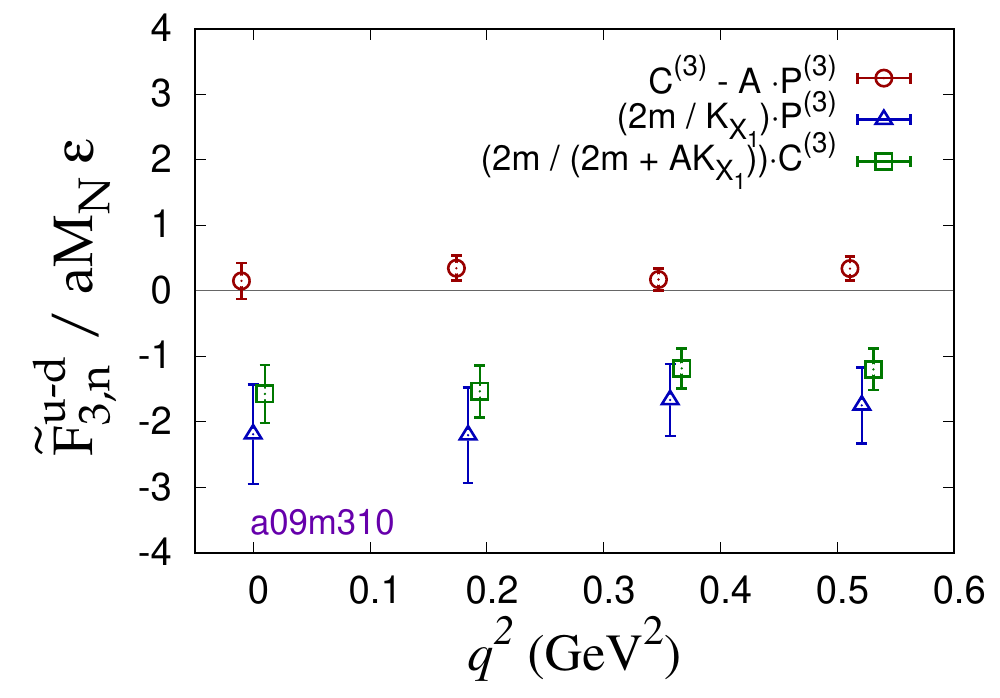}
  \includegraphics[width=0.24\textwidth]{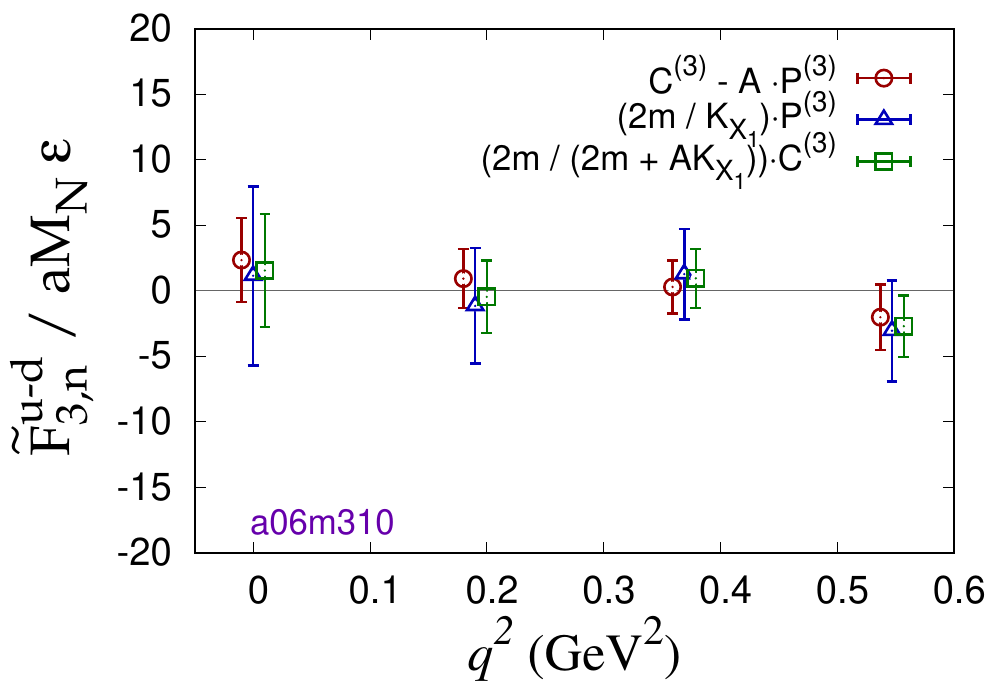}\\
  \includegraphics[width=0.24\textwidth]{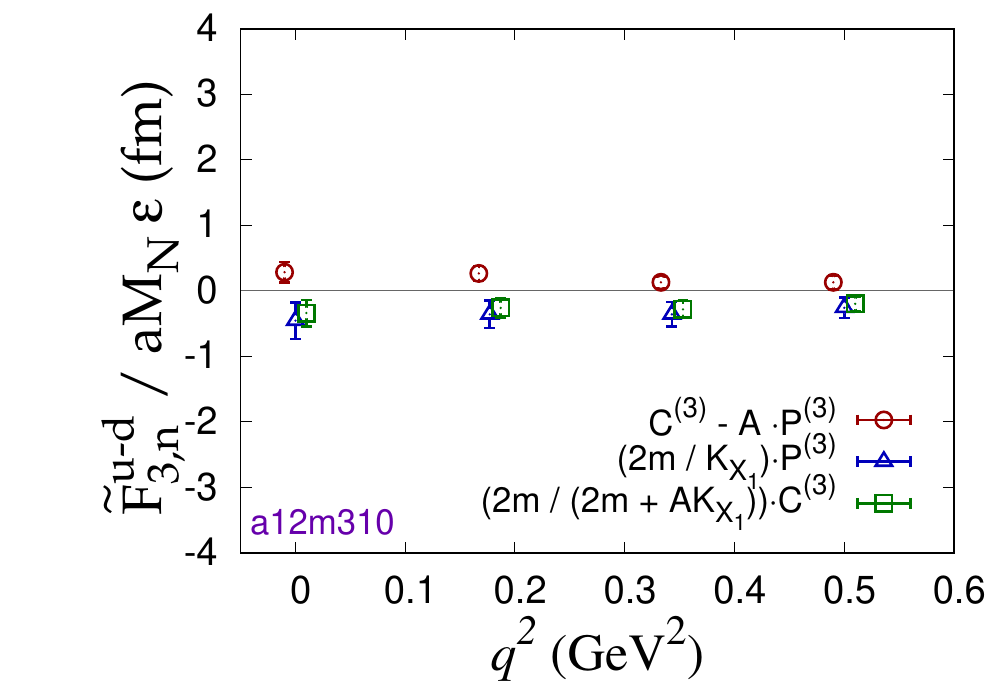}
  \includegraphics[width=0.24\textwidth]{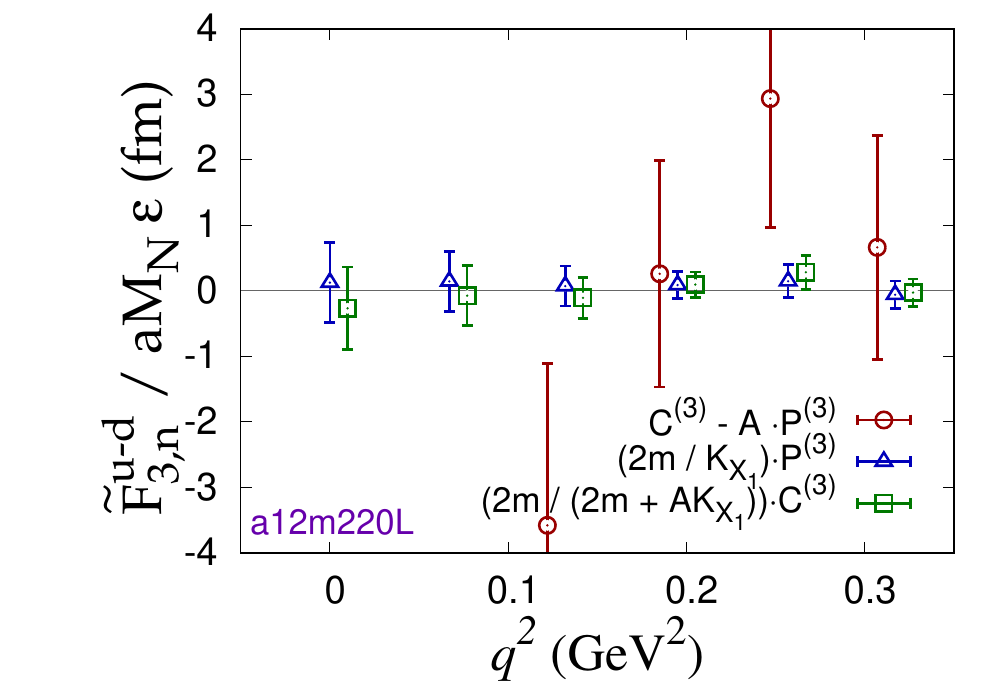}
  \includegraphics[width=0.24\textwidth]{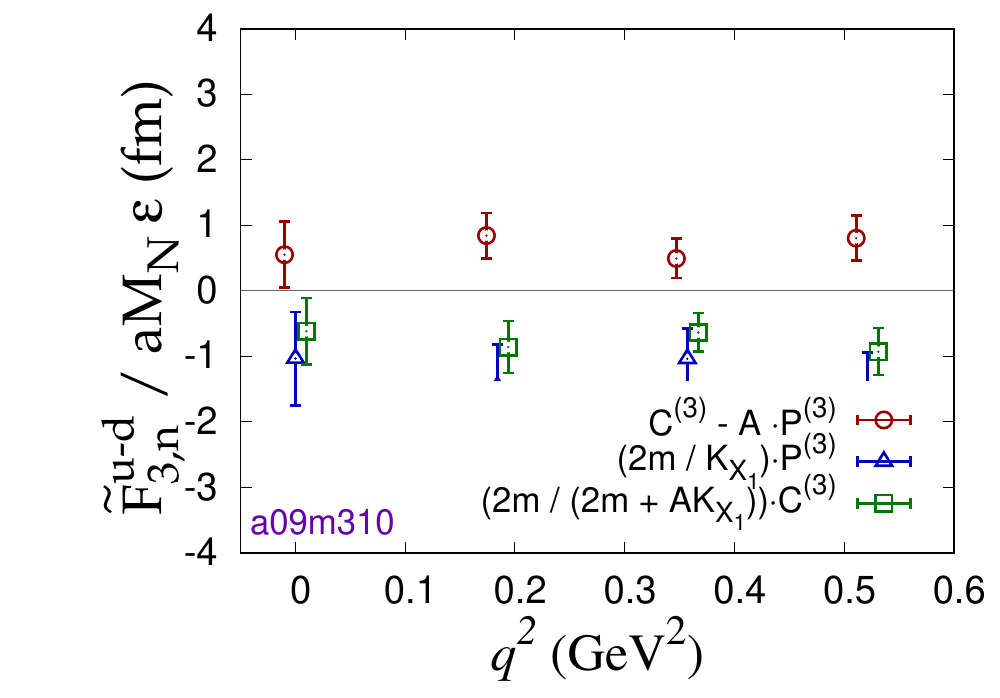}
  \includegraphics[width=0.24\textwidth]{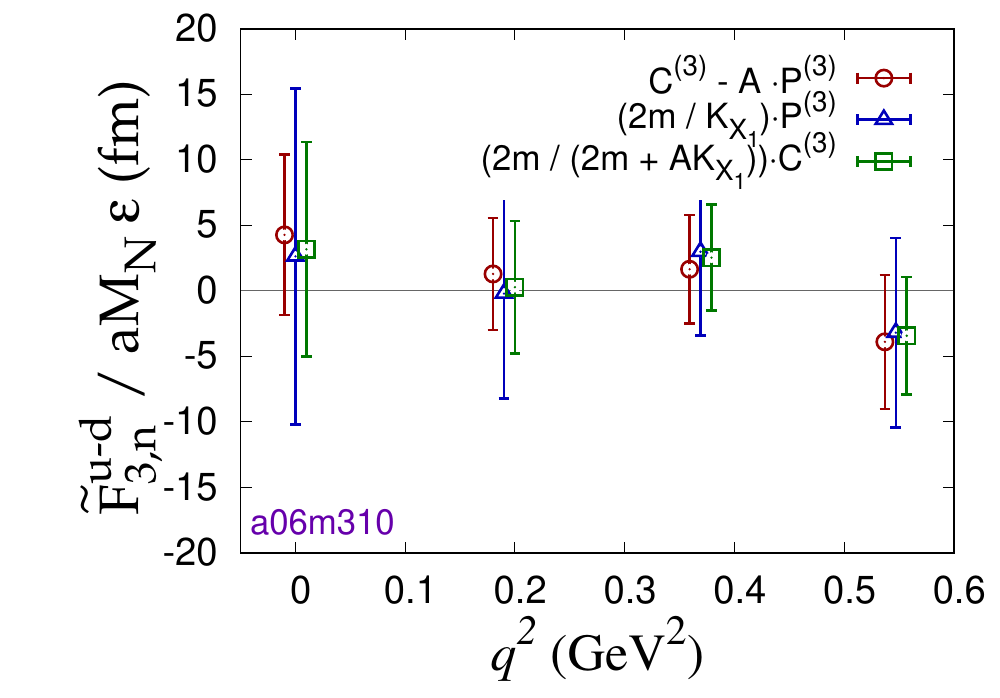}\\
  \caption{The form factor ${\tilde F}_{3,n}^{u-d}/aM_N\varepsilon$ using the subtracted qcEDM operator for the three different subtraction approaches specified in the labels and given in \protect\cref{eq:sub1,eq:P-effect,eq:C-effect}. Top figures are the results from the four ensembles using the ``Standard'' excited state fit, and the bottom figures are the result of using the ``$N\pi$'' excited state fits as explained in the text. Differences between the three approaches specified by the labels are due to residual $O(a)$ artifacts in our action, which vanish in the continuum limit. As explained in the text, we consider estimates using $C^{(3)} - A P^{(3)}$ (red circles) are the least reliable.
  }
  \label{fig:F3Q2}
\end{figure*}

\section{Exploratory numerical calculations}

\relax
\label{sec:Num}
\begin{table*}   
\setlength{\tabcolsep}{10pt}
  \begin{tabular}{|l|l|l|l|l|l|l|}
    \hline
    \multirow{2}{*}{Ensemble}&\multicolumn5{c|}{{\({\tilde F}_3^{\gamma_5}/{\tilde F}_3^{\strut\rm qcEDM}\)}}
       &\multirow{2}{*}{\(\displaystyle\frac{K_{X1}}{2am + A K_{X1}}\)}\\
       \cline{2-6}
    &\multicolumn1{c|}{\(Q^2=1\)}&\multicolumn1{c|}{\(Q^2=2\)}&\multicolumn1{c|}{\(Q^2=3\)}&\multicolumn1{c|}{\(Q^2=4\)}&\multicolumn1{c|}{\(Q^2=5\)}&\\
    \hline
    a12m310  & 0.879(17) & 0.863(14) & 0.867(18) & 0.844(23) & 0.864(13) & 0.694(48)  \\
    a12m220L & 0.81(10)  & 0.769(77) & 0.869(75) & 0.98(18)  & 0.94(11)  & 0.7807(70) \\
    a09m310  & 1.063(35) & 1.042(40) & 1.078(45) & 1.006(58) & 1.039(44) & 0.740(61)  \\
    a06m310  & &&&&
    & 0.859(64)  \\
    \hline
  \end{tabular}
  \caption{{The ratio \( {\tilde F}_3^{\gamma_5} / {\tilde F}_3^{\rm qcEDM}\) for the \(\gamma_5\) and \(\rm qcEDM\) unsubtracted lattice operators for the five smallest values of $Q^2$. As expected, the ratios are, within errors, independent of \(Q^2\) and the quark mass, and close to the \(K_{X1}/(2am + A K_{X1})\) obtained from the pion correlators (last column) using Eq.~\protect\eqref{eq:FitFormula}. We do not find a significant signal in  \( {\tilde F}_3^{\gamma_5} / {\tilde F}_3^{\rm qcEDM}\) with the current data for the $a06m310$. The data for ${\tilde F}_3$ are obtained using the ``standard'' method for removing excited state contamination. }}
  \label{tab:F3rat}
\end{table*}

\begin{figure*}[tp]   
  \centering
  \includegraphics[width=0.24\textwidth]{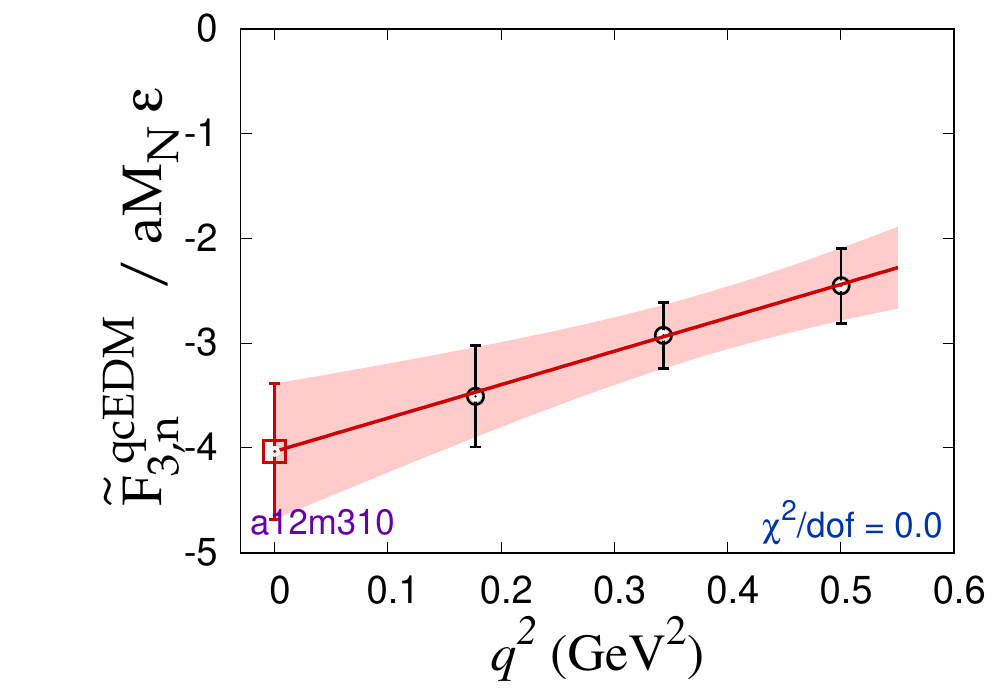}
  \includegraphics[width=0.24\textwidth]{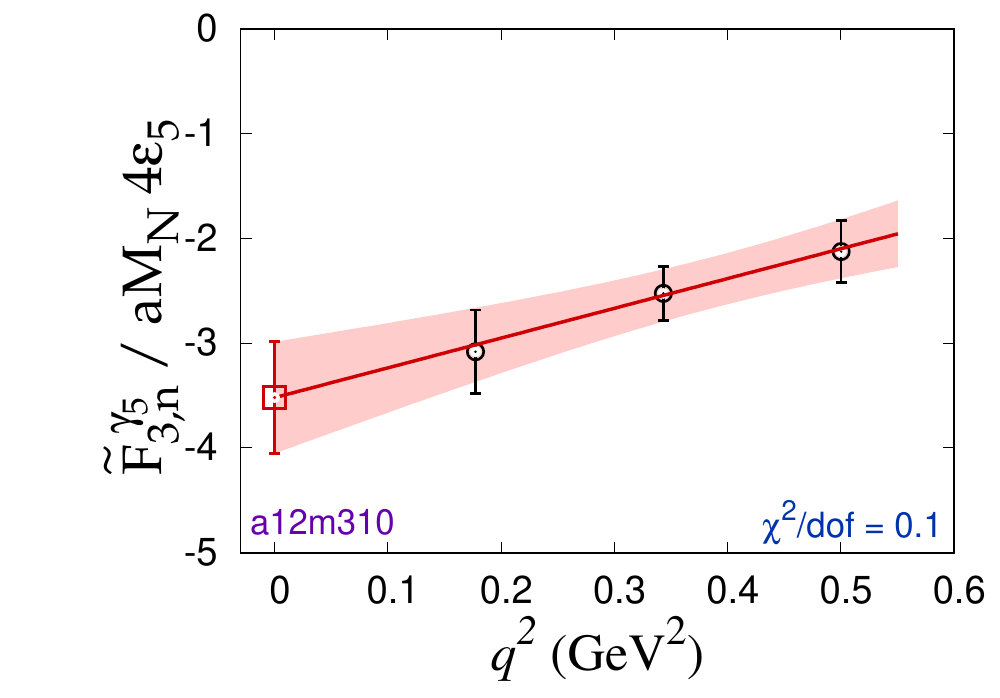}
  \includegraphics[width=0.24\textwidth]{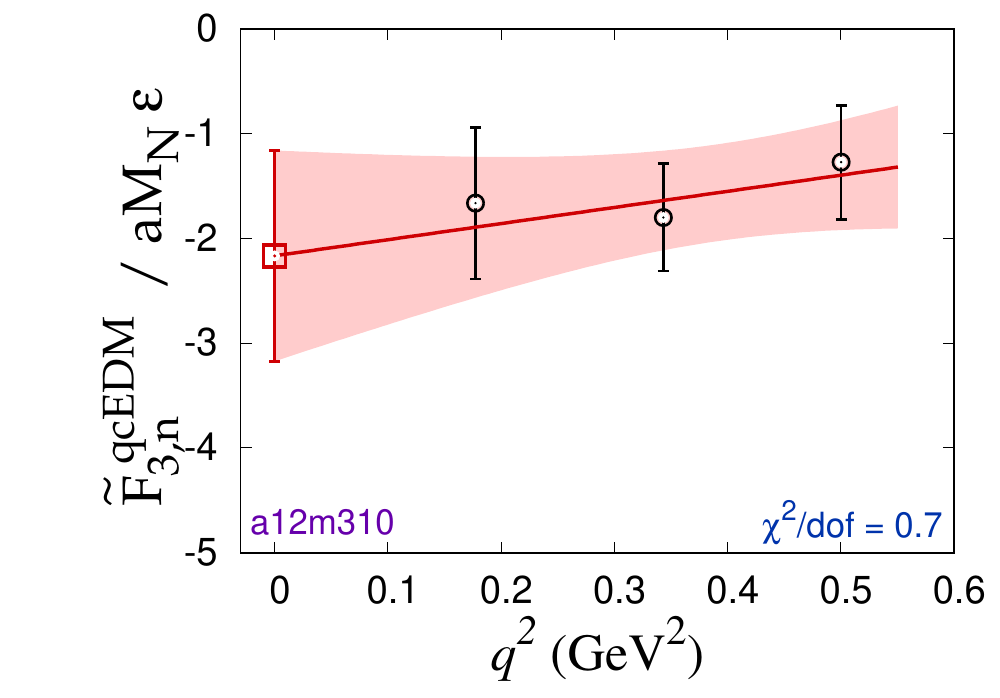}
  \includegraphics[width=0.24\textwidth]{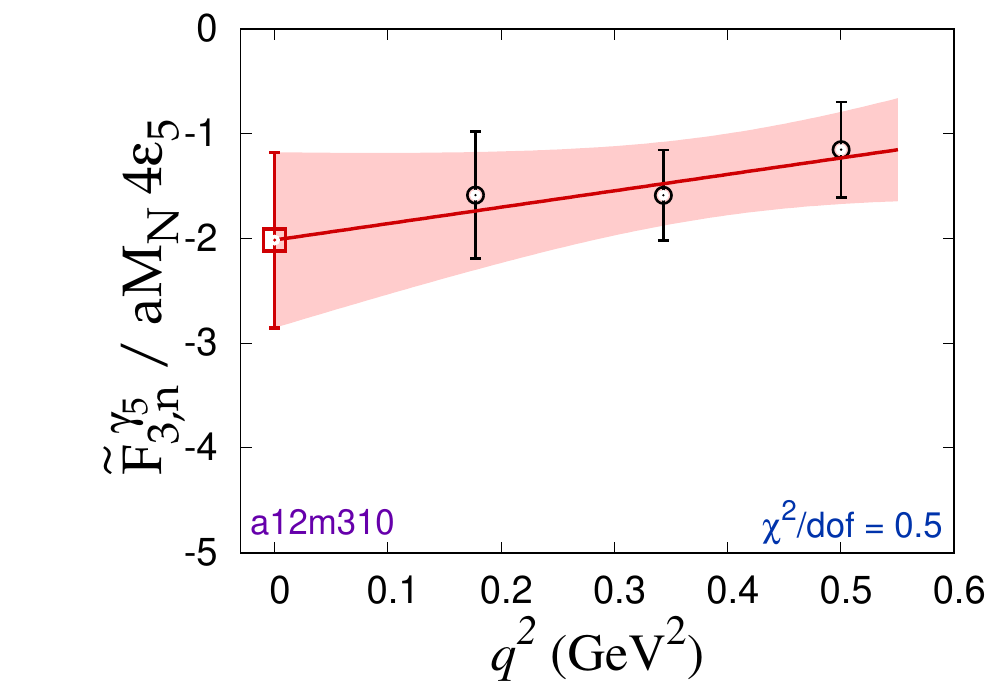}\\
  \includegraphics[width=0.24\textwidth]{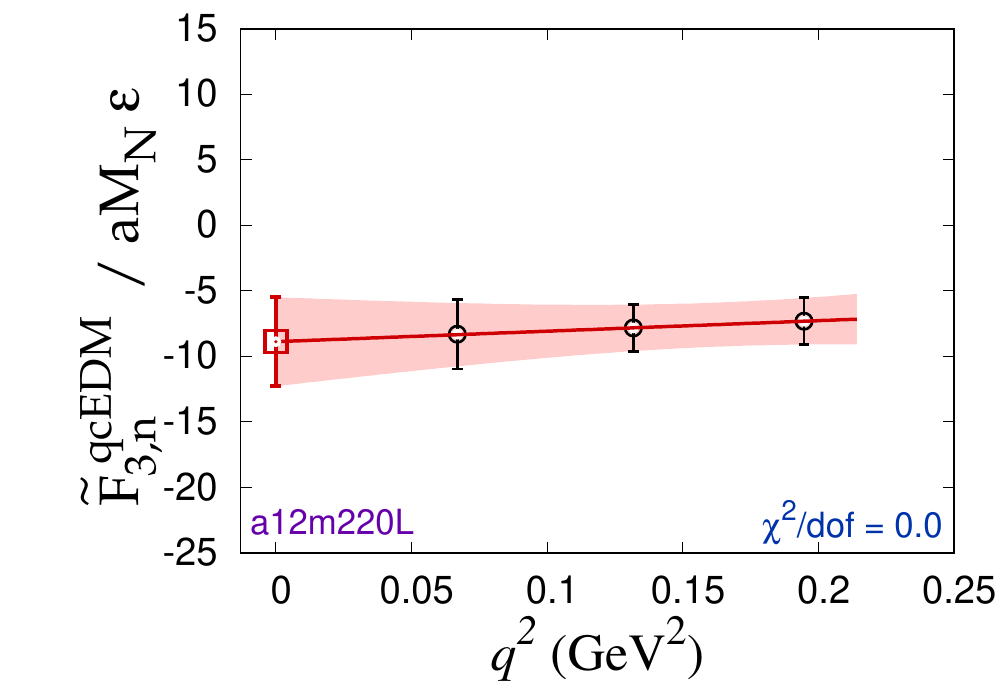}
  \includegraphics[width=0.24\textwidth]{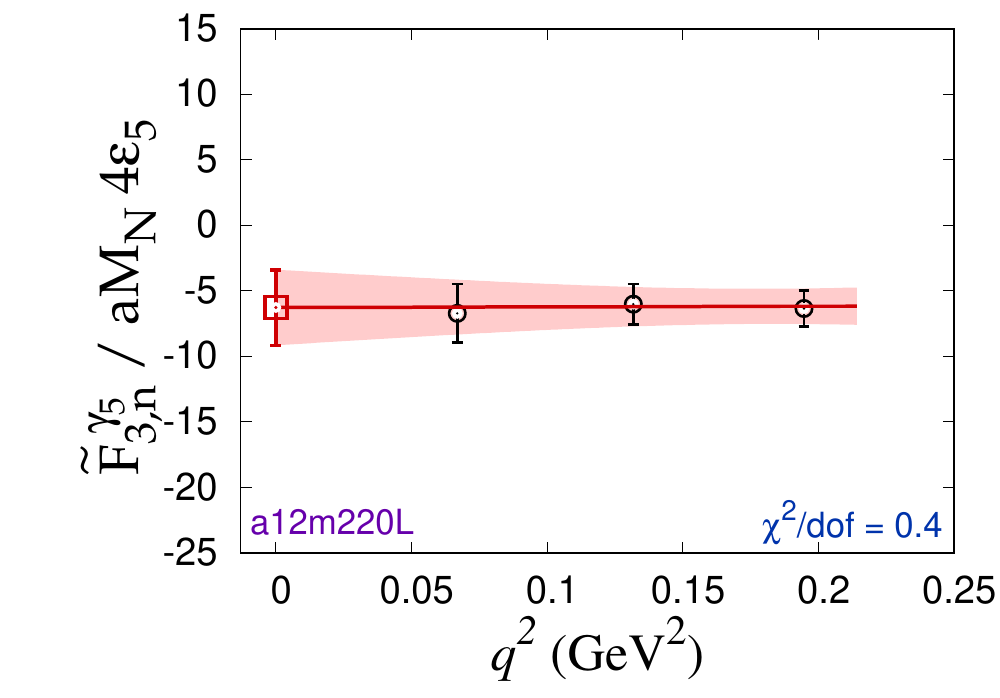}
  \includegraphics[width=0.24\textwidth]{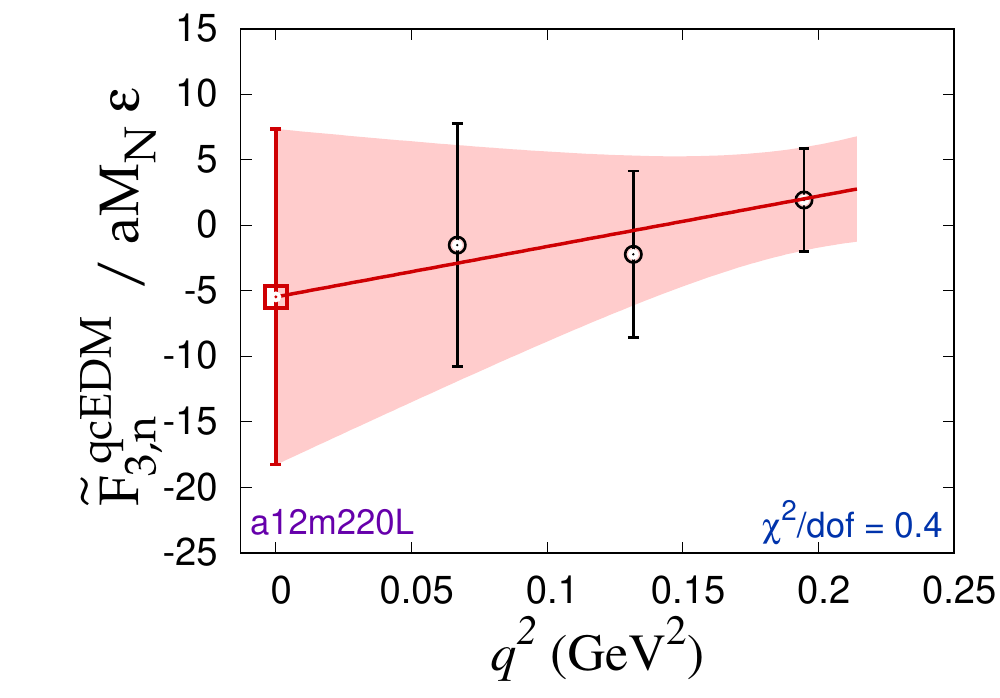}
  \includegraphics[width=0.24\textwidth]{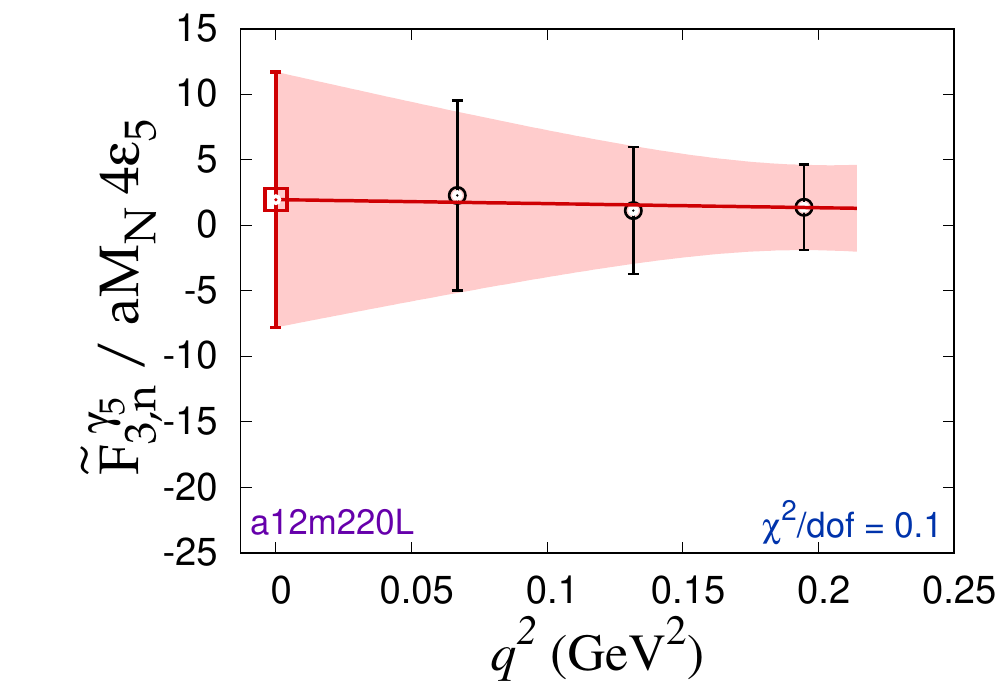}\\
  \includegraphics[width=0.24\textwidth]{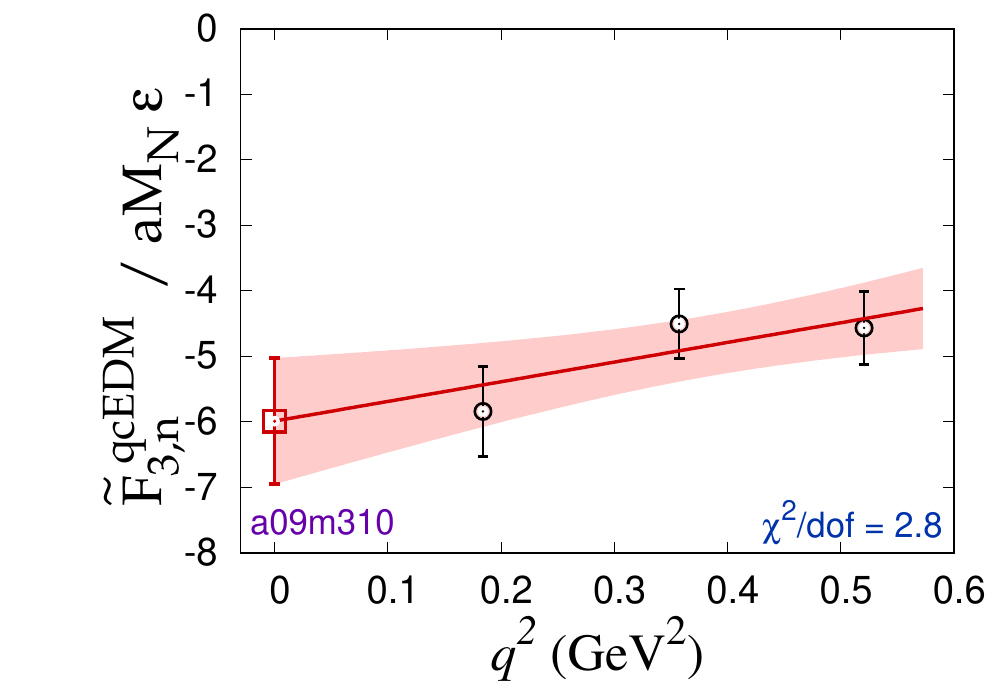}
  \includegraphics[width=0.24\textwidth]{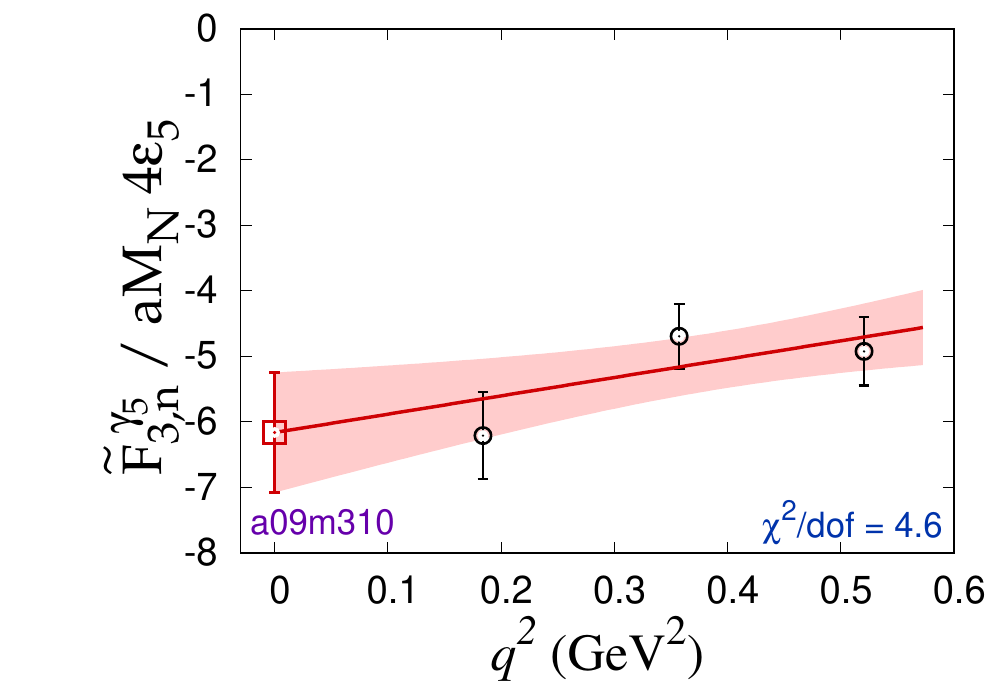}
  \includegraphics[width=0.24\textwidth]{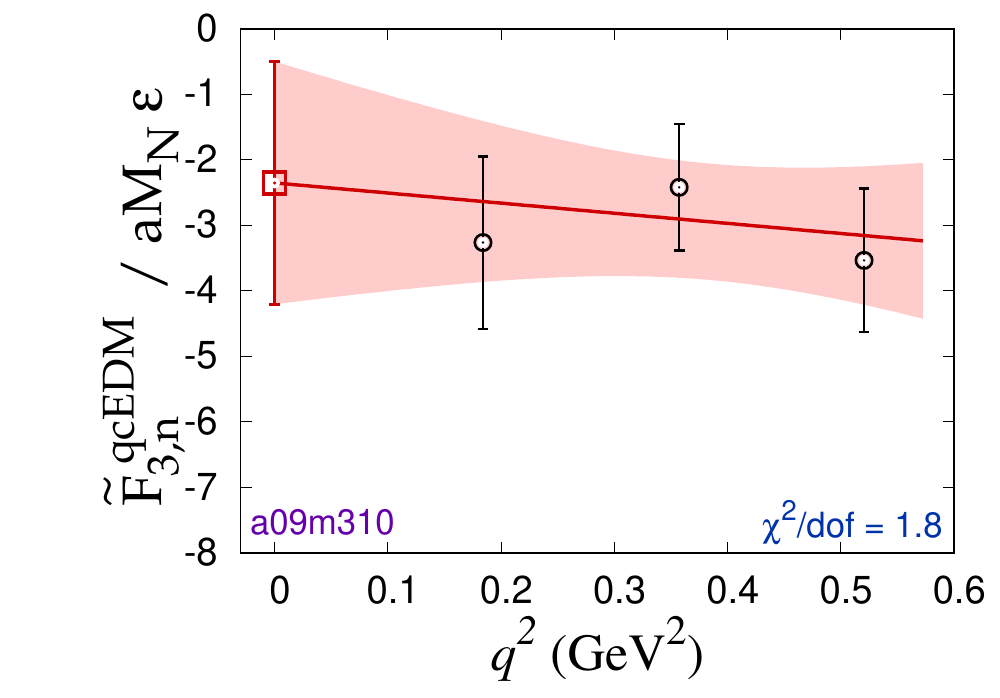}
  \includegraphics[width=0.24\textwidth]{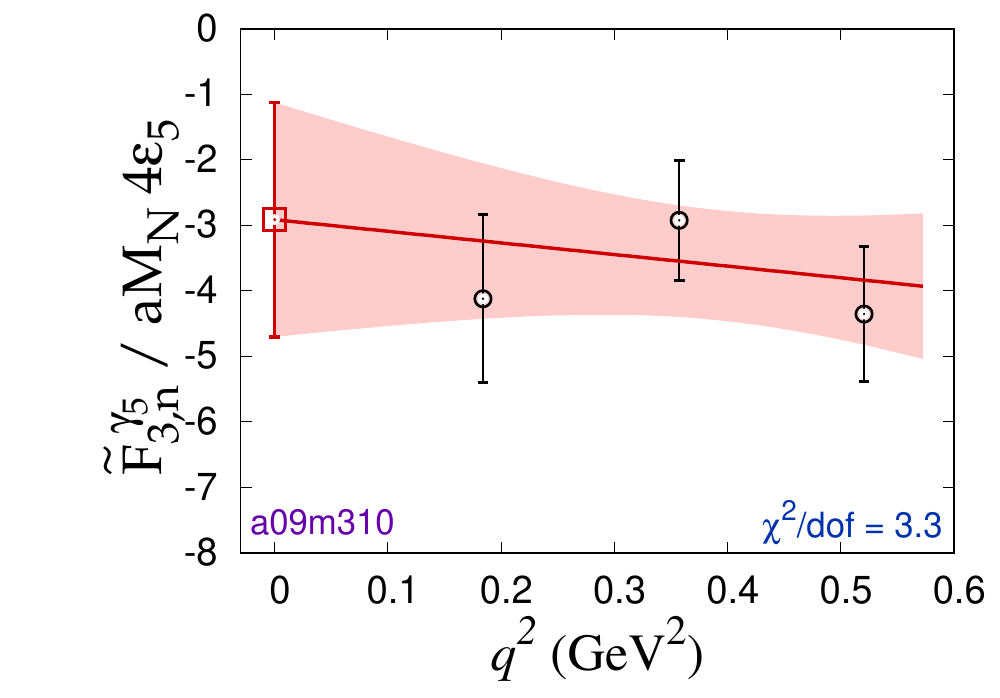}\\
  \includegraphics[width=0.24\textwidth]{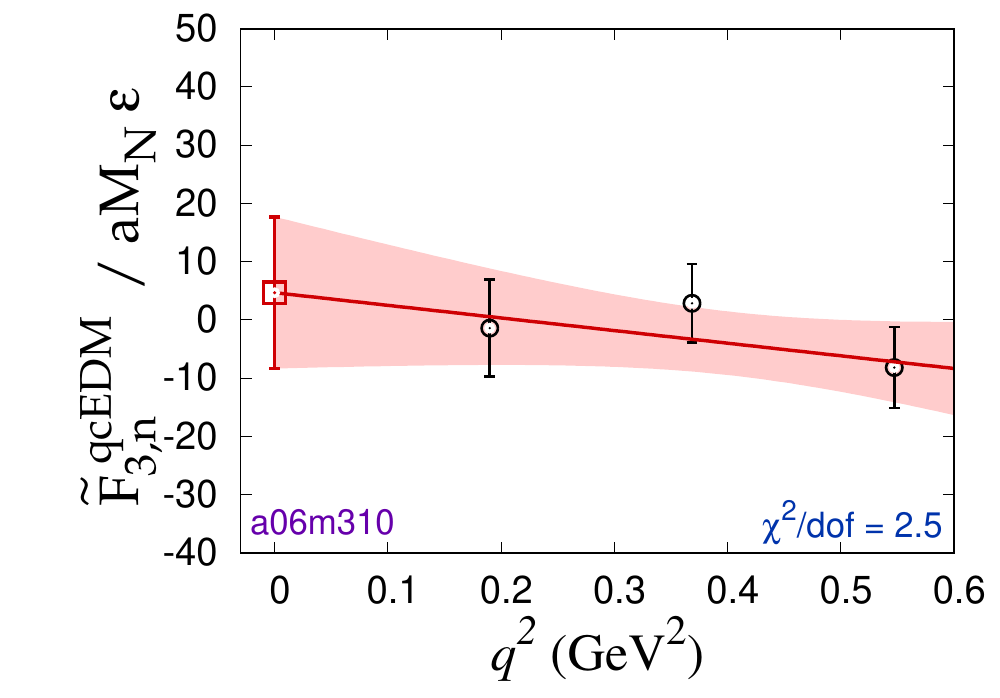}
  \includegraphics[width=0.24\textwidth]{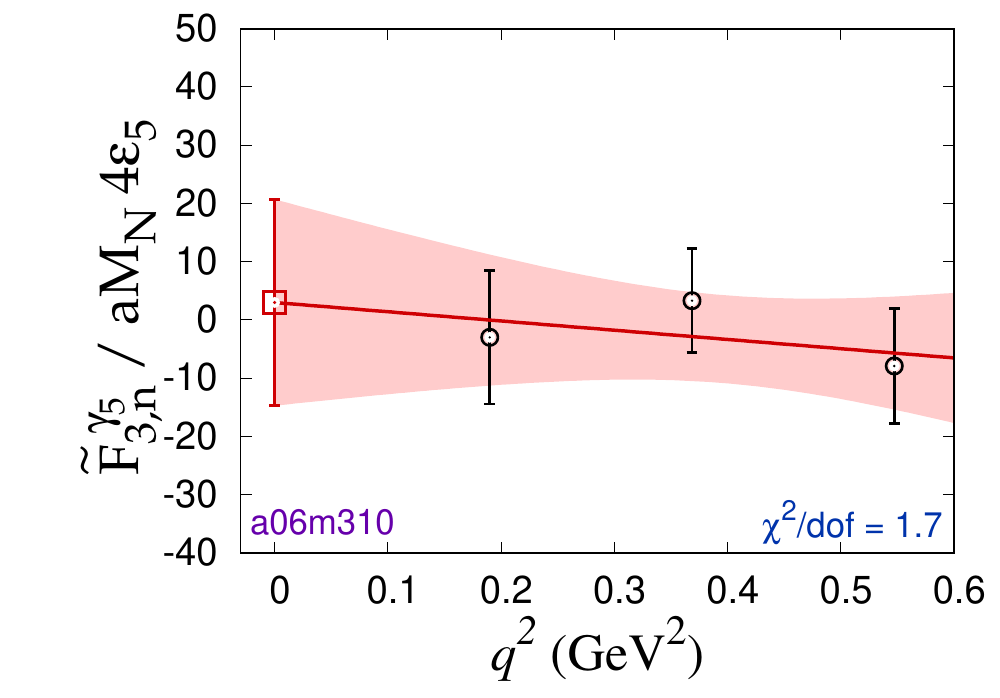}
  \includegraphics[width=0.24\textwidth]{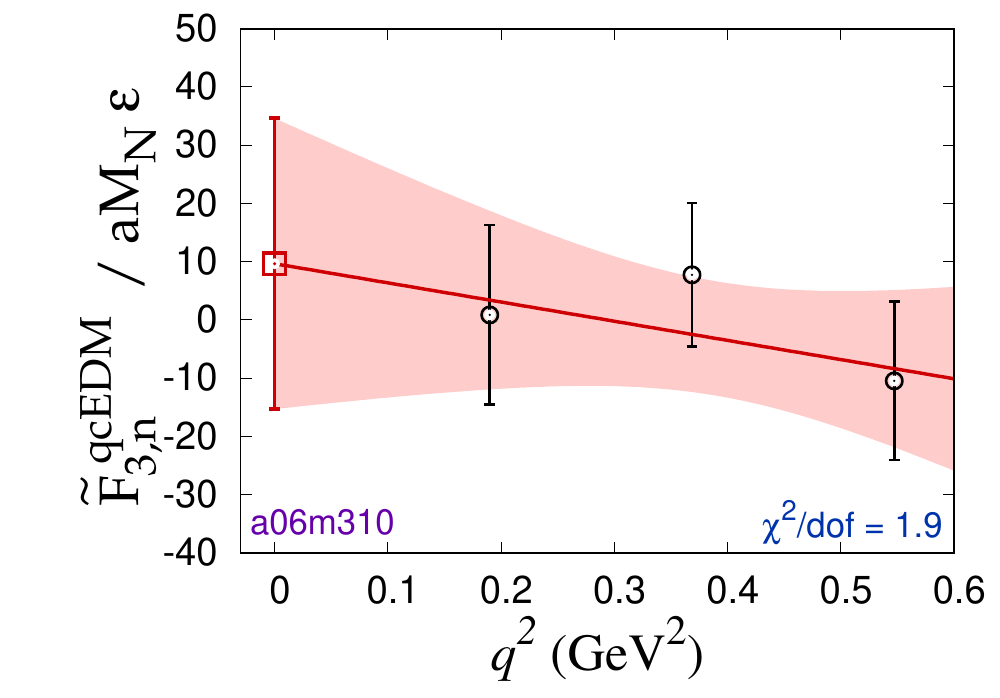}
  \includegraphics[width=0.24\textwidth]{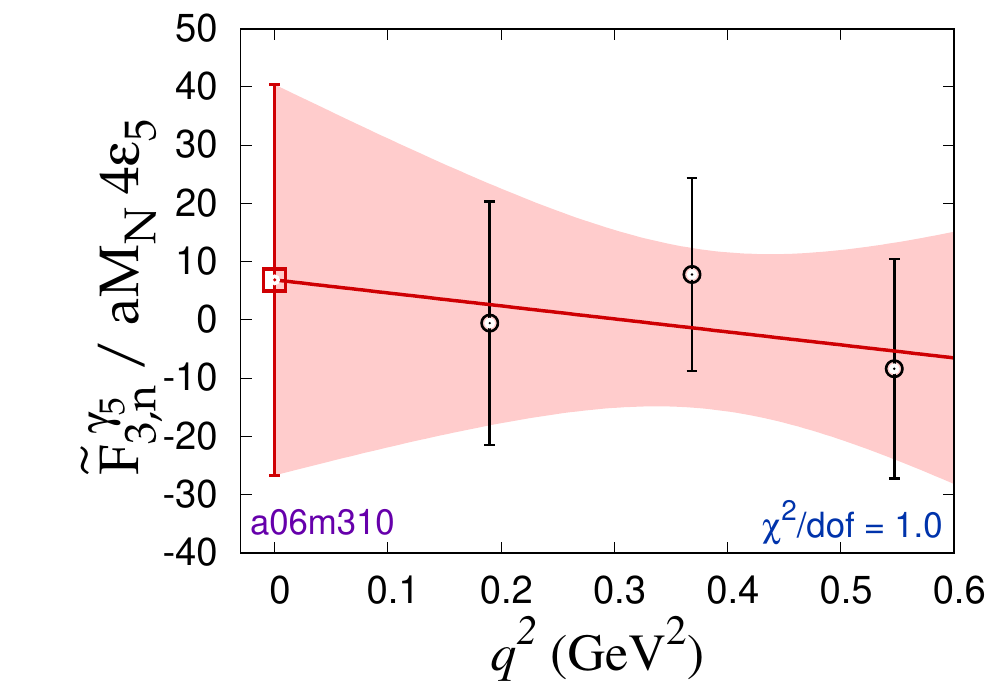}
  \caption{Dependence of \({\tilde F}_3\) on \(Q^2\) for the neutron obtained from the insertion of the 
  qcEDM and $\gamma_5$ operators. Data in the left two columns are  obtained using the ``Standard'' fit to control ESC, and in the right two columns, using the  ``$N\pi$''. The latter have much larger errors. }
  \label{fig:F3Q3}
\end{figure*}

We use the  method previously described in Ref.~\citep{Bhattacharya:2021lol} for calculating the contribution of \CPV\ interactions that are based on extracting the form-factor $F_3$.  In this method, it is essential to remove the ESC in going from 3-point functions to matrix elements as discussed for the \CPV\ $\Theta$-term in Ref.~\citep{Bhattacharya:2021lol}. The methods we use are described in~\cite{Jang:2019vkm,Gupta:2021ahb,Park:2021ypf,Bhattacharya:2021lol}.  

We also caution the reader that, here on, all the data for the \CPV\ form factor $F_3$ will be presented in terms of ${\tilde F}_3$ defined in 
\cite{Bhattacharya:2021lol}. It can be extracted more reliably on the lattice and 
${\tilde F}_3(0)=F_3(0)$ in the limit of interest, $Q^2=0$.

In~\cref{fig:ESCnpi}, we show examples of fits used to remove ESC in correlation functions with the insertion of $V_4$ in the presence of  the qcEDM and $\gamma_5$ operators from which \({\tilde F}_3\) is extracted. In these fits, we consider two possible values for the first excited-state energy as discussed in Ref.~\cite{Bhattacharya:2021lol} for the similar case of the $\Theta$-term:  (i) the ``standard'' fit with the energy  given by the two-point function, and (ii) the ``$N\pi$'' fit using the non-interacting energy of the \(N\pi\) state.  While the results depend sensitively on the excited state spectrum used in the analysis, the fits have similar $\chi^2/dof$, i.e., the fits do not provide an objective selection criteria.  The data in \cref{fig:F3Q2} and the final 
results in \cref{sec:mixing} highlight the size of 
this uncontrolled systematic, which needs to be addressed in future calculations. 

We now discuss the calculation of the parameter $K_{X1}$, which arises if there is residual $O(a)$ chiral symmetry breaking, and the extraction of ${\tilde F}_3$. The data are from the ``standard'' method for removing excited state contamination.

From \cref{eq:P-effect,eq:C-effect}, we note that 
the ratio \({\tilde F}_3(P^{(3)})/F_3(C^{(3)}) = K_{X1}/(2am + A K_{X1})+O(a^2)\).  We, however, notice that the \(O(a^2)\) terms are likely to be large; in fact, the \(O(a^2\Lambda_{\rm QCD}^2)\sim 0.01\hbox{--}0.04\) corrections on the right hand side of \cref{eq:FitFormula} may provide a substantial correction to the leading term, \(O(2am)\sim0.01\). 
We study this by noting that under the assumption that corrections 
are small, all numbers in a given row of \cref{tab:F3rat} should agree. This is roughly true for the different $Q^2\strut$, but the $O(25\%)$ difference between 
 \( {\tilde F}_3^{\gamma_5} / {\tilde F}_3^{\rm qcEDM}\) and  \(K_{X1}/(2am + A K_{X1})\), is indicative 
 of $O(a^2)$ effects. A similar effect is anticipated in  \(\tilde{F}_3(Q^2)\) calculated using the three subtraction schemes defined in \cref{sec:AWI}, and shown in \cref{fig:F3Q2}.  The determinations using \cref{eq:P-effect,eq:C-effect} are close, however, they could both have large \(O(a^2)\) effects coming from the Ward identity. 
 The direct determination, \cref{eq:sub1}, can also have large \(O(a^2)\) effects and is a difference of two numbers of similar size. Furthermore, any difference between estimates using \cref{eq:P-effect,eq:C-effect} will get  magnified by the large factor \(A K_{X1}/2 a m\), to give a large difference from the direct determination, \cref{eq:sub1}, value.  In short, at this stage, we do not have control over \(O(a^2)\) errors. Looking at \cref{fig:F3Q2}, estimates from the direct determination have large fluctuations on the $a12m220L$ ensemble, are consistent with the other two on $a06m310$, and show a large difference on the remaining two. Estimates from \cref{eq:P-effect,eq:C-effect} are consistent. Of these, we choose  the results from \cref{eq:C-effect} in our final analyses because they have lower statistical errors.

The extraction of ${\tilde F}_3$ at  \(Q^2 = 0\) is carried out using an  extrapolation linear in $Q^2$ using data at the three smallest values of \(Q^2\) as shown in \Cref{fig:F3Q3}. To estimate the systematic uncertainty due to choosing the linear ansatz,  we use the difference between the extrapolated \(\tilde{F}_3(0)\) and the \(\tilde{F}_3(Q^2)\) at the smallest nonzero \(Q^2\). 

\begin{table*}   
\begin{tabular}{|l|ll|ll|ll|}
\hline
\multirow{2}{*}{Ensemble}&\multicolumn{2}{c|}{qEDM}&\multicolumn{2}{c|}{\(X_c\), Standard Fit}&\multicolumn{2}{c|}{\(X_c\), \(N\pi\) Fit}\\
\cline{2-7}
&\multicolumn{1}{c}{\(g_T^u\)}&\multicolumn{1}{c|}{\(g_T^d\)}&\multicolumn{1}{c}{Lattice}&\multicolumn{1}{c|}{2 GeV}&\multicolumn{1}{c}{Lattice}&\multicolumn{1}{c|}{2 GeV}\\
\hline
a12m310&\(0.859(12)\)&\(-0.206(7)\)&\(\hphantom{-}0.239(86)\)&\(\hphantom{-}0.205(85)\)&\(\hphantom{-}0.28(16)\)&\(\hphantom{-}0.25(16)\)\\
a12m220L&\(0.846(11)\)&\(-0.203(5)\)&\(-1.2(1.4)\)&\(-1.3(1.4)\)&\(-7.8(5.4)\)&\(-7.8(5.4)\)\\
a09m310&\(0.824(7)\)&\(-0.203(3)\)&\(\hphantom{-}0.15(28)\)&\(\hphantom{-}0.17(28)\)&\(\hphantom{-}0.55(50)\)&\(\hphantom{-}0.57(50)\)\\
a06m310&\(0.784(15)\)&\(-0.192(8)\)&\(\hphantom{-}2.3(3.2)\)&\(\hphantom{-}2.4(3.2)\)&\(\hphantom{-}4.3(6.1)\)&\(\hphantom{-}4.4(6.2)\)\\
\hline
\end{tabular}
\caption{Results for \(X_c \equiv d_N/{\tilde d}\), renormalized in the $\overline{\rm MS}$ scheme at $2$~GeV as explained in the text, are given for the two methods used for removing ESC---without (standard fit) and with a $N\pi$ excited state. 
The results for the matrix elements of the qEDM operator, which  are given by the tensor charges, $g_T^u$ and $g_T^d$, are quoted from our published work~\cite{Gupta:2018qil}.  }
\label{tab:renres}
\end{table*}

\section{Renormalization and Chiral-continuum extrapolation}
\label{sec:mixing}

The subtracted isovector {qcEDM} operator $\tilde C^{(3)}(a)$ is free of power divergences but 
still has logarithmic divergences as $a \to 0$. In the calculation of the nEDM,  it is implicit that one has to work with $QCD+QED$ since the operator  \(\tilde C^{(3)}\) has to be inserted together  with the electromagnetic current in the correlation functions. In this theory, \(\tilde C^{(3)}\) mixes with the quark EDM operator 
$E^{(3)}(a)$ defined in \cref{eq:O3def}. One therefore, needs to calculate the mixing and running of these two operators. Only the anomalous dimension matrix (universal) part of this has been calculated at $O(\alpha_s)$~\cite{Bhattacharya:2015rsa}. In this leading-logarithm approximation (tree-level matching and one-loop running), 
the lattice and $\overline{\rm MS}$ operators are related by 
\begin{align}
\vec{O}_{\overline{\rm MS}} (\mu) &= \nonumber\\*
\span\!\!
U \begin{pmatrix}
\left(\frac{\alpha_s (\mu)}{\alpha_s(a^{-1})} \right)^{\mathrlap{-\gamma_{11}/\beta_0}} & 0 
\\
0 & \left( \frac{\alpha_s (\mu)}{\alpha_s(a^{-1})} \right)^{-\gamma_{22}/\beta_0} 
    \end{pmatrix}
U^{-1} \vec{O} (a) \nonumber\\
\label{eq:renorm}
\end{align}
where 
\begin{align}
\vec{O} (a) &= 
    \begin{pmatrix}
    \tilde C^{(3)} (a) \\
    E^{(3)} (a) 
    \end{pmatrix}\nonumber\\
\vec{O}_{\overline{\rm MS}} (\mu) &= 
    \begin{pmatrix}
     C_{\overline{\rm MS}}^{(3)} (\mu) \\
    E_{\overline{\rm MS}} ^{(3)} (\mu) 
    \end{pmatrix}\nonumber\\
    U &=
    \begin{pmatrix}
1 &  - \frac{\gamma_{12}}{\gamma_{11} - \gamma_{22}}
\\
0 & 1
    \end{pmatrix}\,,
\end{align}
and
\begin{equation}
    \frac{\alpha_s(\mu)}{\alpha_s (a^{-1})} = \frac{1}{1 - \frac{\alpha_s (a^{-1} )}{\pi} \, \beta_0 \, \log (\mu a)}~. 
\end{equation}
The coefficients of the $\alpha_s/\pi$ term in the 
beta function and the anomalous dimension matrix are
\begin{align}
\beta_0 &= \frac{2 N_F - 11 N_C}{6}\,,
\end{align}
\begin{equation}
\gamma_{11} = \frac{5 C_F - 2 C_A}{2} ,\   
    \gamma_{12} = 2 C_F ,\ 
    \gamma_{22} = \frac{C_F}{2} 
\end{equation}
{with}
\begin{equation}
    C_F = \frac{N_C^2 - 1}{N_C}~, \qquad\qquad C_A = N_C
\end{equation}
Using $N_F=4$ we obtain 
\begin{equation}
    -\frac{\gamma_{11}}{\beta_0} = \frac{2}{25}~, \quad 
        -\frac{\gamma_{22}}{\beta_0} = \frac{4}{25}~, \quad 
    U = 
\left(
    \begin{array}{cc}
1 & 8 
\\
0 & 1
    \end{array}
    \right)~.
    \label{eq:renorm_end}
\end{equation}
Because the matching is done at tree level and followed by 1-loop running, the renormalization process is insensitive to the scheme. Consequently the data for $X_c \equiv - \tilde{F}_3(0) / aM_N\varepsilon$ (see \cref{eq:Xc2}) carry an unresolved uncertainty of order $\alpha_s (\mu) /\pi$. At this order, one can, therefore, choose to use either the renormalized or unrenormalized tensor charges. We have chosen to use the renormaized values given in Ref.~\citep{Bhattacharya:2015rsa}.

\begin{figure*}[tp]   
  \centering
  \includegraphics[width=0.47\textwidth]{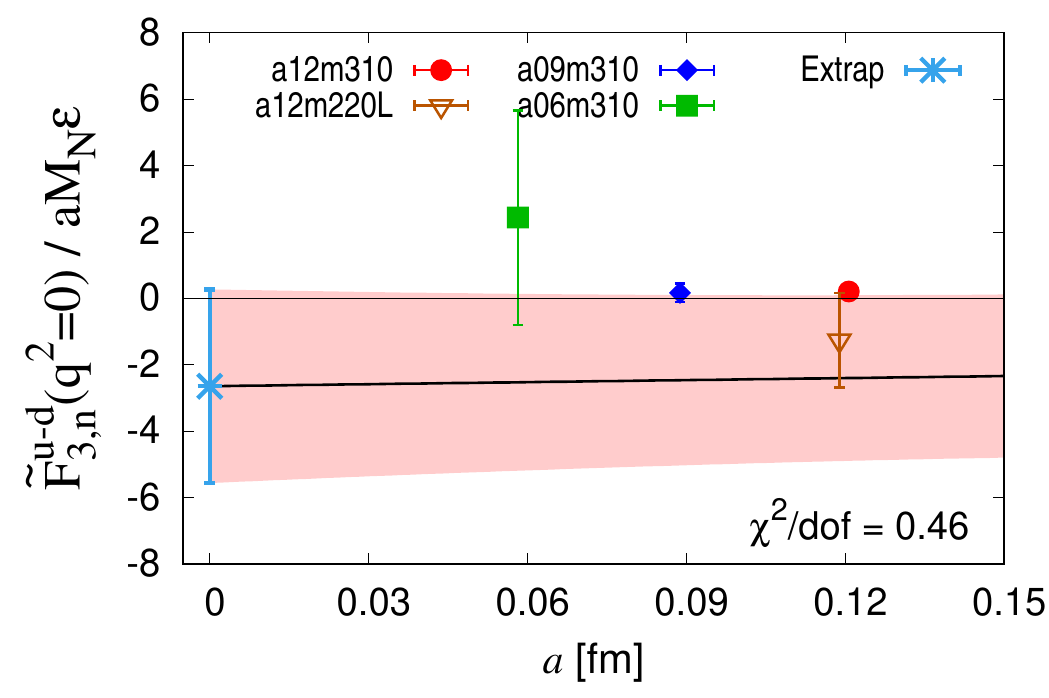}\qquad
  \includegraphics[width=0.47\textwidth]{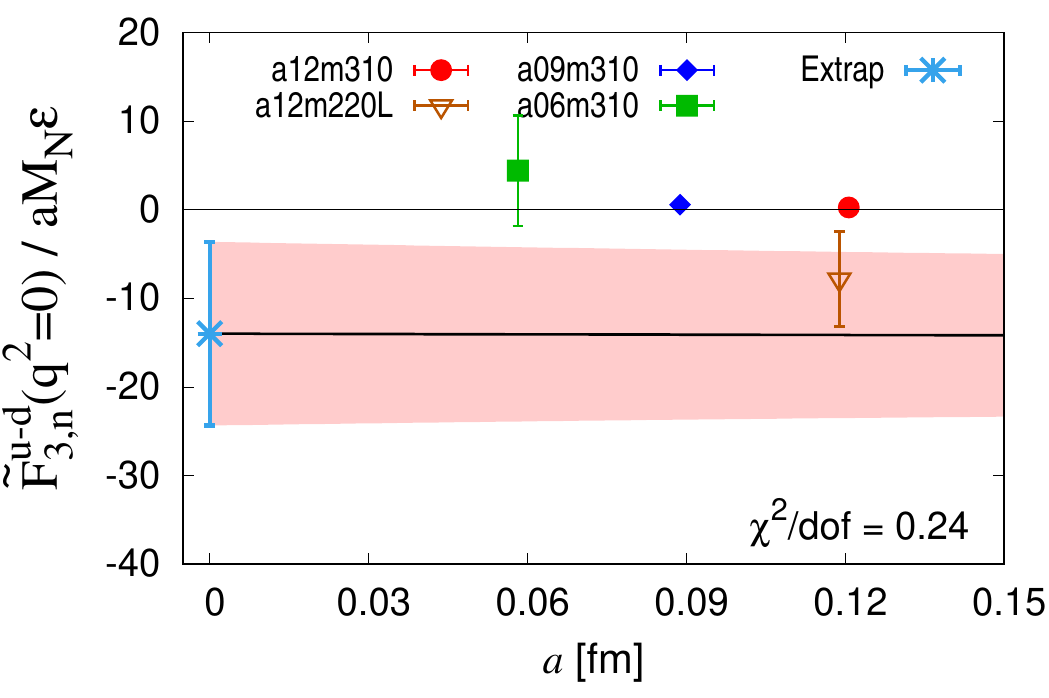}\\
  \includegraphics[width=0.47\textwidth]{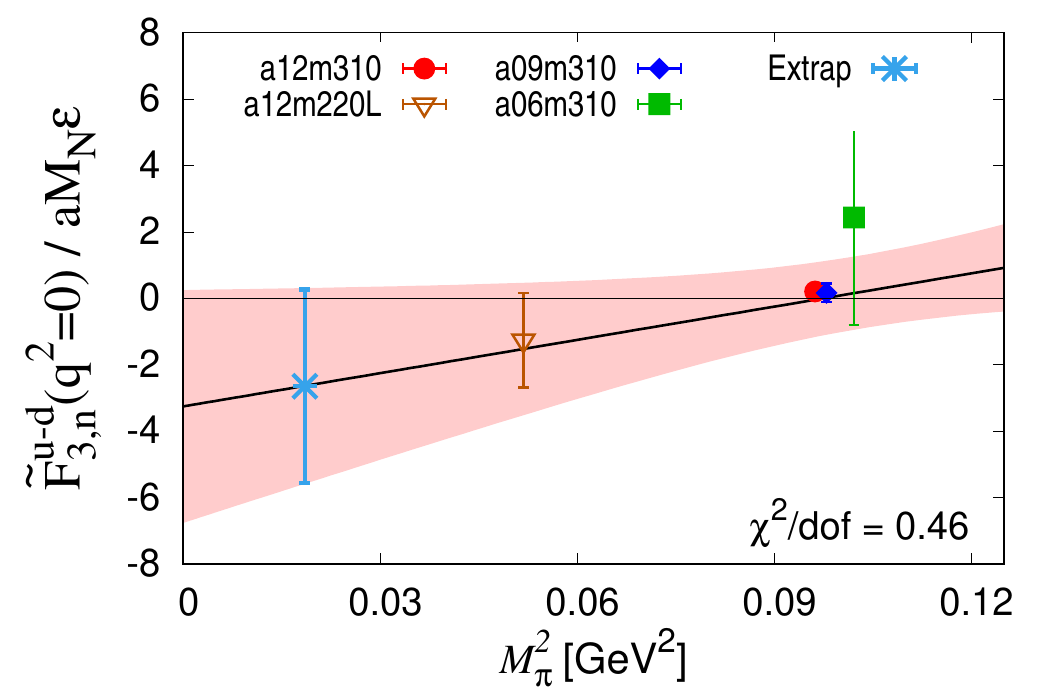}\qquad
  \includegraphics[width=0.47\textwidth]{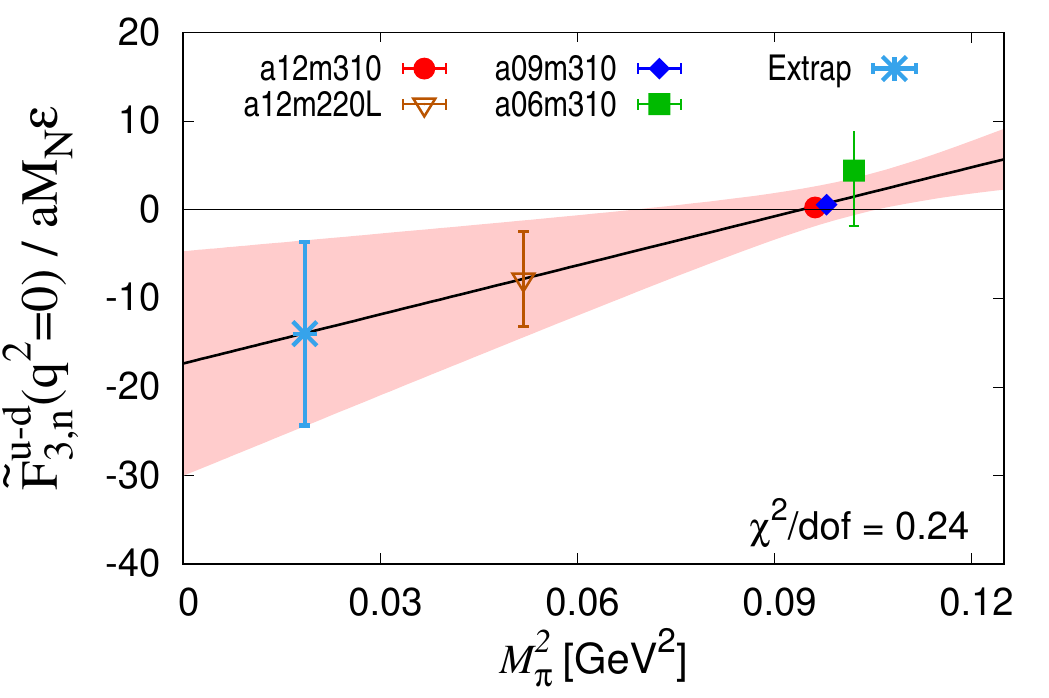}
  
  \caption{Extrapolation to the continuum and physical pion mass limit using the fit ansatz ${c_1 + c_2 M_\pi^2 + c_3 a}$. Results from different ensembles are renormalized in $\overline{\rm MS}$ scheme at $\mu=2$~GeV using \cref{eq:renorm}.  Left column shows the results from the ``Standard'' excited state fit, and the right column shows the results from the ``$N\pi$'' excited state fits. Extrapolated values are ${\tilde{F}_3(0) / aM_N\varepsilon = -2.6(2.9)}$ for the ``Standard'' excited state fit, and ${\tilde{F}_3(0) / aM_N\varepsilon = -14(10)}$ for the ``$N\pi$'' excited state fit.}
  \label{fig:F3cont}
\end{figure*}

The resulting renormalized values for $X_c$  are given in \cref{tab:renres} for two ways of removing ESC--with and without including an $N \pi$ excited state. Their extrapolation, linear in \(a\) and \(M_\pi\), to the physical point is shown in \cref{fig:F3cont}. The data show no significant dependence on the lattice spacing $a$. The dependence on $M_\pi^2$ is much larger with the $N \pi$ analysis, however, it is important to note that this chiral behavior is predicated on a single point, i.e., $a12m220L$. The final results in the continuum limit are $X_c \equiv -{{F}_3(0) / aM_N\varepsilon = 2.6(2.9)}$ for the ``Standard'' excited state fit, and ${14(10)}$ for the ``$N\pi$'' excited state fit, where the quoted errors are statistical. 

\section{Conclusions}
\label{sec:conclusions} 

In the analysis of the contribution of the qcEDM operator to nEDM $d_n$ presented here, we have focused on the issue of the power-divergent mixing of the qcEDM operator with the quark-pseudoscalar operator.  This mixing is independent of the explicit breaking of the chiral symmetry on the lattice, and is generic to all hard cutoff schemes.  For the isovector case, the pseudoscalar operator gives no physical effects in the continuum limit; at finite lattice spacing with Wilson-clover fermions, however, it results in an effect proportional to the qcEDM operator itself. This finite lattice spacing artifact is seen in explicit calculations with the isovector pseudoscalar operator, which gives a large EDM signal on the lattice. More importantly, this proportionality turns the divergent mixing into an extra finite multiplicative renormalization of the qcEDM operator, an effect that survives in the continuum limit if the theory is not fully \(O(a)\) improved.

We further show that the uncertainty in this finite renormalization constant  is substantial if \(c_{\rm SW}\) is tuned using tree-level tadpole improvement, i.e., not fully $O(a)$ improved. Any residual, even small, \(O(a)\) contribution gets divided by \(2am\), an equally small number. We have devised a scheme to determine this finite constant nonperturbatively provided we can ignore \(O(a^2)\) corrections.

Unfortunately, our data show that at close to physical masses and at the values of \(a\) used in this study, the \(O(a^2)\) corrections can be as large as \(\approx25\%\). This leads to an irreducible uncertainty in the results, which is in addition to the uncertainty due to the continuum and chiral extrapolations. Using techniques like gradient-flow that remove chiral symmetry breaking, this source of uncertainty can be avoided.

A $\chi$PT analysis presented in \cref{sec:alphadet} predicts the ratio of the phase $\alpha$ generated on the insertion of the qcEDM and pseudoscalar operators given in~\cref{eq:Acheck}. Our data on the four ensembles (see \cref{fig:alpha_cedm_g5}) show agreement with this analysis within ten percent.

To study excited state contributions in the extraction of ground-state matrix elements of the  qcEDM and pseudoscalar operators, we have made fits with two choices of the mass of the first excited state, that from fits to the nucleon 2-point correlation function and the $N \pi$ state. The fits cannot be distinguished by the $\chi^2/dof$, however, the results differ by a factor of five. 

To obtain a result for the isovector 
contribution of the qcEDM operator to nEDM, future work needs to address two challenges, in addition to issues of chiral and continuum extrapolation, exposed by this study: (i) possibly large $O(a^2)$ effects as discussed in~\cref{sec:Num} and (ii) the difference in estimates between removing ESC with and without including $N\pi$ excited states in the spectral decomposition of the correlation functions as discussed in~\cref{sec:mixing}. 

\vspace{0.5cm}
\noindent {\textbf {Acknowledgements}}: We thank the MILC collaboration for providing the HISQ lattices. The calculations used the  CHROMA software suite~\cite{Edwards:2004sx}. Simulations were carried out at (i) the NERSC supported by DOE under Contract No. DE-AC02-05CH11231;  (ii) the Oak Ridge Leadership Computing Facility, which is a DOE Office of Science User Facility supported under Award No. DE-AC05-00OR22725 through the INCITE program project HEP133, (iii) the USQCD collaboration resources funded by DOE HEP,  and (iv) Institutional Computing at Los Alamos National Laboratory. This work was supported by LANL LDRD program. TB, RG and EM were also supported by the DOE HEP and NP  under Contract No. DE-AC52-06NA25396. 
VC acknowledges support by the U.S. DOE under Grant No. DE-FG02-00ER41132.

\onecolumngrid
\clearpage
\iftwocolumn
\twocolumngrid
\fi

\appendix

\section{Nonsinglet Axial Ward Identity at \texorpdfstring{\(O(a)\)}{O(a)}}
\label{sec:AWI_app}

The starting point of the demonstration that, on-shell, the effect of the isovector pseudoscalar operator is 
proportional to qcEDM is the nonsinglet axial Ward Identity (AWI) obtained by considering the 
axial transformation on the quark fields $\psi^T = (u,d,s,c)$ defined in \cref{eq:axialtrans}.
We will be concerned with a rotation of the \(u\) and \(d\) quarks by equal and opposite amount, {\it i.e.,} $T^a={\rm diag}(1,-1,0,0)/2$, but for now we keep the notation generic. 
Denoting by $O(x_1, ..., x_n)$  any product of local operators, the  nonsinglet AWI reads 
\begin{align}
\big\langle O(x_1,..., x_n)  \times{}\hspace*{0.32\textwidth}\span\nonumber\\
\span
\Big(   \partial_x^\mu A_\mu^a (x) 
-   \bar{\psi} (x)  \{ m_W , T^a \} \gamma_5 \psi  (x)  - X^a  (x)  \Big) \big \rangle  \nonumber\\
\qquad\qquad\qquad\qquad &= - \left \langle
\frac{\delta O (x_1, ..., x_n)}{\delta (i \xi^a (x))} 
\right \rangle ~,
\label{eq:AWI1}
\end{align}
where  
\begin{equation}
A_\mu^{a} (x) =  \bar{\psi}  (x) T^a \gamma_\mu \gamma_5 \psi  (x)  
\end{equation}
and $X^a (x)$ is given by the variation of the Wilson-Clover term~\cite{Karsten:1980wd,Bochicchio:1985xa,Guadagnoli:2003zq}, see 
\cref{eq:DO1}, 
\begin{equation}
\frac{X^a}{2} =  -  a   \bar{\psi}\, T^a  \Big( \frac{r}{2}  D^2 + \kappa_{SW} \sigma \cdot G\Big) \gamma_5 \,  \psi  ~.
\label{eq:id2}
\end{equation}
Insertions of $X^a (x)$ vanish at tree level in the continuum limit, but quantum effects induce power-divergent mixing with lower dimensional operators, 
that has to taken into account when taking the continuum limit.    
This is done by writing ~\cite{Karsten:1980wd,Bochicchio:1985xa,Guadagnoli:2003zq}
\begin{align}
X^a (x) &= a   \tilde{X}^a  (x)  -   \bar \psi (x)  \left\{ T^a, m_{\rm sub} \right\}  \gamma_5 \psi (x)  - {}\nonumber\\
&\quad\qquad\qquad( Z_A-1)   \partial_x^\mu A_\mu^a (x)  ~, 
\label{eq:X1}
\end{align}
where $\tilde{X}^a(x) $  is a  `subtracted'  dimension-five operator,  {\it i.e.,} lower-dimensional operators are subtracted from it so that it is free  of power divergences and the  Green's functions 
of $a \tilde X^a (x)$  with  elementary fields  vanish in the continuum limit\punctfootnote,{This is a convenient definition when the anomalous dimensions are perturbatively small, as in our case.  In general, one could frame the discussion completely in terms of the running of `renormalized' operators, \(Z_X \tilde{X}^a\), defined such that their Green's functions with renormalized elementary fields have finite continuum limits.} and \(m_{\rm sub}\) is a `mass' counterterm for the fermion that arises from the explicit chiral symmetry breaking in Wilson fermions\punctfootnote.{Note that \(m_{\rm sub}\) depends on all parameters in the action including \(m_W\). The critical mass \(m_{\rm crit}\) defining a massless theory is obtained by \(m_{\rm sub}|_{m_W=m_{\rm crit}} = m_{\rm crit}\). One can also show~\cite{Bhattacharya:2000pn} that \(m_W-m_{\rm sub} = Z_P Z_m (m_W-m_{\rm crit})\), which can be used to relate the masses appearing in the Axial and Vector Ward Identities.}
$a \tilde{X}^a(x) $ has no impact on the analysis of the axial WI with elementary fields, 
 though   it induces contact terms in the continuum limit of axial WIs involving composite fields~\cite{Bochicchio:1985xa, Testa:1998ez}.
 $\tilde X^a(x)$ is however essential for our analysis of the {qcEDM} at finite $a$.
Using the above expression in \cref{eq:AWI1}, one arrives at 
\begin{align}
  \Big\langle O(x_1,..., x_n)  \Big(  Z_A  \partial_x^\mu A_\mu^a (x) 
- {}\hspace*{0.15\textwidth}\span\nonumber\\
\span\bar{\psi} (x)  \{ m , T^a \} \gamma_5 \psi  (x)  
- 
a \tilde{X}^a (x) 
 \Big) \Big\rangle
\nonumber \\
\qquad\qquad\qquad\quad&= -   \left \langle   \frac{\delta O (x_1, ..., x_n)}{\delta (i \xi^a (x))}  
\right \rangle\,, 
\label{eq:AWI3n}
\end{align}
with 
\begin{equation}
m = m_W - m_{\rm sub}~.
\end{equation}
Note that we can define the flavor-diagonal matrix that appears in the pseudoscalar operator on the LHS as 
{ \(  \{ m , T^a \} \equiv  Z_A \bar m  + O(a m^2)\)}, where \(\bar m\) is the standard definition~\cite{Bhattacharya:2000pn} of the quark mass from Axial Ward Identity.

Next, we  project the subtracted operator $\tilde X^a$  on the basis of subtracted {hermitian} dimension-5 operators $O_{n}^{(5)}$, 
given in Ref.~\citep{Bhattacharya:2015rsa}, 
\begin{equation}
\tilde X^a = 
{i} \sum_n
K_{Xn}
\tilde O_n^{(5)} 
\label{eq:Xtilde1}
\end{equation}
and  analyze the consequences of \cref{eq:Xtilde1}   for \cref{eq:AWI3n}. 
The basis of unsubtracted dimension-5 operators  $O_n^{(5)}$   appearing on the RHS of \cref{eq:Xtilde1}  for generic nonsinglet generator $T^a$  and generic diagonal quark mass $m$,  is given in Ref.~\cite{Bhattacharya:2015rsa}. As discussed in \cref{sec:operatorbasis}, in our situation, however, only one subtraction coefficient, \(K_{X1}\) is needed: up to corrections of 
$O(a^2 m)$  and $O(a \,  \alpha_{\rm EM}/\pi)$, 
\cref{eq:AWI3n} becomes  
\begin{align}
\Big\langle O(x_1,..., x_n)  
  \Big[
   Z_A \, ( 1 + b_A  m a ) \,  \partial_x^\mu A_\mu^a (x)  - {}\hspace*{0.05\textwidth}\span\nonumber\\
   \span a {Z_A} c_A \, \partial^2  \left( \bar \psi T^a \gamma_5 \psi\right)
 - \bar{\psi} (x)  \{ m , T^a \} \gamma_5 \psi  (x)   - {}
\nonumber \\
\qquad\qquad\qquad\qquad& a i  K_{X1} \ \tilde C^{(a)} 
 \Big] \Big \rangle \nonumber\\
&= - 
\Big \langle   \frac{\delta O (x_1, ..., x_n)}{\delta (i \xi^a (x))}  
\Big \rangle ~. 
\label{eq:AWI4n}
\end{align}
A detailed analysis in \cref{sec:KX1} shows that   
the proportionality coefficient $K_{X1}$ is given by 
\begin{align}
K_{X1} 
&= +\frac{r}{2}  \ 
\Big( 
c_{SW} - 1 - 2 \beta_1^{(5)} (g) 
\Big)\nonumber\\
\beta_1^{(5)} (g)  &= a_2 g^2 + O(g^4)~, 
\label{eq:KX1}
\end{align}
and starts at $O(g^2)$. 
Finally, upon integration over $\int d^4x$, \cref{eq:AWI4n} gives the final result in \cref{eq:AWI4.5n}. 

\section{Dimension-5 Operators}
\label{sec:operatorbasis}

To derive \cref{eq:AWI4n} from \cref{eq:AWI3n}, we need to show that only four 
of the following full list of dimen\-sion-5 CPV operators contribute\footnote{{We provide here the Euclidean version of the basis.}}~\cite{Bhattacharya:2015rsa}: 
\begin{subequations}
\begin{align}
O^{(5)}_1   &\equiv  C^{(a)} 
= i\,  \bar\psi \sigma^{\mu\nu} \gamma_5 G_{\mu\nu}  T^a\psi 
 \label{eq:CEDMdef} \\
O^{(5)}_2
&\equiv   \partial^2 P^{(a)}
=   \partial^2  \left(   \bar\psi i\gamma_5  T^a \psi \right) 
\\
 O^{(5)}_3 
  &\equiv   
 E^{(3)} = 
\frac{i e}{2} \,  \bar\psi\sigma^{\mu\nu}F_{\mu\nu} \{Q, T^a\} \psi 
\label{eq:O3def}
\\
O^{(5)}_4 &=
 \textrm{Tr} \left[ m Q^2  T^a \right] \,
  \frac{{i}}{2} \epsilon^{\mu \nu \alpha \beta}  F_{\mu \nu}  F_{\alpha \beta}
\\
O^{(5)}_5  &=      
 \textrm{Tr} \left[  m  T^a \right] \,
  \frac{{i}}{2} \epsilon^{\mu \nu \alpha \beta}  G^b_{\mu \nu}  G^b_{\alpha \beta}
\\
O^{(5)}_6  &=   
{i}\, 
\textrm{Tr}\left[ m T^a\right]  \partial_\mu  \left( \bar\psi\gamma^\mu\gamma_5   \psi \right)
\\
O^{(5)}_7  &= 
\frac{{i}}{2} \partial_\mu  \left(  \bar\psi\gamma^\mu\gamma_5   \left\{   m, T^a \right\} \psi \right) \nonumber\\
&{}- \frac{{i}}{3} \textrm{Tr}\left[ m T^a\right]  \partial_\mu  \left( \bar\psi\gamma^\mu\gamma_5   \psi \right)
\\
O^{(5)}_8   &=   
\frac{1}{2}  \, \bar\psi i\gamma_5    \left\{  m^2, T^a \right\}   \psi
\\
O^{(5)}_9   &= 
\textrm{Tr} \left[ m^2 \right]  \ \bar\psi i\gamma_5 T^a  \psi
\\
O^{(5)}_{10}   &= 
\textrm{Tr} \left[ m T^a \right]  \ \bar\psi i\gamma_5   m  \psi
\\
O^{(5)}_{11} &=    
i\bar\psi_E\gamma_5 T^a \psi_E 
\\
O^{(5)}_{12} &=
 \partial_\mu[\bar\psi_E\gamma^\mu\gamma_5 T^a  \psi+{}
 \bar\psi\gamma^\mu\gamma_5 T^a  \psi_E]
\\ 
O^{(5)}_{13}  &=
  \bar\psi  \gamma_5  \slashed{\partial} T^a   \psi_E  + {\rm h.c.}
  \\ 
O^{(5)}_{14} &= 
\frac{ i  e}{2} \
\bar\psi    \{ Q, T^a \}  \Aslash^{(\gamma)}  \gamma_5\psi_E  + {\rm h.c.}\,. 
\end{align}
\label{eq:D5ops}
\end{subequations}

Here we have used the notation $\psi_E \equiv (\slashed{D} + m) \psi$.  To simplify the 
discussion, we will start by assuming that the mass matrix is proportional to the identity and  
point out the minor modifications later on. 
Keeping in mind that $O(x_1, ..., x_n)$ has the structure $N(x_1) \, J_{\rm EM}^\mu (x_2) \, \bar N(x_3)$ with $N$ the neutron source and sink operator and $J_{\rm EM}^\mu$ the electromagnetic current, 
the various  $\tilde O_n^{(5)}$  contribute to \cref{eq:AWI3n} as follows:

\begin{itemize}
\item  $O^{(5)}_1$ is the isovector chromo-EDM operator itself and contributes an $O(a)$ term  to the LHS of \cref{eq:AWI3n}. 
In fact, as shown below, this is the leading $O(a)$ contribution. 

\item 
{
Insertions of the  operators  $O^{(5)}_{2,6,7}$ in  \cref{eq:AWI3n} effectively amount to an $O(a)$ shift of the axial current, 
which (up to corrections of $O(a (m_u - m_d))$ for $T^a = T^3$) can be parameterized  as 
\begin{align}
    Z_A  A^a_\mu  &\to(1 + b_A m a) Z_A  \times{}\nonumber\\
\qquad\qquad
\span \left[ A^a_\mu  - a c_A \partial_\mu  \left( \bar \psi T^a \gamma_5 \psi\right)\right] 
\end{align}
{up to \(O(a^2)\)}, where $m$ denotes the light quark mass. In short, the three $K_{X2},K_{X6}$ and $K_{X7}$ are 
reduced to $b_A$ and $c_A$.
}

\item  $O^{(5)}_{3,4}$ involve one and two powers of the electromagnetic field strength. In order to eliminate the photon field 
in the correlation functions in \cref{eq:AWI3n}, one needs electromagnetic loops, making the 
contribution of $O^{(5)}_{3,4}$ to \cref{eq:AWI3n} of $O(a \,  \alpha_{\rm EM}/\pi)$, and thus negligible to the order we are working.

\item $O^{(5)}_5$  
vanishes under the assumption that $m \propto I$. It otherwise provides a  term  of $O(am)$ to the 
LHS of \cref{eq:AWI3n}. 
{In the case of the isovector operator, the 
effect is $O(a (m_u - m_d))$ and hence negligible}.

\item  $O^{(5)}_{8,9}$  become $m^2  \bar\psi i\gamma_5  t^a \psi $ when  $m \propto I$.  Therefore, their 
  contributions have the same form of  the pseudoscalar insertion in  \cref{eq:AWI3n}, but suppressed  by  $O(a m)$.

\item $O^{(5)}_{10}$  vanishes under the assumption $m \propto I$.   In the case of general flavor structure for $m$, 
 $O^{(5)}_{10}$   contribute  terms of  $O(a m^2)$ 
to   \cref{eq:AWI3n}.  
{When considering isovector insertions, 
the new contribution scales as $O(a m (m_u - m_d))$, which 
can be safely neglected.}

\item The operators $O^{(5)}_{11,12,13,14}$ vanish by using the  quark equations of motion but can contribute contact terms to the 
LHS of \cref{eq:AWI3n}. 
However, it turns out that none of them actually contributes to the  order we are working.  
$O^{(5)}_{11}$ contains two equation of motion operators. 
Therefore, when inserted into \cref{eq:AWI3n}, it will always involve a contraction with a quark field in the neutron source or sink operator, 
and thus it will not contribute to the residue of the neutron pole. 
$O^{(5)}_{12}$ is a total derivative and drops out of  \cref{eq:AWI3n}. 
$O^{(5)}_{13}$ is gauge-variant operator and drops out of  \cref{eq:AWI3n} as long as $O (x_1, ... , x_n)$ is a gauge singlet, 
which is the case for  $O(x_1,x_2,x_3)  \propto N(x_1) \, J_{\rm EM}^\mu (x_2) \, \bar N(x_3)$.
$O^{(5)}_{14}$ involves the photon field and can contribute only 
at $O(a \alpha_{\rm EM}/\pi)$ to  \cref{eq:AWI3n}.

\end{itemize}

\section{Origin of the artifact \texorpdfstring{$K_{X1}$}{K\unichar{"2092}\unichar{"2081}}}
\label{sec:KX1}
\label{sec:xydet}

In this appendix, we give a more explicit form of 
 $\tilde{X}^a$ and identify the coefficient $K_{X1}$ given in \cref{eq:KX1}. This is 
done by manipulating the RHS of \cref{eq:id2} and comparing it to  \cref{eq:X1}.   
In order to re-express the first term on the RHS of \cref{eq:id2}, we note the identity
\begin{equation}
\bar{\psi}  T^a D^2 \gamma_5 \psi = 
\bar \psi  T^a \Big[ 
 \gamma_5 D_L^2  - \frac{1}{2} \sigma \cdot G \gamma_5 
\Big] \psi  - a O^a_6~, 
\label{eq:id4}
\end{equation}
where $D_L$ is defined in \cref{eq:DO1} and 
$O^a_6$ is a dimension-six operator 
with tree-level matrix elements of $O(a^0)$. 
We next introduce  subtracted  
operators  $\tilde{O}^a_6$  and ${- i} \tilde O_1^{(5)} \equiv (\bar \psi T^a  \sigma \cdot G \gamma_5 \psi)_{\rm sub}$ by introducing subtraction coefficeints $\beta$ and $\tilde\beta$  
as follows:
\begin{subequations}
\label{eq:id5}
\begin{align}
a O^a_6 &= a \tilde{O}^a_6  
+ \frac{{\beta}_1}{a^2} \bar \psi T^a \gamma_5 \psi 
+ \frac{{\beta}_4}{a}  \partial_\mu A^a_\mu\nonumber\\
&{-i}  \sum_n \, \beta_n^{(5)} \tilde O_n^{(5)} 
\qquad
\\
{-i} O_1^{(5)} 
 &= 
{-i} \tilde O_1^{(5)} 
+ \frac{\tilde{\beta}_1}{a^2} \bar \psi  T^a \gamma_5 \psi 
+ \frac{\tilde{\beta}_4}{a}  \partial_\mu A^a_\mu ~,\nonumber\\
\end{align}
\end{subequations}
where the $\tilde{O}_n^{(5)}$ are the subtracted versions of 
$O_n^{(5)}$ given in \cref{eq:D5ops}. 
By using \cref{eq:id5} in \cref{eq:id4} 
and defining
\begin{equation}
O_{EOM}^a = \, \bar \psi  T^a  \gamma_5 (D_L - m_W) (D_L + m_W)  \psi , 
\end{equation}
one arrives at:
\begin{widetext}
\vspace*{-10pt}
\begin{align}
a \bar \psi  T^a D^2 \gamma_5 \psi &= 
- a^2  \tilde{O}_6^a  
+   \ a \, O_{EOM}^a
+ i a \left( \frac{1}{2} + \beta_1^{(5)} \right) \tilde O_1^{(5)}  
+ i  a \sum_{n \neq 1} \beta_n^{(5)} \tilde O_n^{(5)} 
\nonumber \\  
& \qquad {}-  \left( \beta_4 + \frac{\tilde \beta_4}{2} \right)  \partial_\mu A^a_\mu 
- \frac{1}{a}  \left( \beta_1 + \frac{\tilde \beta_1}{2} - (a m_W)^2 \right)  \ \bar \psi  T^a \gamma_5 \psi ~. 
\label{eq:id6}
\end{align}

Using the above results one can write $X^a$ as follows:
\begin{align}
X^a &=   r a^2 \tilde O_6^a
{{} + i} \frac{r}{2} \left( c_{SW} - 1- 2 \beta_1^{(5)} \right) \, a \tilde O_1^{(5)}
{{}-i} a r \sum_{n \neq 1} \beta_n^{(5)} \tilde O_n^{(5)} 
- a r O_{EOM}^a
\nonumber \\
& 
- \frac{r}{2}  \left( \tilde \beta _4 ( c_{SW} - 1) - 2 \beta_4 \right) \, \partial_\mu A^{a \mu} 
- \frac{r}{2a}  \left( 
\tilde \beta _1 ( c_{SW} - 1) - 2 \beta_1 + 2 (a m_W)^2 
\right)
\bar \psi  T^a \gamma_5 \psi ~. 
\label{eq:id7}
\end{align}
\end{widetext}

Comparing \cref{eq:X1,eq:id7} we see that 
the first line of \cref{eq:id7} provides
an explicit representation for the subtracted  dimension-five operator  $a \tilde{X}^a$ appearing 
in \cref{eq:X1} with the identification $ (r/2) ( c_{SW} - 1- 2 \beta_1^{(5)} ) = K_{X1} $ as in \cref{eq:KX1}.

\section{Determination of \texorpdfstring{\(\alpha_N\)}{\textalpha\_N} using \texorpdfstring{$\chi$PT}{\textchi PT}}
\label{sec:alphadet}

\begin{table}[t] 
    \centering
   \begin{tabular}{|l|rll|}
   \hline
   Ens.~ID&\multicolumn1c{\(\alpha^{\rm chiral}_5/\epsilon_5\)}&\multicolumn1c{\(am\)}&\multicolumn1{c|}{\(\alpha_5 m a /\epsilon_5\)}\\
   \hline
  a12m310&\(-41.260(76)\)&0.012106(92)&\(-0.4995(39)\)\\
  a12m220L&\(-85.19(28)\)&0.006178(46)&\(-0.5263(43)\)\\
  a09m310&\(-55.80(12)\)&0.008472(24)&\(-0.4727(17)\)\\
   a06m310&\(-91.60(52)\)&0.0052860(99)&\(-0.4842(29)\)\\
   \hline
    \end{tabular}
    \caption{Verifying \(\chi\)PT for \(\alpha_5^{\rm chiral}\).}
    \label{tab:chiral_alpha5}
\end{table}

Consider computing 
correlations functions of the local nucleon interpolating field
\begin{equation}
    N_{\rm chiral}(x) = \varepsilon^{a b c} q^{a T}(x) C \gamma_5 i \tau_2 q^b(x) \, q^c(x).
    \label{eq:Nordinary}
\end{equation}
in the presence of \CPV\ isovector pseudoscalar and chromo-electric terms in the Lagrangian 
\begin{equation}
    -\frac{\epsilon_5}{a}   \bar q i \gamma_5 \tau_3 q  - \frac{1}{4} (a r) \epsilon \bar q i \gamma_5 \sigma^{\mu\nu} \gamma_5 G_{\mu \nu} \tau_3 q.    
\end{equation} 
As discussed in \citep{Bar:2015zwa,Nagata:2008zzc},
this field has good chiral properties, and transforms linearly under an isovector axial rotation.
In particular, it could eliminate the pseudoscalar interaction from the Lagrangian. The only effect would be to replace 
\begin{equation}
    N_{\rm chiral}(x) \rightarrow (1 + i \frac{\epsilon_5 }{ 2 \bar m a} \gamma_5 \tau_3) N_{\rm chiral}(x),
\end{equation}
plus $\mathcal O(a)$ corrections. In $\chi$PT, the nucleon field $N_\chi$ 
transforms as~\cite{Bar:2015zwa}
\begin{equation}\label{eq:bar}
    N_{\chi} \to  \left(1 + i \frac{\boldsymbol{\pi} \cdot \boldsymbol{\tau}  }{2 F_\pi }\right) N_\chi(x),
\end{equation}

Similarly, we build the chiral Lagrangian to include the contribution of the pseudoscalar and chromo-electric interactions to the correlation functions. 
At lowest order, these interactions induce pion tadpole terms, of the form
\begin{equation}
\mathcal L_\pi =  m_\pi^2 \left( \frac{\epsilon_5}{m a}  +   \epsilon 
 \frac{r}{4 m a}  
 \tilde r \right) F_\pi \pi_3,   
 \label{eq:tadpole}
\end{equation}
where $\tilde r$ is the ratio of the vacuum matrix elements of the chromo-magnetic operator and scalar density
\begin{align}
     \tilde r &= \frac{a^2 \langle 0 | \bar \psi \sigma^{\mu \nu} G_{\mu \nu} \psi | 0 \rangle }{\langle 0 | \bar \psi \psi | 0\rangle} =      \mathcal O\left( a^2 \Lambda_\chi^2\right) + \mathcal O(\alpha_s).\nonumber\\
     \label{eq:rtilde}
 \end{align}
The $\mathcal O(\alpha_s)$ corrections arise from the power-divergent mixing of the chromo-magnetic and scalar operators and
depend on the regularization and renormalization scheme chosen for the chromomagnetic operator, and typically suffers from renormalon ambiguities when calculated perturbatively. This power-divergent subtraction is present in hard-cutoff schemes like the lattice or gradient-flow, but is not needed in dimensional regularization.  
In the $\overline{\textrm{MS}}$ scheme, $\tilde{r}$ is, therefore, related only to $m_0^2$, the ratio of chromomagnetic and scalar condensates  typically used in QCD sum rules literature \cite{etde_6260715,Ioffe:1983ju,Pospelov:2000bw,Pospelov:2005pr,Gubler:2018ctz} by
$ \tilde r = a^2 m_0^2$ noting that $g\, G^{\mu\nu}_{\rm sum \ rule} = G^{\mu\nu}_{\rm our\ definition}$. The sum rule estimate is  \(m_0^2\approx0.8\ {\rm GeV}^2\). Only preliminary lattice QCD calculations of this ratio in the gradient-flow scheme exist~\cite{Kim:2021qae,Gubler:2018ctz}. In our calculations, we, however, use a subtracted qcEDM operator \({\tilde C}_3\) (see \cref{eq:Adetermination}) which has \(\tilde r|_{{\tilde C}_3}=0\) at leading order in chiral perturbation heory.  For the isovector case, such a subtraction does not change any physical matrix elements in the continuum theory, but it does affect the phase \(\alpha_N\) that depends on the interpolating operator.

The leading contribution to this phase comes from diagrams in which a pion is emitted by the nucleon interpolating field given in \cref{eq:bar}, and annihilated by \cref{eq:tadpole}, leading to
\begin{equation}
    \alpha_N = -\frac{1}{2}  \left( \frac{\epsilon_5}{m a}  +   \epsilon 
 \frac{r}{4 m a}   
 \left(  \tilde r + \mathcal O\left( a^2 m^2_\pi\right) \right)\right),
 \label{eq:chiPT_alpha}
\end{equation}
where the corrections arise from subleading 
pion- and pion-nucleon interactions induced by the chromo-electric operator,
and depend on additional nonperturbative matrix elements of the chromo-electric/chromo-magnetic operators. In Eq. \eqref{eq:chiPT_alpha}, we assumed that the \(O(\alpha_s)\) terms in \cref{eq:rtilde} are smaller of comparable to the \(O(a^2)\) piece. 

As discussed above, in \cref{sec:AWI}, we defined a subtracted  chromo-electric operator by imposing its  matrix element between a pion and vacuum state to vanish. The subtraction leads to an \(\epsilon_5 \equiv \epsilon/A\) in \cref{eq:chiPT_alpha} and the  \(\tilde r|_{\tilde C_3}\) relevant to the  subtracted operator reduces to zero.  This means that the ratio between the phases induced by the subtracted and unsubtracted operators, which we denote by $\tilde\alpha_N$
and $\alpha_N$ respectively, is
\begin{equation}
\frac{\tilde\alpha_N}{\alpha_N} \sim \frac{a^2 m_\pi^2}{\tilde r} = \mathcal O\left(\frac{m_\pi^2}{\Lambda_\chi^2}\right), 
\label{eq:D8}
\end{equation}
so that $\tilde{\alpha}_N$ is a ${}\sim 10\%$ correction to the phase obtained from the subtraction piece alone. 
This expectation is confirmed by the explicit calculation illustrated in \cref{fig:alpha_cedm_g5}.

With just the pseudoscalar operator (first term in \cref{eq:tadpole}),  one gets \(\alpha^{\rm chiral}_5 = -\epsilon_5/2 m a\) from \cref{eq:chiPT_alpha} if the nucleon interpolating operator has the same 
chiral properties as the operator in \cref{eq:Nordinary}.  In 
our calculations, the quark fields $q$ are smeared and the source used is 
\begin{equation}
    N(x) = \varepsilon^{a b c} q^{a T}(x) C \gamma_5 i \tau_2 \frac{1+\gamma_4}2 q^b(x) \, q^c(x)\,,
\end{equation}
which suppresses parity-mixing. Consequently, one expects a smaller \(\alpha_5\).  To check the chiral analysis, we calculated the two-point function with the \(N_{\rm chiral}\) interpolating operator, but with smeared quark fields. Instead of \(r_\alpha\), we also used
\begin{align}
\bar r_\alpha (\tau) &\equiv\frac{\Im \Tr \gamma_5 \langle N(0) \overline N(\tau)\rangle}
         {\Re \Tr \langle N(0) \overline N(\tau)\rangle}\nonumber\\
         &\xrightarrow{\tau\to\infty} \tan 2\alpha_0
\end{align}
for this calculation.
The results in \cref{tab:chiral_alpha5} show that $\alpha_5 ma/\epsilon_5$ remains close to  its value $ -0.5$ in \cref{eq:chiPT_alpha}, i.e., smearing has a small effect on the chiral analysis.

\bibliographystyle{apsrev4-2-errat} 
\showtitleinbib
\bibliography{ref} 
\onecolumngrid
\iftwocolumn\twocolumngrid\fi
\end{document}